\documentclass[11pt]{article}
\usepackage[utf8]{inputenc}
\usepackage[T1]{fontenc}
\usepackage{amsmath,amssymb,amsthm}
\usepackage{graphicx}
\usepackage{booktabs}
\usepackage{array}
\usepackage{xcolor}
\usepackage{listings}
\usepackage[margin=2.4cm]{geometry}
\usepackage{hyperref}

\DeclareMathOperator{\Tr}{Tr}
\newcommand{\C}{\mathbb{C}}
\newcommand{\M}{\mathcal{M}}
\newtheorem{proposition}{Proposition}

\definecolor{lstcomment}{HTML}{7D848C}
\definecolor{lstkeyword}{HTML}{2A78D6}
\newcommand{\repofile}[1]{\href{https://github.com/andre-a-no/von_neumann_type_i/blob/main/man/article/listings/#1}{\texttt{listings/\allowbreak\detokenize{#1}}}}

\newcommand{\ValDevice}{NVIDIA GeForce RTX 5090}
\newcommand{\ValMode}{full}
\newcommand{\ValTorch}{2.14.1+cu130}

\newcommand{\ValEmissionExp}{2.2\times10^{-16}}
\newcommand{\ValEmissionRK}{2.1\times10^{-8}}
\newcommand{\ValNce}{64}
\newcommand{\ValSce}{20000}
\newcommand{\ValSHaar}{400000}
\newcommand{\ValSPage}{50000}
\newcommand{\ValSW}{400000}
\newcommand{\ValSPur}{100000}

\newcommand{\ValLoneOne}{0.022}
\newcommand{\ValLoneTwo}{0.007}
\newcommand{\ValLoneFour}{0.006}
\newcommand{\ValTraceNormTau}{9.0\times10^{-16}}
\newcommand{\ValTraceRows}{%
$\Tr_{\mathrm{blunt}}$ & $0$ & $4.0\times10^{-15}$ & $5.9\times10^{-15}$ & 1.000 \\
$\Tr_{\mathrm{norm}}$ & $0$ & $1.4\times10^{-15}$ & $1.9\times10^{-15}$ & 0.223 \\
$\tau_{\mathrm{vN}}$ & $0$ & $4.4\times10^{-16}$ & $6.1\times10^{-16}$ & 0.074 \\
}
\newcommand{\ValidationRows}{%
$\mathbb E|U_{11}|^4$, $U(n)$ & $n=2$ & 0.333333 & 0.333604 & $4.7\times10^{-4}$ & +0.6 \\
$\mathbb E|U_{11}|^2$, $U(n)$ & $n=2$ & 0.5 & 0.500277 & $4.6\times10^{-4}$ & +0.6 \\
$\mathbb E|U_{11}|^4$, $U(n)$ & $n=4$ & 0.1 & 0.10028 & $2.2\times10^{-4}$ & +1.3 \\
$\mathbb E|U_{11}|^4$, $U(n)$ & $n=8$ & 0.0277778 & 0.0277966 & $7.5\times10^{-5}$ & +0.3 \\
$\mathbb E|U_{11}|^4$, $U(n)$ & $n=16$ & 0.00735294 & 0.00735803 & $2.3\times10^{-5}$ & +0.2 \\
$\mathbb E|U_{11}|^4$, $U(n)$ & $n=32$ & 0.00189394 & 0.00189625 & $6.2\times10^{-6}$ & +0.4 \\
$\mathbb E|U_{11}|^2$, $U(n)$ & $n=32$ & 0.03125 & 0.0312403 & $4.8\times10^{-5}$ & -0.2 \\
$\mathbb E|U_{11}|^4$, $O(n)$ & $n=2$ & 0.375 & 0.375106 & $5.8\times10^{-4}$ & +0.2 \\
$\mathbb E|U_{11}|^2$, $O(n)$ & $n=2$ & 0.5 & 0.500155 & $5.6\times10^{-4}$ & +0.3 \\
$\mathbb E|U_{11}|^4$, $O(n)$ & $n=4$ & 0.125 & 0.125085 & $3.1\times10^{-4}$ & +0.3 \\
$\mathbb E|U_{11}|^4$, $O(n)$ & $n=8$ & 0.0375 & 0.0376573 & $1.3\times10^{-4}$ & +1.2 \\
$\mathbb E|U_{11}|^4$, $O(n)$ & $n=16$ & 0.0104167 & 0.0104046 & $4.3\times10^{-5}$ & -0.3 \\
$\mathbb E|U_{11}|^4$, $O(n)$ & $n=32$ & 0.00275735 & 0.0027558 & $1.2\times10^{-5}$ & -0.1 \\
$\mathbb E|U_{11}|^2$, $O(n)$ & $n=32$ & 0.03125 & 0.0312731 & $6.7\times10^{-5}$ & +0.3 \\
$\mathbb E\,\tau((AUBU^*)^2)$ & $n=4$ & -0.0666667 & -0.0661548 & $6.3\times10^{-4}$ & +0.8 \\
$\mathbb E\,\tau((AUBU^*)^2)$ & $n=8$ & -0.015873 & -0.0159874 & $2.8\times10^{-4}$ & -0.4 \\
$\mathbb E\,\tau((AUBU^*)^2)$ & $n=16$ & -0.00392157 & -0.00414221 & $1.4\times10^{-4}$ & -1.6 \\
$\langle s^2\rangle$, COE & $n=64$ & 1.27324 & 1.28582 & $3.7\times10^{-4}$ & -- \\
$\langle s^2\rangle$, CUE & $n=64$ & 1.1781 & 1.18024 & $2.4\times10^{-4}$ & -- \\
$\langle s^2\rangle$, CSE & $n=64$ & 1.10447 & 1.10392 & $2.0\times10^{-4}$ & -- \\
$\langle S(\rho_A)\rangle$ & $m=2, n=16$ & 0.646516 & 0.646516 & $1.6\times10^{-4}$ & 0.0 \\
$\langle S(\rho_A)\rangle$ & $m=8, n=16$ & 1.83367 & 1.83388 & $1.7\times10^{-4}$ & +1.2 \\
$\langle S(\rho_A)\rangle$ & $m=16, n=16$ & 2.27487 & 2.27495 & $1.4\times10^{-4}$ & +0.6 \\
$\mathbb E\Tr\rho^2$ & $k=4, r=2$ & 0.666667 & 0.666312 & $3.2\times10^{-4}$ & -1.1 \\
$\mathbb E\Tr\rho^2$ & $k=4, r=4$ & 0.470588 & 0.470523 & $2.1\times10^{-4}$ & -0.3 \\
$\mathbb E\Tr\rho^2$ & $k=16, r=4$ & 0.307692 & 0.307726 & $6.4\times10^{-5}$ & +0.5 \\
$\mathbb E\Tr\rho^2$ & $k=16, r=16$ & 0.124514 & 0.124527 & $1.7\times10^{-5}$ & +0.8 \\
}
\newcommand{\SearchDevice}{NVIDIA GeForce RTX 5090}
\newcommand{\SearchMode}{full}

\newcommand{\SearchSamples}{1\,000\,000}
\newcommand{\SearchStarts}{128}
\newcommand{\SearchSteps}{500}
\newcommand{\SearchPairs}{16}
\newcommand{\SearchMaxErr}{3\times10^{-5}}
\newcommand{\SearchRows}{%
2 & 14/16 & 16/16 & $7.1\times10^{-4}$ & $5.6\times10^{-4}$ & $1.2\times10^{-8}$ & $5.4\times10^{-6}$ & 64 & 2.8 \\
4 & 0/16 & 16/16 & $2.3\times10^{-3}$ & $-0.12$ & $4.4\times10^{-8}$ & $3.1\times10^{-6}$ & 66 & 2.8 \\
8 & 0/16 & 16/16 & $5.6\times10^{-3}$ & $-0.48$ & $2.9\times10^{-7}$ & $2.0\times10^{-6}$ & 70 & 2.8 \\
16 & 0/16 & 16/16 & $6.7\times10^{-3}$ & $-0.72$ & $2.4\times10^{-6}$ & $3.9\times10^{-6}$ & 88 & 2.7 \\
32 & 0/16 & 16/16 & $8.9\times10^{-3}$ & $-0.86$ & $7.4\times10^{-6}$ & $1.6\times10^{-5}$ & 1.8e+02 & 6.6 \\
64 & 0/16 & 16/16 & $7.0\times10^{-3}$ & $-0.93$ & $2.0\times10^{-5}$ & $2.8\times10^{-5}$ & 2.7e+02 & 11 \\
}

\newcommand{\ChainDevice}{NVIDIA GeForce RTX 5090}
\newcommand{\ChainMode}{full}

\newcommand{\ChainFFL}{12}
\newcommand{\ChainFFErr}{7.1\times10^{-15}}
\newcommand{\ChainSUErr}{2.9\times10^{-15}}
\newcommand{\ChainDWL}{16}
\newcommand{\ChainDWDim}{12870}
\newcommand{\ChainDWErr}{4.9\times10^{-15}}
\newcommand{\ChainPesL}{14}
\newcommand{\ChainPesErr}{3.8\times10^{-12}}
\newcommand{\ChainReal}{400}
\newcommand{\ChainKDWL}{22}
\newcommand{\ChainKDWDim}{705432}
\newcommand{\ChainKDWErr}{1.7\times10^{-14}}
\newcommand{\ChainMomL}{14}
\newcommand{\ChainMomErr}{3.6\times10^{-14}}

\newcommand{\ChainLossErr}{2.4\times10^{-10}}
\newcommand{\ChainKrylovRows}{%
XX open, $L=16$ & 12870 & -4.9189757237 & -4.9189757237 & $1.8\times10^{-15}$ & 0.14 \\
XX open, $L=20$ & 184756 & -6.1907449998 & -6.1907449998 & $8.9\times10^{-16}$ & 0.07 \\
XX open, $L=24$ & 2704156 & -7.4629855550 & -7.4629855550 & $2.7\times10^{-15}$ & 0.56 \\
Heisenberg ring, $L=16$ & 12870 & -7.1422963606 & -7.142296361 & $<5\times10^{-10}$ & 0.059 \\
}
\newcommand{\ChainRingRows}{%
4 & 6 & -2.000000000 & -2.000000000 & $<5\times10^{-10}$ & -0.50000 \\
6 & 20 & -2.802775638 & -2.802775638 & $<5\times10^{-10}$ & -0.46713 \\
8 & 70 & -3.651093409 & -3.651093409 & $<5\times10^{-10}$ & -0.45639 \\
10 & 252 & -4.515446354 & -4.515446354 & $<5\times10^{-10}$ & -0.45154 \\
12 & 924 & -5.387390917 & -5.387390917 & $<5\times10^{-10}$ & -0.44895 \\
14 & 3432 & -6.263549534 & -6.263549534 & $<5\times10^{-10}$ & -0.44740 \\
16 & 12870 & -7.142296361 & -7.142296361 & $<5\times10^{-10}$ & -0.44639 \\
}
\newcommand{\NumDevice}{NVIDIA GeForce RTX 5090}
\newcommand{\NumMode}{full}

\newcommand{\NumK}{32}
\newcommand{\NumC}{16}
\newcommand{\NumB}{256}
\newcommand{\NumRGOE}{0.5313}
\newcommand{\NumRGUE}{0.5993}
\newcommand{\NumTimeRows}{%
\texttt{float32} & 4 & 16 & 1 & 1 & 1 & 1 & 1 & 1 \\
\texttt{float64} & 8 & 32 & 1.2 & 7.7 & 21 & 3.7 & 2.1 & 3.4 \\
\texttt{complex64} & 8 & 32 & 1.1 & 1.6 & 1.8 & 1.8 & 1.5 & 1.8 \\
\texttt{complex128} & 16 & 64 & 3.2 & 11 & 65 & 13 & 5.4 & 11 \\
}
\newcommand{\NumAccRows}{%
\texttt{float32} & $6.6\times10^{-7}$ & $2.0\times10^{-6}$ & $2.0\times10^{0}$ & $5.3\times10^{-4}$ \\
\texttt{float64} & $1.3\times10^{-15}$ & $4.0\times10^{-15}$ & $6.6\times10^{-9}$ & $1.4\times10^{-12}$ \\
\texttt{complex64} & $8.3\times10^{-7}$ & $2.3\times10^{-6}$ & $2.3\times10^{0}$ & $2.0\times10^{-4}$ \\
\texttt{complex128} & $1.6\times10^{-15}$ & $4.5\times10^{-15}$ & $8.9\times10^{-9}$ & $1.8\times10^{-12}$ \\
}
\newcommand{\NumTimeHeader}{generate & eigh f(A) & SVD |A| & channel & expm & opt. step}
\newcommand{\NumRefTimes}{generate: 918\,$\mu$s, eigh f(A): 7.24\,$\mu$s, SVD |A|: 11.4\,$\mu$s, channel: 1.83\,$\mu$s, expm: 4.96\,$\mu$s, opt. step: 70.9\,$\mu$s}
\newcommand{\BaseDevice}{NVIDIA GeForce RTX 5090}
\newcommand{\BaseMode}{full}

\newcommand{\BaseMaxSigma}{1.2}
\newcommand{\BaseMCRows}{%
1 & 2 & 160 & 2.49 & 1.50 & 124 \\
16 & 16 & 2852 & 690 & 884 & 2000 \\
32 & 16 & 5702 & 1380 & 1688 & 4005 \\
}
\newcommand{\BaseGSRows}{%
16 & 12870 & 1 & 0.0288 & 0.0552 & 6.2e-15 \\
16 & 12870 & 20 & 0.621 & 0.114 & 4.3e-14 \\
18 & 48620 & 1 & 0.115 & 0.0588 & 3.6e-15 \\
20 & 184756 & 1 & 0.540 & 0.0642 & 1.1e-14 \\
}
\newcommand{\BenchDevice}{NVIDIA GeForce RTX 5090}
\newcommand{\BenchMode}{full}
\newcommand{\BenchTorch}{2.14.1+cu130}
\newcommand{\BenchCuda}{13.0}
\newcommand{\BenchCusolver}{12004}

\newcommand{\BenchCpuThreads}{16}
\newcommand{\BenchTargetLog}{22}
\newcommand{\BenchHasGPU}{1}
\newcommand{\BenchmarkRows}{%
generate & 4 & 1 & 4096 & 63.5 & 1.90 & 0.0299 \\
f(A) eigh & 4 & 1 & 4096 & 0.160 & 3.06 & 19.2 \\
|A| SVD & 4 & 1 & 4096 & 0.388 & 5.27 & 13.6 \\
contrast & 4 & 1 & 4096 & 0.0979 & 1.44 & 14.7 \\
channel & 4 & 1 & 4096 & 0.116 & 0.186 & 1.60 \\
exp(tL) & 4 & 1 & 4096 & 2.05 & 34.1 & 16.6 \\
opt. step & 4 & 1 & 4096 & 2.76 & 7.78 & 2.82 \\
generate & 4 & 16 & 4096 & 995 & 28.1 & 0.0283 \\
f(A) eigh & 4 & 16 & 4096 & 0.514 & 44.7 & 86.9 \\
|A| SVD & 4 & 16 & 4096 & 4.17 & 80.7 & 19.4 \\
contrast & 4 & 16 & 4096 & 0.358 & 22.0 & 61.5 \\
channel & 4 & 16 & 4096 & 1.34 & 4.53 & 3.38 \\
exp(tL) & 4 & 16 & 1024 & 49.2 & 556 & 11.3 \\
opt. step & 4 & 16 & 4096 & 9.54 & 80.2 & 8.41 \\
generate & 4 & 256 & 1024 & 15710 & 534 & 0.0340 \\
f(A) eigh & 4 & 256 & 1024 & 9.01 & 742 & 82.3 \\
|A| SVD & 4 & 256 & 1024 & 66.3 & 1283 & 19.4 \\
contrast & 4 & 256 & 1024 & 5.63 & 359 & 63.7 \\
channel & 4 & 256 & 1024 & 20.9 & 65.0 & 3.12 \\
exp(tL) & 4 & 256 & 64 & 7743 & 15365 & 1.98 \\
opt. step & 4 & 256 & 1024 & 133 & 1604 & 12.0 \\
generate & 16 & 1 & 4096 & 60.9 & 17.3 & 0.283 \\
f(A) eigh & 16 & 1 & 4096 & 0.206 & 23.8 & 116 \\
|A| SVD & 16 & 1 & 4096 & 0.780 & 44.5 & 57.0 \\
contrast & 16 & 1 & 4096 & 0.137 & 10.9 & 79.5 \\
channel & 16 & 1 & 4096 & 0.129 & 1.72 & 13.4 \\
exp(tL) & 16 & 1 & 64 & 1149 & 13608 & 11.8 \\
opt. step & 16 & 1 & 4096 & 3.61 & 74.3 & 20.6 \\
generate & 16 & 16 & 1024 & 1032 & 290 & 0.281 \\
f(A) eigh & 16 & 16 & 1024 & 1.76 & 419 & 238 \\
|A| SVD & 16 & 16 & 1024 & 10.7 & 746 & 69.6 \\
contrast & 16 & 16 & 1024 & 1.26 & 182 & 145 \\
channel & 16 & 16 & 1024 & 1.63 & 65.0 & 39.8 \\
exp(tL) & 16 & 16 & 4 & 79221 & 181417 & 2.29 \\
opt. step & 16 & 16 & 1024 & 34.9 & 1505 & 43.1 \\
generate & 16 & 256 & 64 & 18838 & 5937 & 0.315 \\
f(A) eigh & 16 & 256 & 64 & 78.0 & 6736 & 86.3 \\
|A| SVD & 16 & 256 & 64 & 172 & 11860 & 68.9 \\
contrast & 16 & 256 & 64 & 36.9 & 2871 & 77.7 \\
channel & 16 & 256 & 64 & 26.3 & 989 & 37.6 \\
exp(tL) & 16 & 256 & 1 & 2349043 & 3558629 & 1.51 \\
opt. step & 16 & 256 & 64 & 567 & 21657 & 38.2 \\
generate & 64 & 1 & 1024 & 62.0 & 329 & 5.30 \\
f(A) eigh & 64 & 1 & 1024 & 3.47 & 393 & 113 \\
|A| SVD & 64 & 1 & 1024 & 13.1 & 820 & 62.4 \\
contrast & 64 & 1 & 1024 & 2.48 & 211 & 85.0 \\
channel & 64 & 1 & 1024 & 0.859 & 71.2 & 82.9 \\
opt. step & 64 & 1 & 1024 & 34.2 & 1737 & 50.8 \\
generate & 64 & 16 & 64 & 1262 & 4995 & 3.96 \\
f(A) eigh & 64 & 16 & 64 & 59.2 & 6495 & 110 \\
|A| SVD & 64 & 16 & 64 & 211 & 12966 & 61.4 \\
contrast & 64 & 16 & 64 & 41.2 & 3424 & 83.0 \\
channel & 64 & 16 & 64 & 13.7 & 1054 & 76.8 \\
opt. step & 64 & 16 & 64 & 550 & 28154 & 51.2 \\
generate & 64 & 256 & 4 & 132631 & 103301 & 0.779 \\
f(A) eigh & 64 & 256 & 4 & 1753 & 99206 & 56.6 \\
|A| SVD & 64 & 256 & 4 & 3398 & 204740 & 60.2 \\
contrast & 64 & 256 & 4 & 933 & 54526 & 58.4 \\
channel & 64 & 256 & 4 & 217 & 12346 & 57.0 \\
opt. step & 64 & 256 & 4 & 8867 & 448973 & 50.6 \\
}
\newcommand{\RenyiDevice}{CPU (4 threads)}
\newcommand{\RenyiMode}{full}

\newcommand{\RenyiStarts}{64}
\newcommand{\RenyiSteps}{600}

\newcommand{\RenyiTol}{1\times10^{-12}}
\newcommand{\RenyiDigits}{50}

\newcommand{\RenyiDims}{2, 3}
\newcommand{\RenyiSandLow}{0.499980}
\newcommand{\RenyiSandHigh}{0.500000}
\newcommand{\RenyiPetzLow}{2.000000}
\newcommand{\RenyiPetzHigh}{2.000039}
\newcommand{\RenyiSandLowV}{7.8\times10^{-5}}
\newcommand{\RenyiPetzHighV}{1.3\times10^{-4}}

\newcommand{\RenyiRows}{%
0.2 & $3.8\times10^{-1}$ & $3.8\times10^{-1}$ & -- & -- \\
0.25 & $4.3\times10^{-1}$ & $4.3\times10^{-1}$ & -- & -- \\
0.3 & $4.4\times10^{-1}$ & $4.4\times10^{-1}$ & -- & -- \\
0.35 & $4.0\times10^{-1}$ & $4.0\times10^{-1}$ & -- & -- \\
0.4 & $3.1\times10^{-1}$ & $3.1\times10^{-1}$ & -- & -- \\
0.45 & $1.7\times10^{-1}$ & $1.7\times10^{-1}$ & -- & -- \\
0.48 & $7.0\times10^{-2}$ & $7.0\times10^{-2}$ & -- & -- \\
0.49 & $3.5\times10^{-2}$ & $3.5\times10^{-2}$ & -- & -- \\
0.5 & -- & -- & -- & -- \\
0.51 & -- & -- & -- & -- \\
0.52 & -- & -- & -- & -- \\
0.55 & -- & -- & -- & -- \\
0.6 & -- & -- & -- & -- \\
0.7 & -- & -- & -- & -- \\
0.8 & -- & -- & -- & -- \\
0.9 & -- & -- & -- & -- \\
1.1 & -- & -- & -- & -- \\
1.25 & -- & -- & -- & -- \\
1.5 & -- & -- & -- & -- \\
1.75 & -- & -- & -- & -- \\
1.9 & -- & -- & -- & -- \\
1.95 & -- & -- & -- & -- \\
1.98 & -- & -- & -- & -- \\
2 & -- & -- & -- & -- \\
2.02 & -- & -- & -- & $1.7\times10^{-1}$ \\
2.05 & -- & -- & -- & $5.2\times10^{-1}$ \\
2.1 & -- & -- & -- & $9.9\times10^{-1}$ \\
2.25 & -- & -- & -- & $2.9\times10^{0}$ \\
2.5 & -- & -- & -- & $5.3\times10^{0}$ \\
2.75 & -- & -- & -- & $7.6\times10^{0}$ \\
3 & -- & -- & -- & $7.8\times10^{0}$ \\
}

\newcommand{\PMNmax}{80}

\newcommand{\TCDigits}{40}

\newcommand{\TCItems}{25}

\newcommand{\SRL}{24}

\newcommand{\NXYRows}{%
4 & 6 & 70 & 256 & 0 & 4e-7 & 9e-16 & 2.00 & 1.167 & 0.999 & 0.1126 & 0.0254\\
6 & 20 & 924 & 4096 & 1 & 6e-6 & 1e-15 & 2.00 & 1.380 & 0.999 & 0.1573 & 0.0307\\
8 & 70 & 12870 & 65536 & 18 & 2e-5 & 9e-16 & 2.00 & 1.421 & 0.999 & 0.1969 & 0.0328\\
10 & 252 & 184756 & 1048576 & 930 & 5e-5 & 3e-15 & 2.00 & 1.426 & 0.999 & 0.2340 & 0.0345\\
}
\newcommand{\NXYT}{2}
\newcommand{\NXYSubsteps}{40}
\newcommand{\NXYAccErr}{5\times10^{-5}}
\newcommand{\NXYTOnePost}{3\times10^{-15}}
\newcommand{\NXYBatch}{28}
\newcommand{\ArchRows}{%
known spectrum & $\lambda_{\max}$, 256 operators in $M_{32}^{\oplus8}$ & 2.25 & 28.70 & 13\\
eigenvalues only & spectra of the same operators & 28.27 & 93.60 & 3.3\\
spectral cache & $e^{-itA}$ for 50 values of $t$ & 3019.57 & 11717.58 & 3.9\\
size groups & spectra, 256 operators in $M_8^{\oplus64}$ & 19.69 & 73.53 & 3.7\\
trace of a product & $\Tr(\rho A)$, 32 pairs in $M_{256}^{\oplus2}$ & 16.62 & 41.75 & 2.5\\
sparse jumps & Lindblad, 8 qubits, $S^-_i$ and $S^z_i$ & 354.63 & 1010.80 & 2.9\\
}
\newcommand{\ArchThreads}{4}
\newcommand{\ArchJumpDense}{48.4}
\newcommand{\ArchJumpSparse}{0.12}
\newcommand{\ArchJumpDiff}{4\times10^{-15}}
\newcommand{\NCRows}{%
4 & (i) & 2.73 & 1.0000 & 0.941 & 0.929 & 0.987 & 0.795 & 0.948\\
4 & (ii) & 2.66 & 0.9993 & 0.942 & 0.930 & 0.988 & 0.801 & 0.951\\
4 & (iii) & 2.71 & 1.0000 & 0.941 & 0.930 & 0.988 & 0.799 & 0.952\\
6 & (i) & 5.49 & 0.9981 & 0.846 & 0.807 & 0.952 & 0.490 & 0.816\\
6 & (ii) & 4.88 & 0.9889 & 0.860 & 0.821 & 0.953 & 0.535 & 0.846\\
6 & (iii) & 5.14 & 0.9949 & 0.854 & 0.818 & 0.956 & 0.519 & 0.840\\
}

\newcommand{\NCFourGainPost}{1.07}
\newcommand{\NCFourGainRaw}{1.02}

\newcommand{\NCFourPostFactor}{5.6}

\newcommand{\NCSixMinFPostOne}{0.816}

\newcommand{\NCSixMinFPostThree}{0.840}
\newcommand{\NCSixTOne}{5.49}
\newcommand{\NCSixTTwo}{4.88}
\newcommand{\NCSixTThree}{5.14}
\newcommand{\NCSixFPostOne}{0.952}

\newcommand{\NCSixFPostThree}{0.956}
\newcommand{\NCSixFOne}{0.807}
\newcommand{\NCSixFTwo}{0.821}

\newcommand{\NCSixPTwo}{0.860}
\newcommand{\NCSixPThree}{0.854}
\newcommand{\NCSixFZeroOne}{0.9981}
\newcommand{\NCSixFZeroTwo}{0.9889}

\newcommand{\NCSixGainPost}{1.09}
\newcommand{\NCSixGainRaw}{1.07}
\newcommand{\NCSixGainTotal}{4.07}
\newcommand{\NCSixPostFactor}{4.0}
\newcommand{\NCSixRatioLow}{1.29}
\newcommand{\NCSixRatioHigh}{0.90}

\newcommand{\NCSegments}{4}
\newcommand{\NCSubsteps}{2}
\newcommand{\NCEvalSubsteps}{20}
\newcommand{\NCStarts}{32}
\newcommand{\NCWarm}{8}
\newcommand{\NCWarmLarge}{4}
\newcommand{\NCSettings}{12}
\newcommand{\NCSteps}{150}
\newcommand{\NCStepsZero}{400}

\newcommand{\NCLambda}{0.05}
\newcommand{\NCRecheck}{6\times10^{-4}}
\newcommand{\NCBirthDeath}{4\times10^{-6}}
\newcommand{\NCExactDiff}{6\times10^{-6}}
\newcommand{\NCGradErr}{6\times10^{-8}}
\newcommand{\NCSeconds}{658}

\newcommand{\NCThreads}{4}
\newcommand{\NCnMax}{6}
\newcommand{\CLRows}{%
3 & rates & 4 & $1.0{\times}10^{-2}$ & $2.8{\times}10^{-3}$ & $1.5{\times}10^{-3}$ & $4.8{\times}10^{-4}$ & $7.4{\times}10^{-8}$ & $1.7{\times}10^{-3}$ & 1\\
3 & GKSL layer & 280 & $9.0{\times}10^{-2}$ & $3.8{\times}10^{-2}$ & $1.6{\times}10^{-2}$ & $5.9{\times}10^{-3}$ & $4.2{\times}10^{-6}$ & $2.6{\times}10^{-2}$ & 8\\
3 & channel of $\M$ & 1152 & $1.0{\times}10^{-1}$ & $4.0{\times}10^{-2}$ & $1.6{\times}10^{-2}$ & $7.2{\times}10^{-3}$ & $4.6{\times}10^{-4}$ & $3.0{\times}10^{-2}$ & 7\\
3 & channel of $M_{2^n}$ & 8192 & $2.2{\times}10^{-1}$ & $9.0{\times}10^{-2}$ & $4.5{\times}10^{-2}$ & $1.9{\times}10^{-2}$ & $8.4{\times}10^{-4}$ & $9.2{\times}10^{-2}$ & 7\\
\midrule
4 & rates & 4 & $7.4{\times}10^{-3}$ & $2.6{\times}10^{-3}$ & $9.6{\times}10^{-4}$ & $2.6{\times}10^{-4}$ & $3.6{\times}10^{-9}$ & $1.0{\times}10^{-3}$ & 4\\
4 & GKSL layer & 1386 & $1.2{\times}10^{-1}$ & $5.9{\times}10^{-2}$ & $2.2{\times}10^{-2}$ & $7.4{\times}10^{-3}$ & $1.3{\times}10^{-5}$ & $3.8{\times}10^{-2}$ & 51\\
4 & channel of $\M$ & 18432 & $1.4{\times}10^{-1}$ & $6.3{\times}10^{-2}$ & $2.6{\times}10^{-2}$ & $1.0{\times}10^{-2}$ & $6.5{\times}10^{-4}$ & $5.2{\times}10^{-2}$ & 27\\
4 & channel of $M_{2^n}$ & 131072 & $3.1{\times}10^{-1}$ & $1.4{\times}10^{-1}$ & $7.1{\times}10^{-2}$ & $3.3{\times}10^{-2}$ & $1.2{\times}10^{-3}$ & $1.7{\times}10^{-1}$ & 98\\
}
\newcommand{\CLT}{1}
\newcommand{\CLGdown}{0.1}
\newcommand{\CLGup}{0.02}
\newcommand{\CLGphi}{0.05}
\newcommand{\CLnMin}{3}
\newcommand{\CLnMax}{4}
\newcommand{\CLProbes}{84}
\newcommand{\CLProbesMin}{26}
\newcommand{\CLTest}{17}
\newcommand{\CLSettingsAlg}{9}
\newcommand{\CLSettingsDense}{19}
\newcommand{\CLDimAlg}{70}
\newcommand{\CLDimDense}{256}
\newcommand{\CLSetAlg}{4830}
\newcommand{\CLSetDense}{65280}
\newcommand{\CLIdentDense}{17850}
\newcommand{\CLDenseCheck}{1.9{\times}10^{-15}}
\newcommand{\CLFastPath}{5.7{\times}10^{-16}}
\newcommand{\CLAdam}{300}
\newcommand{\CLLbfgs}{400}
\newcommand{\CLStartsRates}{3}
\newcommand{\CLStartsLindblad}{3}
\newcommand{\CLStartsAlgebra}{2}
\newcommand{\CLStartsDense}{2}
\newcommand{\CLRepeats}{6}
\newcommand{\CLRuntime}{14}
\newcommand{\CLWorkers}{4}
\newcommand{\CLHScale}{0.1}
\newcommand{\CLInitConverged}{1}
\newcommand{\CLInitStarts}{3}
\newcommand{\CLInitLoss}{2.7{\times}10^{-3}}
\newcommand{\CLStartSpread}{1.0{\times}10^{-3}}

\newcommand{\CLTDRatesTwo}{7.4{\times}10^{-3}}

\newcommand{\CLTDRatesFour}{9.6{\times}10^{-4}}

\newcommand{\CLParLindblad}{1386}

\newcommand{\CLTDLindbladFour}{2.2{\times}10^{-2}}

\newcommand{\CLTDLindbladExact}{1.3{\times}10^{-5}}

\newcommand{\CLParAlgebra}{18432}

\newcommand{\CLTDAlgebraFour}{2.6{\times}10^{-2}}

\newcommand{\CLTDAlgebraFive}{1.0{\times}10^{-2}}

\newcommand{\CLTDAlgebraExact}{6.5{\times}10^{-4}}

\newcommand{\CLParDense}{131072}

\newcommand{\CLTDDenseFour}{7.1{\times}10^{-2}}

\newcommand{\CLTDDenseExact}{1.2{\times}10^{-3}}

\newcommand{\CLOffDense}{5.3{\times}10^{-2}}

\newcommand{\CLSlopeRates}{0.48}
\newcommand{\CLSlopeLindblad}{0.41}
\newcommand{\CLSlopeAlgebra}{0.38}
\newcommand{\CLSlopeDense}{0.32}
\newcommand{\CLParRatio}{13}
\newcommand{\CLTarget}{6.3{\times}10^{-2}}

\newcommand{\CLNeedLindblad}{8.1{\times}10^{2}}
\newcommand{\CLNeedAlgebra}{1.0{\times}10^{3}}
\newcommand{\CLNeedDense}{1.4{\times}10^{4}}
\newcommand{\CLNeedRatio}{14}
\newcommand{\CLRatesJ}{1.0002}
\newcommand{\CLRatesJSd}{0.0015}

\newcommand{\CLCRJ}{0.0013}
\newcommand{\CLRatesDown}{0.1001}
\newcommand{\CLRatesDownSd}{0.0003}
\newcommand{\CLLindbladDown}{0.1020}
\newcommand{\CLLindbladDownSd}{0.0008}
\newcommand{\CLCRDown}{0.0003}
\newcommand{\CLRatesUp}{0.0200}
\newcommand{\CLRatesUpSd}{0.0001}
\newcommand{\CLLindbladUp}{0.0216}
\newcommand{\CLLindbladUpSd}{0.0005}
\newcommand{\CLCRUp}{0.0001}
\newcommand{\CLRatesPhi}{0.0501}
\newcommand{\CLRatesPhiSd}{0.0015}
\newcommand{\CLLindbladPhi}{0.0419}
\newcommand{\CLLindbladPhiSd}{0.0037}
\newcommand{\CLCRPhi}{0.0011}
\newcommand{\CLRepeatShots}{1{\times}10^{4}}
\newcommand{\CLResidual}{3.1{\times}10^{-2}}
\newcommand{\RMSteps}{2000}
\newcommand{\RMRestarts}{6}
\newcommand{\RMp}{0.2}
\newcommand{\RMDepthMax}{32}
\newcommand{\RMDepths}{1, 2, 4, 8, 16, 32}
\newcommand{\RMSeconds}{281}
\newcommand{\RMWorkers}{4}
\newcommand{\RMMarkovA}{0.2397}
\newcommand{\RMMarkovHS}{0.413}
\newcommand{\RMMarkovTD}{0.195}
\newcommand{\RMDephasedHS}{0.471}
\newcommand{\RMSemigroupKmax}{3}
\newcommand{\RMDampedSemigroupViolation}{0.253}
\newcommand{\RMSemigroupBound}{0}
\newcommand{\RMGkslChannel}{1.5\times10^{-15}}
\newcommand{\RMGkslLindblad}{1.4\times10^{-4}}
\newcommand{\RMGkslResOne}{1.370}
\newcommand{\RMGkslResDeep}{2.5\times10^{-15}}
\newcommand{\RMGkslDet}{0.0184}
\newcommand{\RMDampedChannel}{2.0\times10^{-4}}
\newcommand{\RMDampedLindblad}{0.101}
\newcommand{\RMDampedResOne}{1.561}
\newcommand{\RMDampedResDeep}{0.163}
\newcommand{\RMDampedDet}{0.2500}
\newcommand{\RMPauliChannel}{2.5\times10^{-4}}
\newcommand{\RMPauliLindblad}{0.413}
\newcommand{\RMPauliResOne}{1.848}
\newcommand{\RMPauliResDeep}{0.407}
\newcommand{\RMPauliDet}{-0.0370}
\newcommand{\RMPauliResMin}{0.407}
\newcommand{\RMPauliResMinTD}{0.192}
\newcommand{\RMPauliResTDAll}{0.1916}
\newcommand{\RMPauliResTDAllDepth}{16}
\newcommand{\RMPauliLindbladTD}{0.195}
\newcommand{\RMRestartSpread}{4.0\times10^{-5}}
\newcommand{\RMPauliResDetMin}{4.1\times10^{-4}}
\newcommand{\RMPauliResDetMax}{0.394}
\newcommand{\RMPauliLindbladDet}{0.003}
\newcommand{\RMPauliChannelDet}{-0.0370}

\ifdefined\BaseMinKS\newcommand{\BaseNew}{1}\else\newcommand{\BaseNew}{0}\fi
\providecommand{\ValidationRows}{\multicolumn{6}{c}{\emph{run scripts/validate\_known\_results.py}}\\}
\providecommand{\SearchRows}{\multicolumn{9}{c}{\emph{run scripts/inequality\_search.py}}\\}

\providecommand{\BenchmarkRows}{\multicolumn{5}{c}{\emph{run scripts/benchmark\_library.py}}\\}
\providecommand{\BenchHasGPU}{0}

\providecommand{\NumTimeRows}{\multicolumn{9}{c}{\emph{run scripts/numerics.py}}\\}
\providecommand{\NumAccRows}{\multicolumn{5}{c}{\emph{run scripts/numerics.py}}\\}
\providecommand{\NumTimeHeader}{generate & eigh $f(A)$ & SVD $|A|$ & channel & expm & opt.\ step}
\providecommand{\BaseMCRows}{\multicolumn{6}{c}{\emph{run scripts/baselines.py}}\\}
\providecommand{\BaseGSRows}{\multicolumn{6}{c}{\emph{run scripts/baselines.py}}\\}
\providecommand{\ValTraceRows}{\multicolumn{5}{c}{\emph{run scripts/validate\_known\_results.py}}\\}
\providecommand{\ChainKrylovRows}{\multicolumn{6}{c}{\emph{run scripts/chains.py}}\\}
\providecommand{\BaseMinKS}{--}
\providecommand{\BaseMaxSigma}{--}
\providecommand{\ChainRingRows}{\multicolumn{6}{c}{\emph{run scripts/chains.py}}\\}
\providecommand{\NXYRows}{\multicolumn{12}{c}{\emph{run scripts/noisy\_xy\_mixer.py}}\\}
\providecommand{\ArchRows}{\multicolumn{5}{c}{\emph{run scripts/architecture\_bench.py}}\\}
\providecommand{\RenyiRows}{\multicolumn{5}{c}{\emph{run scripts/renyi\_threshold.py}}\\}

\providecommand{\RenyiMode}{--}\providecommand{\RenyiDevice}{--}\providecommand{\RenyiTol}{--}

\newcommand{\fig}[2]{\includegraphics[width=#2]{#1}}
\newcommand{\G}[3]{#2}

\title{Finite-Dimensional Type~I von Neumann Algebras in PyTorch:\\
A GPU-Accelerated Framework for Random Block-Diagonal Operators}
\author{Irina Nikolaeva \and Andrej Novikov}
\date{}

\begin{document}
\maketitle

\begin{abstract}
Finite-dimensional von Neumann algebras, the direct sums $\bigoplus_cM_{k_c}(\C)$ of matrix algebras, are basic objects
of operator algebra theory and, through the commutant of a symmetry, the natural home of quantum systems with conserved
charges or superselection rules. To our knowledge no numerical software treats them as a data type, with their centre,
their family of traces, conditional expectations, inclusions and functional calculus. \texttt{torch\_vn\_algebra} is a
PyTorch library that does, for batches of operators on CPUs and GPUs and with automatic differentiation. It fills the
gap between computer algebra, which treats algebras as objects but exactly and for small examples, and quantum
simulators, which compute at scale but with matrices on a Hilbert space, and it converts operators, states and
channels to and from QuTiP and dynamiqs.
It is meant for operator algebraists who explore inequalities and extremal problems numerically, for quantum
information theorists with variational problems over states and channels under a symmetry, and for physicists who need
many small symmetric problems at once. It began as a
re-implementation of our earlier NumPy Monte Carlo study of trace inequalities with the Michelson contrast, and this
paper documents it. Its design rests on one thesis: keeping the sector structure inside the data type makes computations
in the algebra correct by construction and cheap. Haar sampling is exact in every block, trace conventions survive maps
between sectors, and all blocks of one size go through one batched kernel. We show that the library reproduces the
distributions of the earlier NumPy computation and is about two orders of magnitude faster than its loop formulation
for small blocks. More important than speed, the operations are differentiable, which turns the sampling of the earlier
study into variational problems over the states, unitaries and channels of the algebra: from block size 4 on, a million
Haar samples miss the violation of a trace inequality in every one of 16 test cases, while batched gradient ascent
finds the exact supremum in all of them in seconds. We measure each architectural decision against the simple
alternative and validate the library against exact results and against QuTiP and dynamiqs. Its generality is
illustrated by a noisy XY mixer of the quantum alternating operator ansatz, whose Hamming-weight sectors form the
algebra: one batched run gives, for all noise models and rates, the exact effect of post-selection on the weight, and
optimising the controls for the post-selected fidelity shows that the measurement of the conserved charge, not the
noise-aware design, removes most of the error. Finally, channels between algebras become layers of trainable networks. Learning
the noise of the register from measurement counts, models that respect the sectors reach a given accuracy with about
fourteen times fewer shots than a dense model, and a model with the known jump operators estimates the rates close to
the Cramér--Rao bound; residual networks of channels fit Markovian evolutions but provably not a channel with negative
determinant, so their fit error witnesses non-Markovianity.
\end{abstract}

\section*{Program summary}
\begin{description}\itemsep0pt
\item[Program title:] \texttt{torch\_vn\_algebra}
\item[Developer's repository link:] \url{https://github.com/andre-a-no/von_neumann_type_i}
\item[Licensing provisions:] MIT
\item[Programming language:] Python ($\ge3.8$), PyTorch ($\ge1.13$), NumPy ($\ge1.21$)
\item[Nature of problem:] Numerical experiments in finite-dimensional Type~I von Neumann algebras
$\bigoplus_cM_{k_c}(\C)$ and in the quantum systems with superselection rules or conserved charges that they describe:
random operators with prescribed spectra, several trace functionals, states, channels within and between sectors,
symmetric tensor products, spin chains, Hamiltonian and Lindblad dynamics with jumps between sectors (noisy circuits
with a conserved charge), extremal problems under spectral constraints, inclusions of algebras, symmetry reduction,
and quantum Rényi divergences.
\item[Solution method:] Batched block-diagonal tensors processed by PyTorch linear algebra on CPUs and GPUs, one
call per block size; lazy operators that carry their properties, with caches for spectra; exact Haar sampling;
spectral functional calculus with Daleckii--Krein derivatives; Kraus representations; Lanczos and Krylov methods for
sparse sector Hamiltonians; sparse jumps between sectors; batched multistart gradient optimisation; run-time and
memory estimates before expensive steps; inclusion matrices; randomised Wedderburn decomposition; highest-weight bases
and exact Clebsch--Gordan coefficients.
\item[Additional comments:] more than 800 unit tests, run in continuous integration; every number, table and figure of
this paper is produced by the scripts and examples in the repository and regenerated by \texttt{scripts/run\_all.sh}.
The library is the software used in~\cite{abed2024} and in the companion papers~\cite{nn_trace,nn_cs,nn_btc}.
\end{description}

\section{Introduction}\label{sec:intro}

\paragraph{The algebra as the object.} A finite-dimensional von Neumann algebra is, up to $*$-isomorphism, a direct
sum of full matrix algebras~\cite{takesaki2002},
\begin{equation}\label{eq:M}
  \M=\bigoplus_{c=1}^{C}M_{k_c}(\C)\quad\text{acting on}\quad\mathcal H=\bigoplus_{c=1}^C\C^{k_c},
\end{equation}
where $M_k(\C)$ is the algebra of complex $k\times k$ matrices and the $C$ summands are called sectors. Small as they
are, these algebras carry the structure that operator algebra theory studies. Their centre is spanned by the sector
projections. Instead of one trace they have a $C$-parameter family of faithful traces. They have conditional
expectations onto subalgebras, inclusions with an index~\cite{jones1983,ghj1989}, and a Jordan product on their
self-adjoint part; their elements have spectral invariants such as the Michelson contrast, a normalised spread of the spectrum
(Section~\ref{sec:math}). Questions about these
objects are often explored on examples first: which inequalities between traces of products characterise the trace or
the centre~\cite{gardner1979,petz1988,novikov2015,virosztek2017}, and how strongly they fail for functionals that are
not traces~\cite{abed2024}. Such explorations are usually written as ad hoc matrix code. To our knowledge there is no
numerical software in which the algebra~\eqref{eq:M} is the data type: one declares the sector sizes and then samples,
transforms and optimises elements, states and maps, with the centre and the traces built in (Section~\ref{sec:related}).

\paragraph{From an earlier NumPy study to a library.} This library grew out of such an exploration. In~\cite{abed2024}
and in the first version of this work~\cite{v1} we studied trace inequalities with the Michelson contrast by Monte
Carlo sampling in NumPy. The computation draws many positive operators with prescribed contrast and Haar-random
eigenvectors in every block, evaluates a trace functional, and looks at its distribution. Written directly, such code
loops over samples and sectors, restates the trace convention in every script, and has to get the Haar measure right
block by block.\footnote{The first version~\cite{v1} contained errors of this kind in its samplers; they are
corrected and tested here.} \texttt{torch\_vn\_algebra} re-implements that modelling as a structured PyTorch library.
Section~\ref{sec:reproduce} shows that it reproduces the earlier computation: the NumPy formulations and the library
sample the same distribution. For small blocks the library is about two orders of magnitude faster than the loop
formulation, and for blocks of size 16 about three times faster, and the same code runs on a GPU.

\paragraph{From sampling to optimisation.} Speed is the smaller gain. The numerical operations of the library are
differentiable (the one exception is the unitary factor of the polar decomposition), including spectral functions (by the Daleckii--Krein formula), channels between sectors and the time
steps of Hamiltonian and Lindblad dynamics, so a question asked by sampling can be asked again as a variational
problem: maximise a functional over operators with a prescribed spectrum, over states, over unitaries of the algebra or
over all channels, by batched gradient ascent with many starts in one call.

\paragraph{Extremal problems in their own right.} The variational form is not a detour: in operator algebra, quantum information and
physics, many of the central questions are extremal problems. In operator algebra, a trace inequality is settled by its
sharp constant~\cite{carlen2010}, and how strongly a characterisation of the trace~\cite{gardner1979,petz1988} fails
for other functionals is measured by an envelope over unitaries and over spectra with prescribed
invariants~\cite{abed2024,nn_trace}. The Connes--Størmer entropy
of two subalgebras is a supremum over finite partitions of unity~\cite{connes1975,pimsner1986}, and the
Pimsner--Popa constant of a conditional expectation is the best constant in an operator inequality, tied to the
index~\cite{pimsner1986,frank1998}. In quantum information, data processing and the parameter ranges in which it
holds, channel capacities and minimal output entropies, and optimal recovery are optimisation problems. The convex
ones, such as the diamond norm~\cite{watrous2009} or channel-adapted recovery~\cite{fletcher2007}, are semidefinite
programs and are best solved by SDP solvers, which also return dual certificates. Many important ones are not
convex: the minimal output entropy, whose additivity was disproved with random channels in
high dimension~\cite{hastings2009}, suprema over unitaries or over spectra with prescribed contrast, the Rényi and
sandwiched quantities outside the ranges in which they are monotone~\cite{mullerlennert2013,berta2017}, and entropies
maximised over partitions. In physics, optimal control of open systems~\cite{khaneja2005,koch2016}, variational state
preparation~\cite{cerezo2021} and protocols designed to tolerate noise are optimisation problems over unitaries, states
and channels; when the controls and the noise conserve a charge (Section~\ref{sec:noisy}), all of their variables lie
in an algebra~\eqref{eq:M}.

\paragraph{Sampling is no substitute.} Asking such a question by sampling changes the answer, not only the time. The
Monte Carlo of~\cite{abed2024,v1} saw no violation of Gardner's inequality for blocks of size 16, and we now
know why: by concentration of measure, the unitaries near the maximiser have Haar measure that decays rapidly with the
dimension. From block size 4 on, a million Haar samples miss the violation in every test case, while gradient ascent
finds the exact supremum in all of them, to a relative error of $3\times10^{-5}$ and in seconds
(Section~\ref{sec:optimisation}). The violation envelopes of the companion paper~\cite{nn_trace}, the thresholds of
the Rényi divergences in Section~\ref{sec:renyi} and the partitions in~\cite{nn_cs} are all found this way; every
reported extremal value is then recomputed at the optimiser in 40- to 50-digit arithmetic (mpmath, in the paper
scripts), so it is a certified lower bound for the supremum.

\paragraph{Thesis.} The design rests on one thesis: keeping the sector structure inside the data type makes
computations in~\eqref{eq:M} both correct by construction and cheap. By sector structure we mean the blocks, the centre,
the several traces and the maps between algebras with different sectors. Correctness follows because a random element
is sampled block by block from the right measure, a density is always stored with respect to one trace and converted
explicitly, and a map between sectors carries its own dual. Cheapness follows because the library knows which blocks
have equal size and stacks them into one batched kernel call, knows from property flags which kernel a result needs,
and never forms the zero blocks. Section~\ref{sec:design} describes the architecture built on this thesis and measures
each decision against the simple alternative.

\paragraph{Where the algebra appears in physics.} The same algebra is the home of quantum systems with a symmetry. If a
group, or a set of conserved quantities, acts unitarily on $\mathcal H\cong\bigoplus_\lambda
V_\lambda\otimes\C^{m_\lambda}$, with irreducible representations $V_\lambda$ of multiplicity $m_\lambda$, the
operators commuting with the action form the algebra $\bigoplus_\lambda \mathbb 1_{V_\lambda}\otimes
M_{m_\lambda}(\C)\cong\bigoplus_\lambda M_{m_\lambda}(\C)$. For one conserved charge $Q$ (particle number,
magnetisation, Hamming weight) the sectors are the eigenvalues of $Q$ and the centre consists of the functions of $Q$;
a Hamiltonian that conserves $Q$, the evolution it generates and every observable compatible with a superselection
rule for $Q$ lie in this algebra. Examples are superselection rules~\cite{wick1952,bartlett2007} and
decoherence~\cite{zurek2003}, many-body Hamiltonians diagonalised sector by sector~\cite{sandvik2010,weinberg2017},
and symmetry-resolved entanglement~\cite{goldstein2018,xavier2018}. Others are decoherence-free subspaces and
operator-algebra error correction~\cite{zanardi1997,knill2000,beny2007}, the XY mixers of the quantum alternating operator
ansatz~\cite{hadfield2019,wang2020}, random matrices decomposed by symmetry~\cite{dyson1962,mehta2004}, and
equivariant layers in machine learning, which lie in such commutants by Schur's
lemma~\cite{cohen2016,finzi2021,fulton1991}. A library built for trace inequalities therefore serves these
applications once it has dynamics and maps that move probability between sectors. We added them (spin chains with
$U(1)$ and SU(2) sectors, Lindblad dynamics with jumps between sectors, inclusions, a numerical Wedderburn
decomposition), and Section~\ref{sec:noisy} uses them for a question that the original NumPy code could not address:
symmetry verification in a noisy XY mixer.

\paragraph{Why not one large matrix?} $\M$ embeds into $M_D(\C)$, $D=\sum_ck_c$, as the block-diagonal matrices, and
every matrix package handles that embedding exactly. The question is what it costs and what has to be supplied by hand.
\begin{itemize}\itemsep0pt
\item \emph{Cost.} A product or a decomposition costs $O(D^3)$ time and $O(D^2)$ memory in $M_D$, but
$\sum_ck_c^3$ and $\sum_ck_c^2$ in $\M$: for $C$ equal blocks a factor $C^2$ in time and $C$ in memory.
\item \emph{Measures.} The unitary group of $\M$ is $\prod_cU(k_c)$, for $C\ge2$ a null set in $U(D)$. A Haar-random
element of $U(D)$ is not in $\M$; random elements, random states and gradient steps have to respect the blocks.
\item \emph{Traces.} The faithful traces of $\M$ form the family $\sum_cw_c\Tr_c$. Densities, ``trace preserving''
and the dual of a map between algebras with different sectors depend on the weights, which the embedding forgets.
\item \emph{Centre.} Conditioning on a sector, post-selection and sector-resolved quantities are operations with the
centre; in $M_D$ they have to be reconstructed from block indices.
\end{itemize}
For $C=1$ the library reduces to batched linear algebra in $M_k(\C)$, and all its functions accept that case.

\paragraph{What is new and what is re-implemented.} None of the numerical methods is new. Haar unitaries come from the
QR decomposition of Gaussian matrices with Mezzadri's phase correction~\cite{mezzadri2007}. Gradients of matrix
functions use the Daleckii--Krein formula~\cite{daleckii1965,bhatia1997}. Singular values come from batched SVD or,
on GPUs for large blocks, from the Hermitian dilation, holomorphic functions from contour integrals~\cite{trefethen2014}, and the Wedderburn decomposition from
the randomised algorithms of Murota and coauthors~\cite{murota2010,maehara2011}. Block-sparse storage of
charge-conserving tensors is standard in tensor-network libraries~\cite{fishman2022,hauschild2018,rams2025}. Lazy
linear operators that dispatch on their structure are the basis of GPyTorch's \texttt{linear\_operator}~\cite{gardner2018,linearoperator}
and of CoLA~\cite{potapczynski2023}. Our contribution is their combination around the algebra~\eqref{eq:M} and its
conventions: a batched, differentiable data type whose elements know their sector structure, trace convention and
spectral properties, with states, channels between algebras, inclusions and dynamics defined on it.

\paragraph{What the paper shows, and for whom.} Section~\ref{sec:reproduce} reproduces the earlier NumPy study and
compares speed. Section~\ref{sec:design} explains where the speed comes from, with one measurement per design decision
(Table~\ref{tab:arch}), and Table~\ref{tab:challenges} lists seven implementation problems and the sections that
resolve them. Monte Carlo estimates agree with exact Haar, Weingarten, Page and purity values within their standard
errors (Section~\ref{sec:validation}), and batched optimisation finds extremes that sampling misses
(Section~\ref{sec:optimisation}). Sections~\ref{sec:chains} and~\ref{sec:symmetries} add spin chains, inclusions and
symmetry reduction, and Section~\ref{sec:noisy} shows what this generality buys.
Section~\ref{sec:control} turns the noisy register into a design problem: it optimises the controls of a Lindblad
evolution for the post-selected fidelity, over a batch of noise models at once.
Section~\ref{sec:ecosystem} places the library among existing software and says for whom it is made and for
which tasks another package is the better choice. This paper describes the software used in~\cite{abed2024} and
in three companion papers~\cite{nn_trace,nn_cs,nn_btc}; their results are summarised in Section~\ref{sec:usecases}
and not repeated.

\section{Place in the ecosystem and intended users}\label{sec:ecosystem}

\begin{figure}[t]
\centering\fig{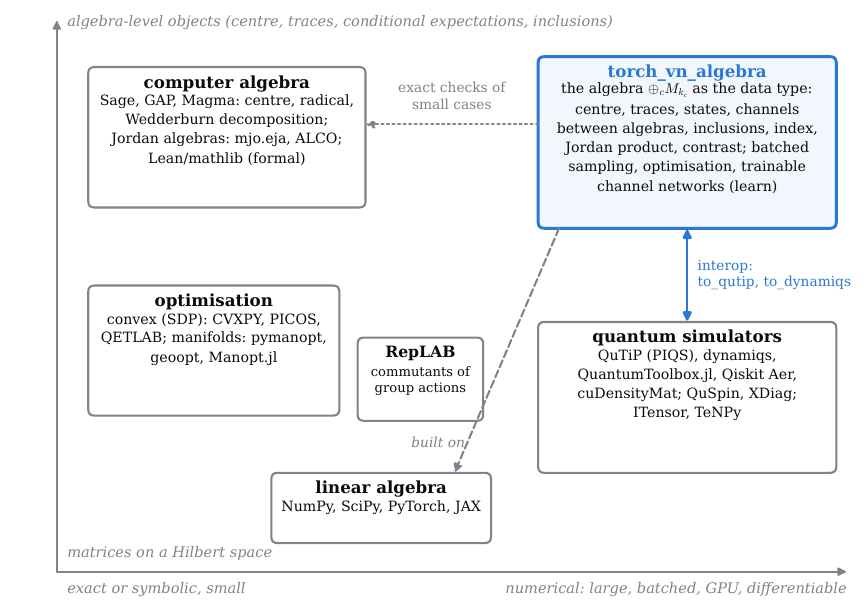}{0.95\textwidth}
\caption{Software for computations with the algebra~\eqref{eq:M}, ordered by two properties: exact and small versus
numerical, batched and differentiable (horizontal), and matrices on a Hilbert space versus objects of the algebra
itself (vertical). The upper right corner, where the algebra is the data type of a numerical code, was empty; the
arrows show what the library builds on, where its results can be checked exactly, and the conversions to the
simulators (Section~\ref{sec:related}).}
\label{fig:ecosystem}
\end{figure}

\paragraph{The gap.} Software that touches finite-dimensional von Neumann algebras sits in two places
(Figure~\ref{fig:ecosystem}). Computer-algebra systems treat algebras as objects, with centres, radicals, Wedderburn
decompositions and Jordan structures, but in exact arithmetic and for small examples. Simulators, exact-diagonalisation
and tensor-network codes compute numerically and at scale, but their object is a matrix or a tensor on a Hilbert space;
blocks appear as storage, not as an algebra with a centre, a family of traces and maps between algebras. Generic linear
algebra (NumPy, SciPy, PyTorch) is below both, and solvers for semidefinite and manifold problems handle convex or
geometric constraints without the structure of~\eqref{eq:M}. What was missing is the numerical counterpart of the first
group: a code in which one declares the sector sizes and then samples, transforms, optimises and learns elements,
states and channels of the algebra in batches, on CPUs and GPUs, with derivatives and certified values. This is the
place of \texttt{torch\_vn\_algebra}. It is built on PyTorch, its test suite checks small cases against exact
results, and it exchanges states, operators, channels and Lindblad data with QuTiP and dynamiqs through the module
\texttt{interop}, so that a computation can move between the corners of the figure.

\paragraph{For whom.} We have four groups of users in mind, each with a typical workflow.
\begin{description}\itemsep2pt
\item[Operator algebraists] who explore inequalities, extremal problems and invariants numerically before proving
them: sharp constants of trace inequalities and their envelopes in terms of the Michelson contrast, the
Connes--St{\o}rmer entropy as a supremum over partitions, conditional expectations, inclusions and their index, and
Jordan products. The library supplies the constraint sets as exact parametrisations and batched multistart search;
the paper scripts recompute extremal values in 40- to 50-digit arithmetic (mpmath), so that a numerical observation
comes with a certified lower bound (Sections~\ref{sec:optimisation} and~\ref{sec:symmetries}; the companion papers~\cite{nn_trace,nn_cs}).
\item[Quantum information theorists] with variational problems over states and channels that respect a symmetry or a
superselection rule: data processing and its thresholds, recovery maps, divergences resolved by sector, and learning a
channel or a Lindblad generator from measurement data. Channels within and between algebras and Lindblad generators are
available as parameters in which every value is physical, and the module \texttt{measurement} provides POVMs,
sampling, likelihoods, instruments that record their outcome in the centre, and Fisher information
(Sections~\ref{sec:renyi} and~\ref{sec:optimisation}).
\item[Physicists with conserved charges] who need many small symmetric problems at once rather than one large one:
disorder realisations, sweeps over noise models and rates, symmetry verification and post-selection, and noise-aware
design of circuits that conserve a charge (Sections~\ref{sec:chains}, \ref{sec:noisy} and~\ref{sec:control}). A system
prepared here can be passed to QuTiP or dynamiqs for further simulation.
\item[Teachers and students] of operator algebras and quantum information: the repository contains a course with
runnable tutorials in which every construction of the lectures (traces, conditional expectations, the basic
construction, channels, Lindblad dynamics) is an object one can sample and test.
\end{description}
A fifth group is emerging: researchers who treat quantum channels as layers of trainable networks. For them the
module \texttt{learn} provides channel, unitary and Lindblad layers between algebras, residual blocks, post-selection
and feed-forward through the centre as \texttt{torch.nn} modules (Section~\ref{sec:learn}).

\paragraph{When to use something else.} The library is not a circuit simulator, a tensor-network code or a large-scale
exact-diagonalisation package. For the dynamics of one large system, QuTiP, dynamiqs or QuantumToolbox.jl are faster and
more complete; for ground states of long chains, QuSpin, XDiag or tensor networks; for pulse-level control, QuTiP-QOC;
for convex problems over channels, an SDP solver, which also returns a certificate of optimality; for exact structure of
small algebras, a computer-algebra system. The interoperability module exists so that these tools and the library can
be used in one workflow.

\section{Related software}\label{sec:related}

Software that touches the algebra~\eqref{eq:M} falls into groups that store matrices or sector blocks for a purpose,
usually physics, rather than treating the algebra itself as the object. Table~\ref{tab:software} summarises the
features relevant here.

\paragraph{Linear algebra.} NumPy and SciPy~\cite{harris2020,virtanen2020}, with ARPACK~\cite{lehoucq1998} for sparse
eigenproblems, handle dense and sparse matrices; stacked arrays give batching, and any block structure is maintained by
hand. GPyTorch's \texttt{linear\_operator}~\cite{gardner2018,linearoperator} and CoLA~\cite{potapczynski2023} go
further. They represent structured matrices (diagonal, Kronecker, block-diagonal, low-rank, sums and products) lazily
and choose solvers and decompositions by structure, on GPUs and with automatic differentiation. The lazy operators of
Section~\ref{sec:lazy} follow the same idea. What these packages do not provide is the operator-algebraic layer: the
centre, several traces, states, and maps between algebras.

\paragraph{Quantum dynamics.} QuTiP~\cite{johansson2013,lambert2026} and its Julia counterpart
QuantumToolbox.jl~\cite{mercurio2025} represent states, operators and superoperators on one Hilbert space and solve
Schrödinger, Lindblad and stochastic equations. QuantumToolbox.jl runs on GPUs and supports automatic
differentiation; QuTiP~5 does so through optional JAX and CuPy data backends. QuTiP includes PIQS~\cite{shammah2018},
which uses permutational invariance to block-diagonalise the Lindblad dynamics of $N$ two-level systems. PIQS is the
closest existing analogue of the sector-wise computations here, specialised to one symmetry.
\texttt{dynamiqs}~\cite{dynamiqs} provides batched, differentiable Schrödinger and Lindblad solvers on GPUs, built on
JAX. Circuit simulators propagate density matrices under noise: Qiskit Aer~\cite{qiskit2019,qiskitaer} with noise
models on CPUs and GPUs, QuEST~\cite{quest2019} on CPUs, GPUs and clusters, and NVIDIA's cuQuantum
SDK~\cite{bayraktar2023}, whose cuDensityMat library~\cite{cudensitymat} integrates master equations for batched
operators on GPUs. PennyLane~\cite{bergholm2018} differentiates hybrid quantum-classical circuits, including mixed-state
simulation. All of these work on one Hilbert space; a conserved charge, if it is used, is exploited by the user.

\paragraph{Symmetry sectors.} Exact-diagonalisation packages build symmetry-reduced bases and sparse sector
Hamiltonians: QuSpin~\cite{weinberg2017,weinberg2019} for spins, bosons and fermions, and XDiag~\cite{wietek2026},
written in C++ with a Julia interface and distributed over many cores. Tensor-network libraries store
charge-conserving tensors block-sparsely: ITensor~\cite{fishman2022}, TeNPy~\cite{hauschild2018},
TensorKit.jl~\cite{devos2025}, and YASTN~\cite{rams2025}, which, like this library, runs on PyTorch. These packages
solve large single problems much better than ours. They do not offer batches of random elements, several traces or
maps between algebras.

\paragraph{Algebra software.} 
To our knowledge, no existing numerical library treats finite-dimensional von Neumann algebras $\bigoplus_cM_{k_c}$
as a data type with a centre, traces and states, conditional expectations, inclusions with their index, and functional
calculus. Computer-algebra systems provide exact structure theory for small algebras. Examples are Euclidean Jordan
algebras with spectral and Peirce decompositions in Orlitzky's SageMath package~\cite{orlitzky2022}, the Jordan
algebras of SageMath, the GAP package ALCO~\cite{alco}, an R package~\cite{hankin2023}, and centres, radicals and
algebras with involution in SageMath, GAP and Magma~\cite{brooksbank2012}. Numerical work block-diagonalises matrix
$*$-algebras as a preprocessing step for semidefinite
programming~\cite{murota2010,maehara2010,deklerk2011,permenter2020}. In the same vein, RepLAB~\cite{rosset2021}
decomposes representations of compact groups by sampling their commutant, and SymbolicWedderburn.jl~\cite{symbolicwedderburn}
computes Wedderburn decompositions of group-symmetric problems exactly. RepLAB's commutant object is the closest
numerical analogue; it has no centre or state interface, inclusions, index or functional calculus. Conic solvers
hard-code the Jordan products of specific symmetric cones~\cite{goulart2026,coey2023}. Subfactor software is
combinatorial, working with principal graphs and planar algebras~\cite{morrison2012,jones2014}. Proof assistants
formalise von Neumann algebras and continuous functional calculus without computing with them~\cite{dedecker2025}.
\texttt{torch\_vn\_algebra} is numerical, batched, GPU-capable and differentiable, which is what Monte Carlo sampling
and optimisation over the algebra need; it takes the block structure as input, and its own Wedderburn decomposition
(Section~\ref{sec:symmetries}) can also supply it.

\paragraph{Optimisation.} Several packages differentiate quantum dynamics. QuTiP~5 does so through its JAX data layer,
QuTiP-JAX, which gives gradients of Lindblad and Monte Carlo solutions with respect to Hamiltonian and counting-field
parameters~\cite{lambert2026}. \texttt{dynamiqs} differentiates its Schrödinger, Lindblad and stochastic solvers with
respect to arbitrary parameters and batches over Hamiltonians, initial states and jump operators~\cite{dynamiqs}.
QuantumToolbox.jl has experimental forward- and reverse-mode differentiation of its master-equation
solver~\cite{mercurio2025}, Qiskit Dynamics uses JAX for differentiable simulation and control
optimisation~\cite{puzzuoli2023}, and PennyLane differentiates noisy circuits on its mixed-state simulator by
backpropagation~\cite{bergholm2018}. All of these differentiate a trajectory in the full Hilbert space with respect
to physical parameters. QuTiP-QOC optimises control pulses with GRAPE, CRAB, GOAT and the JAX-based JOPT, also for open
systems through their Liouvillians~\cite{lambert2026}. Convex problems over states and channels, such as diamond
norms~\cite{watrous2009} and channel-adapted recovery~\cite{fletcher2007}, are semidefinite programs. The modelling
languages CVXPY~\cite{diamond2016} and PICOS~\cite{sagnol2022}, both with complex matrix variables, and the MATLAB
toolbox QETLAB~\cite{qetlab} solve them to the global optimum with a dual certificate. Riemannian optimisation
libraries supply manifolds: Pymanopt~\cite{townsend2016} includes the unitary group and its powers $U(n)^m$,
geoopt~\cite{kochurov2020} and McTorch~\cite{meghwanshi2018} add Stiefel manifolds to PyTorch, and
Manopt.jl~\cite{bergmann2022} works on the real and complex manifolds of Manifolds.jl~\cite{axen2023}.

What \texttt{torch\_vn\_algebra} adds lies between these groups. Its constraint sets are those of the algebra,
parametrised exactly and batched: the unitary group $\prod_cU(k_c)$ with blocks of different sizes, positive and
self-adjoint elements with a prescribed spectrum or Michelson contrast, channels within one algebra and between
algebras with different sectors (\texttt{StinespringParam}) and Lindblad generators (\texttt{LindbladParam}); states
and partitions of unity are parametrised in the paper scripts (Sections~\ref{sec:renyi} and~\ref{sec:cs}). Objectives can use the centre, since sector probabilities, post-selection on a
charge and conditioning on a sector are operations of the data type. Independent starts, parameter values and models
are elements of one batch, so an envelope over a grid, or a threshold scan over all $\alpha$ of
Section~\ref{sec:renyi}, is one call. The paper scripts recompute reported extremal values at the optimiser in 40- to
50-digit arithmetic, so they are certified lower bounds for suprema; they are not upper bounds. Other packages are better in
three respects. A convex problem belongs to an SDP solver, which returns the global optimum with a certificate where
our local search returns a lower bound. A single large system belongs to the simulators above. Pulse-level control,
with waveform constraints and optimisation of the evolution time, is the purpose of QuTiP-QOC. The library is meant
for problems that are non-convex, live in an algebra~\eqref{eq:M}, and come in many instances.

\begin{table}[t]
\centering\footnotesize
\setlength{\tabcolsep}{2.2pt}
\begin{tabular}{lcccccccccc}
\toprule
 & \rotatebox{90}{NumPy/SciPy} & \rotatebox{90}{QuTiP 5 (PIQS)} & \rotatebox{90}{QuantumToolbox.jl} &
 \rotatebox{90}{\texttt{dynamiqs}} & \rotatebox{90}{Qiskit Aer, QuEST} & \rotatebox{90}{cuDensityMat} &
 \rotatebox{90}{QuSpin, XDiag} & \rotatebox{90}{ITensor, TeNPy, YASTN} & \rotatebox{90}{lin.\ operator, CoLA} &
 \rotatebox{90}{this work}\\
\midrule
GPU execution & -- & ($\checkmark$) & $\checkmark$ & $\checkmark$ & $\checkmark$ & $\checkmark$ & -- & ($\checkmark$) & $\checkmark$ & $\checkmark$\\
batched over many problems & ($\checkmark$) & -- & ($\checkmark$) & $\checkmark$ & -- & $\checkmark$ & -- & -- & $\checkmark$ & $\checkmark$\\
automatic differentiation & -- & ($\checkmark$) & $\checkmark$ & $\checkmark$ & -- & ($\checkmark$) & -- & ($\checkmark$) & $\checkmark$ & $\checkmark$\\
noisy (Lindblad) dynamics & -- & $\checkmark$ & $\checkmark$ & $\checkmark$ & $\checkmark$ & $\checkmark$ & -- & -- & -- & $\checkmark$\\
symmetry sectors & -- & ($\checkmark$) & -- & -- & -- & -- & $\checkmark$ & $\checkmark$ & ($\checkmark$) & $\checkmark$\\
centre, several traces, cond.\ expectations & -- & -- & -- & -- & -- & -- & -- & -- & -- & $\checkmark$\\
maps between algebras of different sectors & -- & ($\checkmark$) & -- & -- & -- & -- & -- & -- & -- & $\checkmark$\\
Haar/COE/CSE and prescribed spectra per sector & ($\checkmark$) & ($\checkmark$) & -- & -- & -- & -- & -- & -- & -- & $\checkmark$\\
optimisation over the constraint sets of $\M$ & -- & -- & -- & -- & -- & -- & -- & -- & -- & $\checkmark$\\
large single systems & $\checkmark$ & $\checkmark$ & $\checkmark$ & $\checkmark$ & $\checkmark$ & $\checkmark$ & $\checkmark$ & $\checkmark$ & $\checkmark$ & ($\checkmark$)\\
\bottomrule
\end{tabular}
\caption{Features relevant to computations in the algebra~\eqref{eq:M}, to the best of our knowledge in October 2026
($\checkmark$: supported; ($\checkmark$): partly, through optional backends or plug-ins, for one symmetry (PIQS), or with
user code; --: not supported). QuTiP~5 uses GPUs and automatic differentiation through its JAX and CuPy backends;
PIQS handles the sectors of permutational symmetry and the maps between them. Qiskit Aer runs density-matrix
simulations with noise models on GPUs, QuEST on GPUs and clusters. Optimisation over the constraint sets of $\M$
means unitaries of $\prod_cU(k_c)$ and elements with prescribed spectra as parameters; pulse optimisation (QuTiP-QOC),
SDP solvers and manifold libraries are compared in the text. The last row is where the specialised packages are
the better choice.}
\label{tab:software}
\end{table}

\paragraph{Working with QuTiP and dynamiqs.} The library is meant to complement the simulators, not to replace them.
The module \texttt{interop} converts operators, states and channels to and from \texttt{qutip.Qobj} and dynamiqs
arrays, and Lindblad jump operators to collapse operators; for a spin chain it places the sectors in the computational basis with
QuTiP's conventions. A map between algebras is a map into the algebra, so it is converted together with the pinching
onto the output sectors, and a dense matrix outside the algebra is rejected on the way back. Through this module the
Lindblad solvers of Section~\ref{sec:noisy} are checked against \texttt{qutip.mesolve} and \texttt{dynamiqs.mesolve}
in the test suite: on a four-qubit register with damping and dephasing the exact stepper agrees with both to better than
$10^{-9}$, and the CPTP splitting used in Section~\ref{sec:noisy} converges to them at second order. A computation
can move to whichever package suits its next step, for example to QuTiP for a large single system after the sector
structure has been used to set it up.

\section{Mathematical setting}\label{sec:math}

\paragraph{Algebra and traces.} Let $\M=\bigoplus_{c=1}^C M_{k_c}$ act on $\mathcal H=\bigoplus_c\C^{k_c}$; an element
$A\in\M$ is block diagonal, $A=\bigoplus_c A_c$. Every trace on $\M$ is $\operatorname{tr}_w(A)=\sum_c w_c\Tr A_c$
with weights $w_c\ge0$, faithful iff all $w_c>0$. The library implements three of them,
\[
  \Tr_{\mathrm{blunt}}:\ w_c=1,\qquad \Tr_{\mathrm{norm}}:\ w_c=\tfrac1{k_c},\qquad
  \tau_{\mathrm{vN}}:\ w_c=\tfrac1{Ck_c}.
\]
$\Tr_{\mathrm{blunt}}$ is the plain trace of the block-diagonal matrix; we call it ``blunt'' because it ignores the block
sizes. $\Tr_{\mathrm{norm}}$ normalises each block, and $\tau_{\mathrm{vN}}$ is the faithful tracial state that gives
every sector the same weight. Empty sectors ($k_c=0$) are allowed, carry no weight, and are not counted in $C$ for
$\tau_{\mathrm{vN}}$. The library also accepts nominal factor sizes $n_c\ge k_c$ and then works in the corner
$p\M_0p\cong\bigoplus_cM_{k_c}$ of $\M_0=\bigoplus_cM_{n_c}$ ($p_c$ of rank $k_c$); all traces refer to the corner, and
in most applications $k_c=n_c$. For $A\in\M$, $|A|=(A^*A)^{1/2}$. For $A\ge0$, $A\ne0$, the \emph{Michelson contrast}
is $\Delta(A)=(\lambda_{\max}-\lambda_{\min})/(\lambda_{\max}+\lambda_{\min})\in[0,1]$, with the extreme eigenvalues
taken over all sectors; it is the operator version of the fringe visibility of optics, and $\Delta(A)=0$ iff $A$ is a
multiple of the identity (the library returns $\Delta(0)=0$).

\paragraph{States.} A state $\omega$ on $\M$ is represented by its density $\rho\ge0$ with respect to
$\Tr_{\mathrm{blunt}}$, $\omega(A)=\Tr(\rho A)$, $\sum_c\Tr\rho_c=1$. Its restriction to the centre is the classical
distribution $p_c=\Tr\rho_c$ over sectors. With respect to $\operatorname{tr}_w$ the same state has the density
$\rho^{(w)}_c=\rho_c/w_c$; the library converts between conventions explicitly.

\paragraph{Maps between sectors.} Let $\M_{\mathrm{in}}=\bigoplus_c M_{k_c}$ and $\M_{\mathrm{out}}=\bigoplus_d
M_{m_d}$. Every normal completely positive map $\Phi:\M_{\mathrm{in}}\to\M_{\mathrm{out}}$ (Schrödinger picture) is a
sum of completely positive maps between blocks,
\[
  \Phi(\rho)_d=\sum_c\sum_i K^{dc}_i\,\rho_c\,(K^{dc}_i)^*,\qquad K^{dc}_i:\C^{k_c}\to\C^{m_d}.
\]
It preserves $\Tr_{\mathrm{blunt}}$ iff $\sum_{d,i}(K_i^{dc})^*K_i^{dc}=1_c$ for all $c$. For an input state with
$\Tr\rho_c>0$, $T_{dc}=\Tr\Phi_{dc}(\rho_c)/\Tr\rho_c$ are the transition probabilities between sectors. They do not
depend on $\rho$ iff every $\sum_i(K_i^{dc})^*K_i^{dc}$ is a multiple of $1_c$; by default the library uses the
maximally mixed state of each input sector. The dual with respect to $\operatorname{tr}_w$ on both sides is
$\Phi^{*,w}(A)_c=\sum_{d,i}(w^{\mathrm{out}}_d/w^{\mathrm{in}}_c)(K_i^{dc})^*A_dK_i^{dc}$. For sector-preserving maps on
one algebra the weights cancel and all three traces give the same dual.

\paragraph{Tensor products and fusion of sectors.} $\bigl(\bigoplus_cM_{k_c}\bigr)\otimes\bigl(\bigoplus_dM_{m_d}\bigr)
=\bigoplus_{(c,d)}M_{k_cm_d}$ has sectors labelled by pairs. If the sectors carry additive charges $q_c$, $q'_d$, the
operators that conserve only the \emph{total} charge form the larger algebra
\[
  \M_1\mathbin{\hat\otimes}\M_2=\bigoplus_q M_{n_q},\qquad n_q=\sum_{q_c+q'_d=q}k_cm_d .
\]
It contains $\M_1\otimes\M_2$ and also operators such as $S^+_iS^-_j$ that move charge between the factors. If distinct
sectors of $\M_1$ carry distinct charges, the partial trace $(\Tr_2X)_c=\sum_d\Tr_{m_d}X_{(c,d),(c,d)}$ still lands in
$\M_1$. In both cases it is a map between sectors, and its dual with respect to $\Tr_{\mathrm{blunt}}$ is $A\mapsto
A\otimes1$.

\paragraph{Spin chains and dynamics.} For $L$ spins $1/2$ with conserved $S^z_{\mathrm{tot}}$, iterating the fused
product of single sites $M_1\oplus M_1$ (charges $0,1$) gives $\M=\bigoplus_{N=0}^LM_{\binom LN}$, $N$ the number of up
spins. It contains every $S^z_{\mathrm{tot}}$-conserving Hamiltonian, e.g.\ the XXZ chain
$H=\sum_{\langle ij\rangle}\bigl[\tfrac J2(S^+_iS^-_j+S^-_iS^+_j)+J\Delta_zS^z_iS^z_j\bigr]+\sum_ih_iS^z_i$. For
$H=H^*\in\M$ and jump operators $L_j\in\M$ the Lindblad equation
$\dot\rho=-i[H,\rho]+\sum_j\gamma_j(L_j\rho L_j^*-\tfrac12\{L_j^*L_j,\rho\})$ preserves every sector, so the sector
probabilities $p_c$ are conserved. Jumps between sectors, such as particle loss, are maps between sectors as above.

\section{Reproducing the earlier NumPy study}\label{sec:reproduce}

The Monte Carlo experiments of~\cite{abed2024} (Sect.~5, a single block $M_n(\mathbb R)$, $n\le5$) and of~\cite{v1}
(several sectors) sample trace functionals such as
\begin{equation}\label{eq:exp1}
  z=|\Tr(XUY)|-\Tr(XY),\qquad X,Y\ge0,\quad U\in\M\ \text{unitary},
\end{equation}
with $\Tr=\Tr_{\mathrm{blunt}}$, the operators normalised to $\lambda_{\max}=1$ and their contrasts as parameters
(``Exp.~1'' in Section~\ref{sec:optimisation}). Listing~\ref{lst:mc} is the complete library version: it declares a real
algebra, samples $10^5$ pairs with prescribed contrast and Haar-orthogonal eigenvectors in every block, and evaluates
$z$ for the whole batch.

\begin{lstlisting}[float=t, caption={The Monte Carlo of~\eqref{eq:exp1} in $M_4\oplus M_4\oplus M_8$ over $\mathbb R$:
$10^5$ samples in one batch. \texttt{operator\_from\_eigenvalues} draws independent Haar eigenvectors in every block,
\texttt{trace} is $\Tr_{\mathrm{blunt}}$.}, label=lst:mc]
import torch
from torch_vn_algebra import TypeIAlgebra

alg = TypeIAlgebra([4, 4, 8], [4, 4, 8], complex_valued=False,  # M_4 + M_4 + M_8, real,
                  precision='double')                           # as in Table baselines
B = 100_000                                     # Monte Carlo samples: one batch

def spectrum(dim, delta=0.6):                   # lambda_max = 1, Michelson contrast delta
    lo = (1 - delta) / (1 + delta)
    lam = lo + (1 - lo) * torch.rand(B, dim, dtype=torch.float64)
    lam[:, 0], lam[:, 1] = 1.0, lo
    return lam

X = alg.operator_from_eigenvalues(spectrum, batch_size=B, force_positive=True)
Y = alg.operator_from_eigenvalues(spectrum, batch_size=B, force_positive=True)
U = alg.random_unitary_operator(B)              # Haar-orthogonal in every block
z = (X @ U @ Y).trace.abs() - (X @ Y).trace     # z = |Tr(XUY)| - Tr(XY)
\end{lstlisting}

To check that nothing was lost in the re-implementation, the same computation was written three ways
(\texttt{scripts/baselines.py}). The loop formulation draws one sample and one sector at a time with
\texttt{scipy.stats.ortho\_group}, as code written directly from the definition does. The hand-vectorised NumPy version
stacks QR decompositions with the phase correction. The library version is Listing~\ref{lst:mc}. The three versions
sample the same distribution: their means of $z$ differ by at most \BaseMaxSigma\ standard
errors\ifnum\BaseNew=1, and a two-sample Kolmogorov--Smirnov test of the library against vectorised NumPy gives
$p\ge\BaseMinKS$\fi.

Table~\ref{tab:baselines} shows the times. Against the loop formulation the library gains about two orders of
magnitude for blocks of size~2, where the Python overhead dominates, and about a factor of three for 16 or 32 blocks of
size 16, where both spend their time in the same LAPACK routines. Against carefully vectorised NumPy the library is
not systematically faster on a CPU, because both call the same kernels. Its value there is that the vectorised,
correctly sampled, block-aware version is the default rather than an expert rewrite, and that the same code runs on a
GPU and can be differentiated. GPU timings depend on the precision: the consumer GPU of this run executes FP64 at
1/64 of its FP32 rate, and in double precision it does not beat the CPU on this workload. Section~\ref{sec:design}
explains where the speed comes from, and Section~\ref{sec:optimisation} shows that the batched optimisation finds
extremes of~\eqref{eq:exp1} that no amount of sampling reaches.

\begin{table}[t]
\centering\small
\ifnum\BaseNew=1
\begin{tabular}{rrrrrrr}
\toprule
$C$ & $k$ & NumPy loop & NumPy vectorised & library (CPU) & library (GPU) & GPU, FP32\\
\else
\begin{tabular}{rrrrrr}
\toprule
$C$ & $k$ & NumPy loop & NumPy vectorised & library (CPU) & library (GPU)\\
\fi
\midrule
\BaseMCRows
\bottomrule
\end{tabular}
\caption[]{Time per Monte Carlo sample [$\mu$s] of~\eqref{eq:exp1} for $C$ real sectors of size $k$, in double
precision unless stated (median of up to five runs after a warm-up\G{baselines}{; \BaseMode\ run on \BaseDevice}{}).}
\label{tab:baselines}
\end{table}

\section{Design and implementation}\label{sec:design}

\begin{figure}[t]
\centering\fig{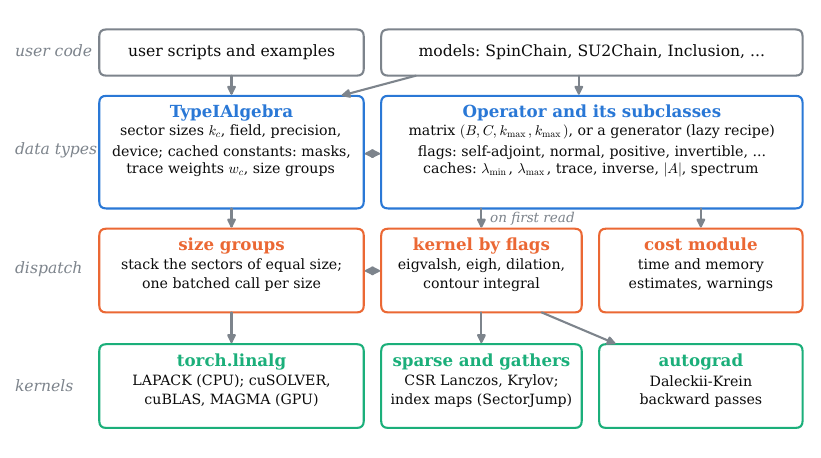}{0.8\textwidth}
\caption{Architecture. User code and models create operators in a \texttt{TypeIAlgebra}; operators are matrices or lazy
generators that carry property flags and caches. When a result is read, the dispatcher groups sectors of equal size
and chooses the kernel from the flags; the cost module estimates time and memory first. The work ends in batched
\texttt{torch.linalg} calls, sparse products or index gathers, and is differentiated by autograd with custom backward
passes.}
\label{fig:architecture}
\end{figure}

\subsection{Software stack and workflow}\label{sec:stack}

The library is pure Python on top of PyTorch~\cite{paszke2019} ($\ge1.13$) and NumPy~\cite{harris2020}
(Figure~\ref{fig:architecture}). Dense linear algebra goes through \texttt{torch.linalg}, i.e.\ LAPACK (MKL or
OpenBLAS) on CPUs and cuSOLVER, cuBLAS and MAGMA on NVIDIA GPUs; sparse products use PyTorch's CSR tensors; automatic
differentiation is PyTorch's reverse mode. The validation and paper scripts also use SciPy, pandas, Matplotlib and
mpmath. The test suite runs in continuous integration on Python~3.8--3.13 and with the oldest supported versions
(Python~3.8, PyTorch~1.13, NumPy~1.21). The paper scripts record the device, the matmul precision, the PyTorch, CUDA and
cuDNN versions and the commit in a JSON file next to their tables; files written by earlier versions of a script lack
some of these fields.

A computation proceeds in four steps. (1) An algebra is declared by its sector sizes, and optionally sector charges,
number field, precision and device (\texttt{TypeIAlgebra}, or a model such as \texttt{SpinChain} that creates it).
(2) Objects are created in it: random operators with prescribed spectra, states, channels, Hamiltonians, or
differentiable parametrisations. (3) They are combined lazily: products, functional calculus, channels, tensor products,
time evolution. (4) Functionals are evaluated for the whole batch at once and reduced to statistics, or fed into an
optimiser that updates the parameters of step~(2). Every step keeps the batch dimension, so the same code serves one
operator and $10^5$ samples. Table~\ref{tab:challenges} lists the implementation problems this raises.

\begin{table}[t]
\centering\footnotesize
\begin{tabular}{>{\raggedright\arraybackslash}p{0.25\textwidth}>{\raggedright\arraybackslash}p{0.58\textwidth}l}
\toprule
problem & solution in the library & section\\
\midrule
(i) blocks of different size on batched kernels & padded $(B,C,k_{\max},k_{\max})$ layout; one batched call per
distinct block size; single sectors and sparse sector Hamiltonians where padding would dominate & \ref{sec:layout}, \ref{sec:speed}\\
(ii) conventions between sectors & states stored as $\Tr_{\mathrm{blunt}}$-densities with explicit conversions; duals
computed with the weights of the chosen trace & \ref{sec:math}\\
(iii) exact random ensembles & Mezzadri's QR algorithm with phase correction; $W^TW$ and $W^RW$ for COE/CSE; checked
against Weingarten moments and spacing statistics & \ref{sec:calculus}, \ref{sec:validation}\\
(iv) singular gradients at degenerate spectra & Daleckii--Krein backward passes; parametrisations that build operators
from their spectra & \ref{sec:calculus}\\
(v) extremes missed by sampling & batched multistart optimisation with exact contrast constraints & \ref{sec:optimisation}\\
(vi) time and memory budgets & probing of batched kernels with time estimates; memory checked against the free
device memory & \ref{sec:safeguards}\\
(vii) field and precision & two switches per algebra, conversions, measured trade-offs & \ref{sec:numerics}\\
\bottomrule
\end{tabular}
\caption{Implementation problems and where they are resolved.}
\label{tab:challenges}
\end{table}

\subsection{Data layout and objects}\label{sec:layout}

\begin{figure}[t]
\centering\fig{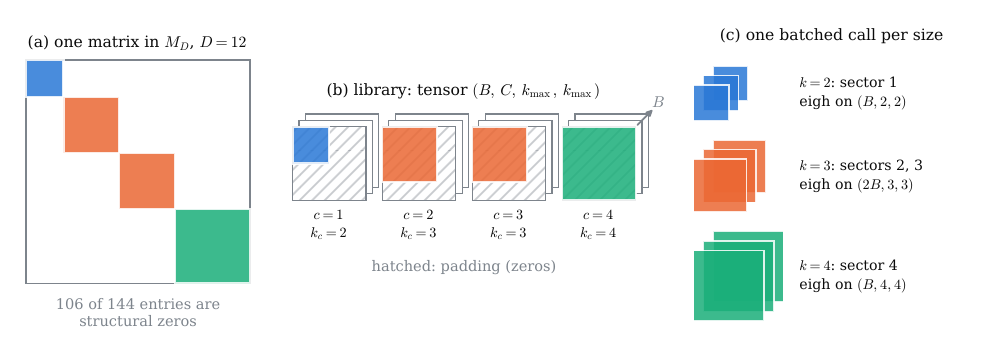}{0.88\textwidth}
\caption{Data layout for $\M=M_2\oplus M_3\oplus M_3\oplus M_4$. (a) As one block-diagonal matrix most entries are
structural zeros. (b) The library stores a batch of $B$ operators as a $(B,C,k_{\max},k_{\max})$ tensor, each block in
the top-left corner of its slice. (c) Decompositions stack the blocks of equal size, so the two sectors of size 3 and
all samples go into one batched call.}
\label{fig:layout}
\end{figure}

An operator is a tensor of shape $(B,C,k_{\max},k_{\max})$: $B$ samples, $C$ sectors, block $c$ in the top-left
$k_c\times k_c$ corner and zeros elsewhere (Figure~\ref{fig:layout}). Linear algebra is applied to the batch of active
blocks, so one call to a batched LAPACK or cuSOLVER routine serves all samples. Padding costs memory when the sector
sizes differ strongly: for a chain of 14 spins, 77\,\% of a padded operator is zeros. The library warns in that case,
and \texttt{SpinChain} lets one work in single sectors $M_{\binom LN}$ without padding, which is what dense
diagonalisation needs anyway. \texttt{Operator}, with the subclass \texttt{DensityMatrix} for states,
\texttt{Channel} (sector preserving) and \texttt{InterSectorChannel} (between sectors or algebras) are thin wrappers
around such tensors; \texttt{SectorJump} is a sparse jump operator for the Lindblad solvers (Section~\ref{sec:speed}).
Table~\ref{tab:modules} lists the modules.

\begin{table}[t]
\centering\footnotesize
\begin{tabular}{lp{0.7\textwidth}}
\toprule
module & content \\
\midrule
\texttt{algebra} & \texttt{TypeIAlgebra}, \texttt{Operator}: random operators, unitary ensembles, traces,
functional calculus\\
\texttt{states} & \texttt{DensityMatrix}, random and Gibbs states, Lüders rule, entropies, fidelity\\
\texttt{channels} & \texttt{Channel}, \texttt{InterSectorChannel}, \texttt{SectorJump}, Choi matrices, standard and random channels\\
\texttt{composite} & plain and fused (charge-conserving) tensor products, \texttt{kron}, partial traces\\
\texttt{chains}, \texttt{krylov} & \texttt{SpinChain}: $U(1)$ and momentum sectors, XXZ Hamiltonians with batched disorder,
jumps between sectors; sparse sector Hamiltonians, batched Lanczos and Krylov\\
\texttt{dynamics} & exact Schrödinger/von Neumann evolution, RK4, Lindblad equation (RK4, completely positive
splitting, $e^{t\mathcal L}$)\\
\texttt{functional} & functional calculus: self-adjoint, normal, singular values, holomorphic (contour); spectral cache\\
\texttt{optimize}, \texttt{cost} & parametrisations with prescribed contrast, channels (\texttt{StinespringParam}) and
Lindblad generators (\texttt{LindbladParam}); batched multistart \texttt{extremize} and \texttt{fit}; run-time and memory
estimates\\
\texttt{measurement} & POVMs (sectors, bases, rotated and randomised settings), sampling, likelihoods, instruments
recording the outcome in the centre, Fisher information\\
\texttt{inclusions} & \texttt{Inclusion}: inclusion matrices, conditional expectations, Markov trace, Jones,
Watatani and Pimsner--Popa indices, basic construction, spin-chain tower\\
\texttt{decompose}, \texttt{su2} & numerical Wedderburn decomposition, commutants, real/complex/quaternionic types;
\texttt{SU2Chain}: SU(2) multiplets, Clebsch--Gordan coefficients, Wigner--Eckart theorem\\
\texttt{divergences}, \texttt{jordan} & Petz and sandwiched Rényi divergences, recovery maps, $L^p$ norms; Jordan
product, quadratic representation, spin factors\\
\texttt{permutational} & $N$ two-level systems with permutation symmetry: total-spin sectors, collective operators,
Dicke states, local jumps between sectors (\texttt{PermutationalLindbladian})\\
\texttt{interop} & conversion to and from dense matrices, QuTiP and dynamiqs (chain layout, pinching of maps)\\
\texttt{learn} & channels as trainable networks: channel, unitary and Lindblad layers, residual blocks, post-selection,
feed-forward through the centre; losses (Section~\ref{sec:learn})\\
\bottomrule
\end{tabular}
\caption{Modules of \texttt{torch\_vn\_algebra}.}
\label{tab:modules}
\end{table}

\subsection{Lazy evaluation and caches}\label{sec:lazy}

\paragraph{Operators are recipes until used.} An \texttt{Operator} holds either its matrix or a \emph{generator}, a
closure that computes it. Sums, products, adjoints, functions $f(A)$, channel images $\Phi(\rho)$ and maps between
sectors return new operators that hold only a generator. The matrix is formed on the first access to
\texttt{.matrix}: the generator runs, its result is stored, and the generator is dropped so that it no longer keeps the
parent matrices alive. A long expression therefore costs exactly the products needed for the requested result, and
branches that are never read are never computed. The batch size of a lazy operator is known without computing it, so
shapes and memory estimates are available before any work is done. Laziness defers and skips work but does not fuse
kernels. Where fusion matters, the library uses direct formulas instead: $\Tr(\rho A)=\sum_{ij}\rho_{ij}A_{ji}$ in
$O(k^2)$ per block without forming $\rho A$, and the sector probabilities of a channel image as one contraction
$\sum K_{mi}\rho_{ij}\overline{K_{mj}}$ without forming the image.

\paragraph{Properties travel with the recipe.} Every operator carries flags (self-adjoint, normal, positive,
invertible, projection) that are known, unknown, or checked. Operations set the flags that hold mathematically: sums of
self-adjoint operators are self-adjoint, products of invertible operators invertible, and an operator built from its
spectrum is self-adjoint, and positive if the spectrum is. Flags select the kernel. A self-adjoint operator uses the
Hermitian eigensolver for its norm, trace norm, inverse and $|A|$, never an SVD, and $\lambda_{\max}$ of an operator
built from its spectrum is read from the spectrum. An unknown flag is computed once, on demand, and stored. Flags passed
by the user are trusted on eager operators and validated when a lazy operator is materialised, with one host
synchronisation for all checks.

\paragraph{Caches.} The extreme eigenvalues (from one eigenvalue call, without eigenvectors), the trace, the inverse
and $|A|$ are cached on the operator. The eigendecomposition itself is kept only on request (\texttt{cache\_spectrum()},
or the context \texttt{caching()}); afterwards every function of the operator, for instance $e^{-itH}$ for many $t$,
costs two matrix products. The cache is keyed by the identity and version counter of the stored tensor, so an in-place
change invalidates it. The algebra caches its own constants on the device (block sizes, masks, identities, trace
weights, index tensors of the size groups), so no Python list is copied to the GPU inside a loop.

\subsection{Where the speed comes from}\label{sec:speed}

Three principles account for the speed, and Table~\ref{tab:arch} measures several of the decisions behind them by
computing the same result with and without each. The alternatives are deliberately the simple ones a direct
implementation would use, so the factors measure the value of each decision, not a comparison with other software.

The first principle is to give every kernel as much independent work as possible. Every operation acts on the whole
batch, so a Monte Carlo study of $10^5$ samples is one tensor program, not $10^5$ calls. Decompositions are applied to
the stacked active blocks of all sectors of equal size, one batched call per distinct size
(Figure~\ref{fig:layout}c); contiguous sectors are selected as views, and an unpadded algebra needs no scatter.

The second principle is to compute only what is read. One batched Hermitian eigensolver serves spectra, $f(A)$ and
normal operators, and on GPUs also $|A|$, singular values and polar decompositions of blocks larger than 32
(Section~\ref{sec:calculus}); it is fast and, unlike batched SVD on GPUs, not limited in block size. Where only the spectrum is needed (extreme eigenvalues, contrasts, entropies),
eigenvectors are not formed. Where the spectrum is known from the construction, no decomposition is run at all, and a
cached eigendecomposition turns many functions of one operator into products.

The third principle is to keep the device busy and the memory small. Inner loops do not call \texttt{.item()} or
branch on device values, so checks synchronise once per call, and the Lindblad steppers prepare constant propagators
once per step size. Sparse structure is used where it exists. Sector Hamiltonians of spin chains are CSR matrices for
batched Lanczos and Krylov solvers (Section~\ref{sec:chains}). Jumps that map basis vectors to multiples of basis
vectors ($S^\pm_i$, $S^z_i$, particle loss) are stored as \texttt{SectorJump}, with a target sector per sector and a
target position and coefficient per basis vector. The Lindblad solvers then apply $L\rho L^*$ as a weighted permutation
of the entries of $\rho$, $L^*ZL$ as a gather and $L^*L$ as a diagonal, in $O(\sum_ck_c^2)$ operations against
$O(C^2k_{\max}^3)$ for a dense map between sectors. A batch of coefficients gives one jump per sample, so a sweep over
noise rates is one batched evolution.

\begin{table}[t]
\centering\small
\begin{tabular}{llrrr}
\toprule
mechanism & computation & library [ms] & without [ms] & speed-up\\
\midrule
\ArchRows
\bottomrule
\end{tabular}
\caption[]{The same result with and without one design decision (best of several runs, double precision, CPU with
\G{architecture}{\ArchThreads}{several} threads; \texttt{scripts/architecture\_bench.py}).
``Without'': eigenvalues by \texttt{eigvalsh} instead of the known spectrum; \texttt{eigh} instead of
\texttt{eigvalsh}; one decomposition per time instead of the spectral cache; a Python loop over sectors; forming
$\rho A$; dense jumps between sectors. For the last row the eight jumps $S^-_i$ take
\G{architecture}{\ArchJumpDense}{--}\,MiB as dense maps and \G{architecture}{\ArchJumpSparse}{--}\,MiB as
\texttt{SectorJump}s, and the two evolutions agree to \G{architecture}{$\ArchJumpDiff$}{round-off}; its time is
dominated by the dense Hamiltonian, which both versions share.}
\label{tab:arch}
\end{table}

\subsection{Random unitaries, functional calculus and derivatives}\label{sec:calculus}

\paragraph{Random unitaries.} Haar unitaries on $U(n)$ and $O(n)$ come from the QR decomposition of a matrix with
i.i.d.\ complex (real) Gaussian entries after fixing the phases (signs) of $\operatorname{diag}R$~\cite{mezzadri2007};
$SU(n)$ and $SO(n)$ follow by multiplying with $\det^{-1/n}$ or flipping the sign of one column. The circular
orthogonal and symplectic ensembles are $W^TW$ and $W^RW$ with $W$ Haar on $U(n)$ ($n$ even for $W^RW$, $W^R$ the
quaternion dual)~\cite{mehta2004}. Section~\ref{sec:validation} checks all of them.

\paragraph{Functional calculus.} The module \texttt{functional} provides $f(A)$ at four levels of generality, each
with batched kernels and one call per block size~\cite{higham2008}. (1) For self-adjoint $A$, $f(A)=Vf(\Lambda)V^*$
from the Hermitian eigensolver. (2) For normal $A=H+iK$, $H$ and $K$ commute, and the eigenvectors of $H+\alpha K$
diagonalise $A$ for all but finitely many $\alpha$. The library tries a few irrational values, refines nearly tied
eigenvectors with a second coefficient and corrects the remaining residual to first order, so normal operators also
need only the Hermitian eigensolver. (3) Singular values, $|A|$, the pseudo-inverse and the polar decomposition come
from batched SVD, except on GPUs for blocks larger than 32, where batched SVD falls back to a loop; there they come
from the Hermitian eigensolver of the dilation $\left(\begin{smallmatrix}0&A\\A^*&0\end{smallmatrix}\right)$, whose
eigenvalues are $\pm s_i$. (4) For arbitrary $A$ and $f$ holomorphic near the spectrum,
$f(A)=(2\pi i)^{-1}\oint f(z)(z-A)^{-1}dz$ is evaluated by the trapezoidal rule on a circle around the
spectrum~\cite{trefethen2014}, i.e.\ by batched LU solves, doubling the nodes until two results agree; a result whose error
estimate stays above the square root of the tolerance raises an error, and one between the two triggers a warning.

\paragraph{Differentiation.} All numerical operations are differentiable, except the unitary factor of the polar
decomposition, which is returned without gradient. Gradients propagated naively through eigenvectors
contain factors $1/(\lambda_i-\lambda_j)$ and fail at degenerate spectra, which are typical here (projections,
symmetric and extremal operators). The functional calculus therefore has its own backward pass: by the Daleckii--Krein
formula~\cite{daleckii1965,bhatia1997}, the derivative of $f$ at self-adjoint $A$ is $dF=V\,(L\circ V^*dA\,V)\,V^*$,
with the Loewner matrix $L_{ij}=(f(\lambda_i)-f(\lambda_j))/(\lambda_i-\lambda_j)$ of divided differences, replaced by
$f'$ for coinciding eigenvalues, which is finite whenever $f$ is differentiable on the spectrum. Backward passes through
the eigenvectors stay finite at degenerate spectra only for losses that do not mix a degenerate
eigenspace~\cite{kasim2020}, and they then miss the entries $f'(\lambda)$ inside the degenerate blocks. The same
backward pass serves $\sqrt A$, $\log A$, $A^p$, $|A|$ (as $\sqrt{A^*A}$, with $L_{ij}=1/(s_i+s_j)$) and inverses of
self-adjoint operators (the pseudo-inverse of a general operator has a constant-rank backward pass of its own);
normal $A$ use complex divided differences, and the contour integral is differentiated through its solves.

\subsection{Number field and precision}\label{sec:numerics}

Every algebra has two switches, \texttt{complex\_valued} and \texttt{precision} (\texttt{'single'} or
\texttt{'double'}), giving \texttt{float32}, \texttt{float64}, \texttt{complex64} and \texttt{complex128}; everything
built on the algebra inherits them. \texttt{alg.like(...)} creates the same sector structure with other numerics and
\texttt{op.cast(alg2)} moves an operator there, for instance to explore in single precision and re-evaluate the
interesting cases in double precision.

\paragraph{The field is part of the model.} Real matrices change the symmetry group and the random ensembles, not only
the speed. Haar matrices of a real algebra are orthogonal ($\mathbb E|U_{11}|^4=3/(n(n+2))$ instead of $2/(n(n+1))$,
Table~\ref{tab:validation}). Real symmetric and complex Hermitian random matrices have mean spacing ratios $0.5307$
(GOE) and $0.5996$ (GUE)~\cite[Table~I]{atas2013}, against \G{numerics}{\NumRGOE\ and \NumRGUE}{(not generated)} in our
samples. Real arithmetic loses nothing when the problem is real, e.g.\ for ground states of Hamiltonians without complex
hoppings, or for the supremum of $|\Tr(XUY)|$ over unitaries for real $X,Y\ge0$, which is attained by an orthogonal
matrix. Complex matrices are required for $e^{-iHt}$ and Lindblad dynamics with a Hamiltonian (the library refuses to
run them in a real algebra), for Haar unitaries, CUE/CSE and random pure states, and for systems without an antiunitary
symmetry squaring to $+1$.

\paragraph{Precision.} Single precision resolves about $10^{-7}$ relative to the largest entry, enough for Monte
Carlo statistics and exploratory optimisation. Double precision is needed when the quantity of interest is small
compared with the operators, as Table~\ref{tab:numerics-acc} shows: the entropy of an almost pure state is wrong by more
than $100\,\%$ in single precision, and the round-off floor of a trace inequality at equality is several orders of
magnitude higher.

\paragraph{Cost.} An entry takes 4, 8, 8 or 16 bytes (\texttt{alg.bytes\_per\_operator(B)} reports the memory of a
batch). A complex multiply--add costs about four real ones; for eigensolvers the measured ratio of complex to real is
smaller, while double against single precision costs a factor that depends on the device
(Table~\ref{tab:numerics-time}, measured on the GPU of the full run). Data-centre GPUs execute FP64 at half the FP32
rate, consumer GPUs at 1/32--1/64; the full-size runs of this paper used a consumer GPU. FP32 products may run in the
TF32 format on recent GPUs; PyTorch keeps true FP32 by default, the library does not change this, and the paper scripts
record the setting. We recommend the smallest type that is exact for the model and the precision that the size of the
effect requires.

\begin{table}[t]
\centering\small
\begin{tabular}{lrrrrrrrr}
\toprule
type & B/entry & MB/batch & \NumTimeHeader\\
\midrule
\NumTimeRows
\bottomrule
\end{tabular}
\caption[]{Bytes per entry, memory of one batch [MB], and time relative to \texttt{float32} for
$C=\G{numerics}{\NumC}{8}$ sectors of size $k=\G{numerics}{\NumK}{16}$ and a batch of \G{numerics}{\NumB}{B}
operators\G{numerics}{ (\NumMode\ run on \NumDevice; float32 times per sample: \NumRefTimes)}{}.}
\label{tab:numerics-time}
\end{table}

\begin{table}[t]
\centering\small
\begin{tabular}{lcccc}
\toprule
type & $\max\|UU^*-1\|$ & $\|\sqrt A^2-A\|/\|A\|$ & rel.\ error $S(\rho)$, $\rho\approx$ pure & $|z|$ at equality\\
\midrule
\NumAccRows
\bottomrule
\end{tabular}
\caption[]{Accuracy of the same computations in the four storage types: Haar unitaries ($n=64$), SVD square
root, entropy of $\rho=(1-\varepsilon)|0\rangle\langle0|+\varepsilon\,1/n$ with $\varepsilon=10^{-7}$, and the
round-off floor of $z=\Tr(Y|X|Y)-\Tr|YXY|$ for commuting $X,Y$ (exactly $0$)\G{numerics}{; \NumMode\ run on \NumDevice}{}.}
\label{tab:numerics-acc}
\end{table}

\subsection{Safeguards}\label{sec:safeguards}

Batched computations make it easy to request hours of work or more memory than the device has. The \texttt{cost}
module therefore times the first few items of every large batched decomposition and matrix exponential and warns with
the expected total time when it exceeds a threshold (10\,s by default). It compares the estimated peak memory of large
allocations with the free device memory, warning above 50\,\% and raising \texttt{InsufficientMemoryError} above
90\,\%. After the second step of a time-stepping or optimisation loop it warns with the expected duration. All
thresholds can be changed, and the checks can be switched off locally.

\section{Usage by example}\label{sec:usage}

The repository contains short complete programs in \texttt{man/article/listings}; the test suite runs them, so they
stay in sync with the code. Each runs on a CPU in about half a minute or less; the nine programs \texttt{ex*.py} use a
GPU when one is available.
Table~\ref{tab:examples} lists the nine introductory examples; programs of later sections are described where they are
used.

\begin{table}[t]
\centering\scriptsize
\begin{tabular}{>{\raggedright\arraybackslash}p{0.25\textwidth}>{\raggedright\arraybackslash}p{0.39\textwidth}%
>{\raggedright\arraybackslash}p{0.28\textwidth}}
\toprule
program & what it shows & key API\\
\midrule
\repofile{ex1_random_operators.py} & $10^4$ positive operators in $M_2\oplus M_3\oplus M_4$ with contrast $0.6$ and
Haar-random eigenvectors in every sector; the three traces and spectral functions &
\texttt{TypeIAlgebra}, \texttt{operator\_from\_eigenvalues}, \texttt{michelson\_contrast}, \texttt{Tr\_blunt},
\texttt{tau\_vN}, \texttt{apply\_function}\\
\repofile{ex2_states_measurement.py} & Lüders conditioning on a projective measurement moves all probability into one
sector; expectation values do not depend on the trace that represents the state &
\texttt{states.random\_density\_matrix}, \texttt{condition\_on}, \texttt{sector\_probabilities}, \texttt{density}\\
\repofile{ex3_channels.py} & particle loss of two fermionic modes as a map between the sectors $N=0,1,2$; unital
and non-unital channels and the contrast & \texttt{InterSectorChannel.from\_blocks},
\texttt{transition\_matrix}, \texttt{random\_mixed\_unitary\_channel}\\
\repofile{ex4_dynamics.py} & spontaneous emission of a two-level atom by RK4 and by the exponential of the
generator, against $P_e(0)e^{-\gamma t}$ and $|\rho_{eg}(0)|e^{-\gamma t/2}$ & \texttt{dynamics.lindblad\_evolve},
\texttt{dynamics.lindblad\_channel}\\
\repofile{ex5_page_entropy.py} & mean entanglement entropy of random pure states on $\C^m\otimes\C^n$ against
Page's formula~\cite{page1993} & \texttt{tensor\_product}, \texttt{partial\_trace},
\texttt{states.von\_neumann\_entropy}\\
\repofile{ex6_inequality_search.py} & Haar sampling against batched gradient ascent for
$\sup_U(|\Tr(XUY)|-\Tr(XY))=\|YX\|_1-\Tr(XY)$ (Section~\ref{sec:optimisation}) &
\texttt{random\_unitary\_operator}, \texttt{expm}, \texttt{trace\_norm}\\
\repofile{ex7_spin_chain.py} & Heisenberg ring of ten spins: the sector of the ground state in a uniform field,
the mean level-spacing ratio of 100 disorder realisations in one batch, entanglement entropies &
\texttt{SpinChain}, \texttt{random\_field\_heisenberg}, \texttt{level\_spacing\_ratio},
\texttt{entanglement\_entropy}\\
\repofile{ex8_large_and_open_chains.py} & Lanczos ground state of a sector of dimension $12870$, Krylov evolution
of a domain wall, momentum sectors, particle loss through jumps between sectors & \texttt{xxz\_sparse},
\texttt{krylov.ground\_state}, \texttt{krylov.evolve}, \texttt{xxz\_momentum}\\
\repofile{ex9_qaoa_xy_mixer.py} & XY mixer on a ring of six qubits with amplitude damping, dephasing and
post-selection on the Hamming weight & \texttt{SpinChain.xxz}, \texttt{lowering}, \texttt{algebra.central},
\texttt{condition\_on}\\
\bottomrule
\end{tabular}
\caption{The introductory programs in \texttt{man/article/listings} of the repository.}
\label{tab:examples}
\end{table}

Several examples check exact statements. In \texttt{ex3}, the Kraus operators $\sqrt{1-gN}$, $\sqrt g\,a_1$, $\sqrt
g\,a_2$ ($0\le g\le\frac12$) map sector $N$ to $N$ and $N-1$, and the classical part is the transition matrix of a pure
death process, $T=\bigl(\begin{smallmatrix}1&g&0\\0&1-g&2g\\0&0&1-2g\end{smallmatrix}\bigr)$ (columns: from
$N=0,1,2$). A unital positive map cannot increase the Michelson contrast, while the non-unital loss channel increases
it for most samples. In \texttt{ex7} the ground state moves through the sectors until the chain is fully polarised above the
saturation field $h_s=2J$; \texttt{ex9} is the starting point of Section~\ref{sec:noisy}.

\section{Validation against exact results}\label{sec:validation}

\begin{figure}[t]
\centering
\fig{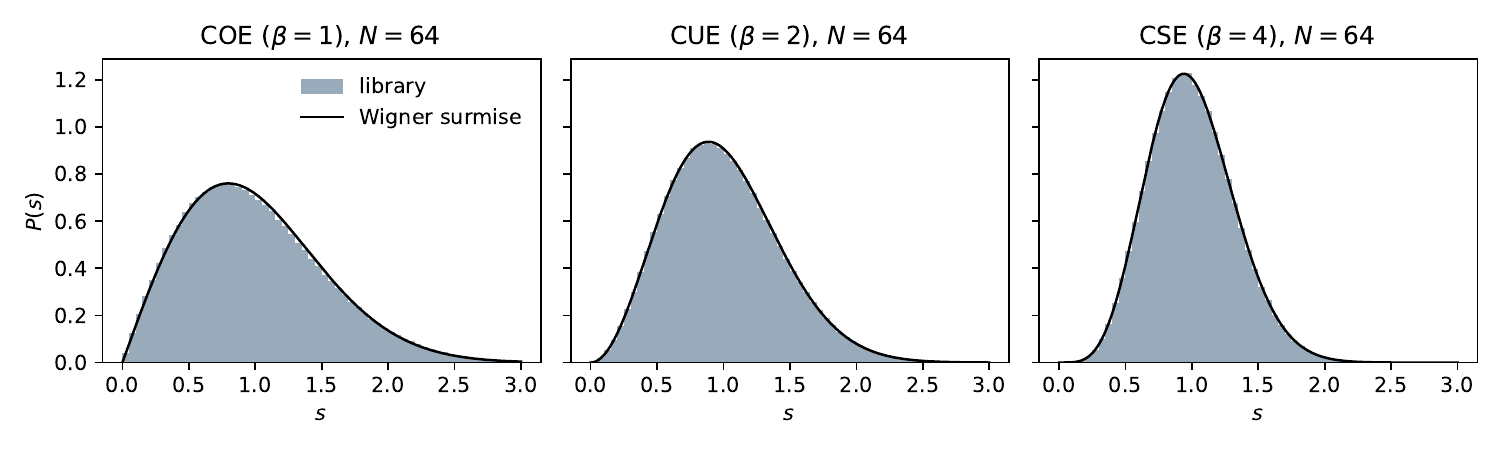}{0.8\textwidth}\\[1mm]
\fig{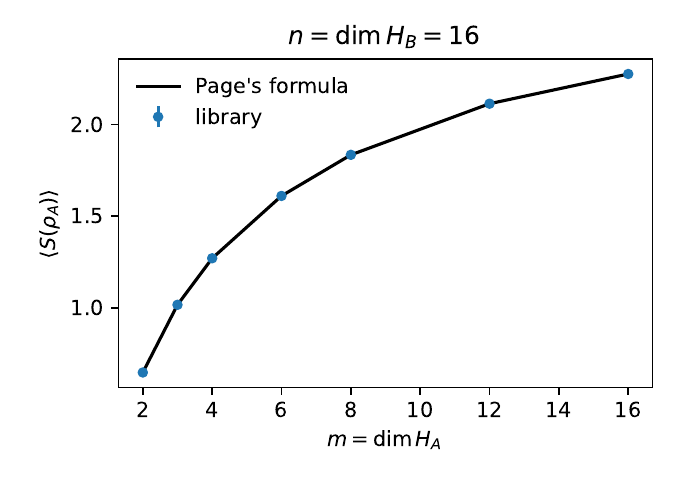}{0.34\textwidth}
\caption[]{Top: nearest-neighbour spacings of COE, CUE and CSE eigenphases\G{validation}{ ($n=\ValNce$, \ValSce\
matrices per ensemble)}{} against the Wigner surmise. Bottom: mean entanglement entropy of random pure states against
Page's formula\G{validation}{ (\ValSPage\ states per point)}{} (error bars: two standard errors).}
\label{fig:spacings}
\end{figure}

Table~\ref{tab:validation} compares Monte Carlo estimates (double precision) with exact values. The rows cover fourth
moments of Haar unitaries and orthogonal matrices, and for traceless involutions $A=B$ and Haar $U$ the value
$\mathbb E\,\tau\bigl((AUBU^*)^2\bigr)=-1/(n^2-1)$ of the Weingarten calculus~\cite{collins2006}. They also cover Page's
formula for $\langle S(\rho_A)\rangle$~\cite{page1993,foong1994,sanchezruiz1995} and the mean purity $(k+r)/(kr+1)$ of
random states of rank $r\le k$~\cite{lubkin1978,zyczkowski2001}. The last column is the deviation in standard errors; for
a correct sampler these should look like standard normal deviates. For the circular ensembles the Wigner surmise only
approximates the exact spacing distribution, so no $z$-score is given; Figure~\ref{fig:spacings} shows the
distributions, with $L^1$ distances \G{validation}{\ValLoneOne, \ValLoneTwo\ and \ValLoneFour\ for
$\beta=1,2,4$}{(not generated)} from the surmise and level repulsion $P(s)\sim s^\beta$. The spontaneous-emission
example agrees with the analytic solution to \G{validation}{$\ValEmissionExp$ (exact exponential) and $\ValEmissionRK$
(RK4)}{(not generated)}. The three traces satisfy their defining identities on random operators to round-off
(Table~\ref{tab:traces}), and $\Tr_{\mathrm{norm}}=C\,\tau_{\mathrm{vN}}$ holds to
\G{validation}{$\ValTraceNormTau$}{round-off}.

\begin{table}[t]
\centering\small
\begin{tabular}{llrrrr}
\toprule
quantity & parameters & exact & Monte Carlo & std.\ error & $z$\\
\midrule
\ValidationRows
\bottomrule
\end{tabular}
\caption[]{Monte Carlo estimates versus exact values (for $\langle s^2\rangle$ the Wigner surmise, an approximation,
hence no $z$)\G{validation}{; samples: \ValSHaar\ per Haar moment, \ValSW\ per Weingarten row, \ValSce\ matrices per
circular ensemble, \ValSPage\ states per Page row, \ValSPur\ per purity row (\ValMode\ run on \ValDevice, PyTorch
\ValTorch)}{}.}
\label{tab:validation}
\end{table}

\begin{table}[t]
\centering\small
\begin{tabular}{lcccc}
\toprule
trace & error of $\operatorname{tr}(1)$ & $\max|\operatorname{tr}(AB)-\operatorname{tr}(BA)|$ &
$\max|\operatorname{tr}(UAU^*)-\operatorname{tr}(A)|$ & $\min\operatorname{tr}(A^*A)/\|A\|_F^2$\\
\midrule
\ValTraceRows
\bottomrule
\end{tabular}
\caption[]{Identities of the three traces on 2000 random non-Hermitian operators in $M_2\oplus M_3\oplus M_5$
(double precision): normalisation ($\sum_ck_c$, $C$, $1$), tracial property, unitary invariance, and faithfulness;
the last column is bounded below by $\min_cw_c$\G{validation}{; \ValMode\ run on \ValDevice}{}.}
\label{tab:traces}
\end{table}

\section{From sampling to optimisation}\label{sec:optimisation}

The NumPy study~\cite{abed2024,v1} sampled three trace functionals:
\begin{align*}
  \text{Exp.~1:}\quad & z=|\Tr(XUY)|-\Tr(XY), && X,Y\ge0,\ U\in\M\ \text{unitary},\\
  \text{Exp.~2:}\quad & z=\Tr|XY|-\Tr(|X|\,|Y|), && X=UX_0,\ Y=VY_0,\ X_0,Y_0\ge0,\ U,V\in\M\ \text{unitary},\\
  \text{Exp.~3:}\quad & z=\Tr(Y|X|Y)-\Tr|YXY|, && X=X^*,\ Y\ge0,
\end{align*}
with $\Tr=\Tr_{\mathrm{blunt}}$, operators normalised to $\lambda_{\max}=1$, and the contrasts of the operators as
parameters (of $|X|$ and $|Y|$ for Exp.~2 and 3). Two observations motivate a different approach.

\begin{figure}[t]
\centering\fig{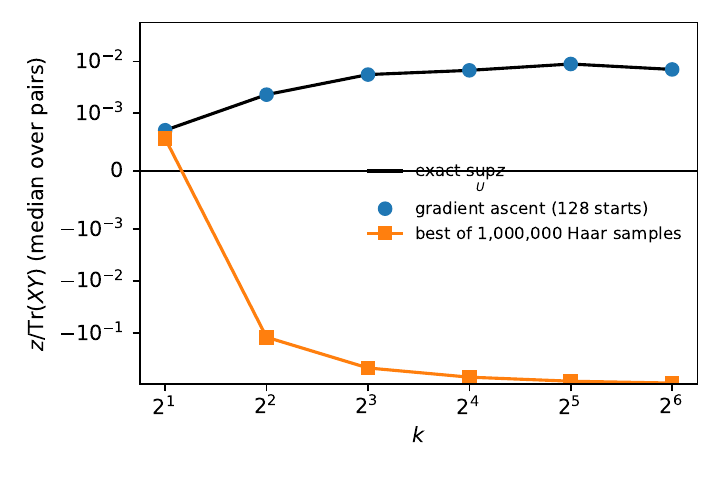}{0.45\textwidth}
\caption{Best $z$ found by sampling and by gradient ascent, relative to $\Tr(XY)$, against the exact
supremum (symmetric log scale).}
\label{fig:search}
\end{figure}

\paragraph{Sampling misses the extremes.} For fixed $X,Y$ the supremum in Exp.~1 is known exactly, $\sup_U
z=\|YX\|_1-\Tr(XY)>0$. The best of \G{search}{\SearchSamples}{many} Haar samples falls short of it for every $k$, misses
the violation for all pairs from $k=4$ on, and moves further away as $k$ grows (Table~\ref{tab:search},
Figure~\ref{fig:search}). Batched gradient ascent (\G{search}{\SearchStarts\ starts, \SearchSteps\ steps}{multistart})
finds the supremum to a relative error of at most \G{search}{$\SearchMaxErr$}{about $10^{-5}$}. This is concentration of
measure: the unitaries close to the maximiser form a set whose Haar measure decays rapidly with the dimension. The same
effect explains why the Monte Carlo samples of~\cite{v1} showed no violations for $k=16$: it is a property of the
sampling, not of the inequality.

\begin{table}[t]
\centering\footnotesize\setlength{\tabcolsep}{4pt}
\begin{tabular}{rcccrccrr}
\toprule
$k$ & \multicolumn{2}{c}{violation found} & $\sup z/\Tr XY$ & best sample & \multicolumn{2}{c}{ascent rel.\ error}
& \multicolumn{2}{c}{time [s]}\\
& sampling & ascent & & $z/\Tr XY$ & median & max & sampl. & ascent\\
\midrule
\SearchRows
\bottomrule
\end{tabular}
\caption[]{Exp.~1 for fixed random $X,Y$ in $M_k(\C)$ (eigenvalues uniform in $[0.2,1.2]$, double precision): Haar
sampling versus gradient ascent (medians over \G{search}{\SearchPairs}{several}
pairs\G{search}{; \SearchMode\ run on \SearchDevice}{}).}
\label{tab:search}
\end{table}

\paragraph{Constrained optimisation.} The natural objects are envelopes such as
$\overline z(\delta)=\sup\{z:\Delta(X)=\delta\}$ and $\underline z(\delta)=\inf\{z:\Delta(X)=\delta\}$, taken over all
remaining variables. The module \texttt{optimize} parametrises the constraint set exactly. The spectrum is
$\lambda_{\max}=1$, $\lambda_{\min}=(1-\delta)/(1+\delta)$, and the other eigenvalues
$\lambda_{\min}+(1-\lambda_{\min})(1+\sin\theta)/2$ in between; unlike a sigmoid, this reaches both ends at finite
$\theta$, where extremal spectra often lie. The positions of $\lambda_{\max}$ and $\lambda_{\min}$ are drawn per start,
in the same sector or in different ones. The eigenvectors are $U=U_0e^{G-G^*}$ with a Haar-random $U_0$ per start and a
chart that is re-centred after every round. Discrete variables, such as the signs of the eigenvalues of a self-adjoint
$X$ (Exp.~3), are drawn per sector and improved by single sign flips between gradient rounds. Every iterate is
admissible, and every grid point and start is one element of the batch, so each envelope is one batched optimisation.

Two implementation details matter. (a) In a real algebra a chart $U_0e^{G-G^*}$ never leaves the connected component of
$U_0$, and an extremiser with determinant $+1$ in every sector is reached from a Haar-random $U_0$ only with probability
$2^{-C}$. Half of the starts therefore begin at a structured point (diagonal $X$, $Y$ rotated in pairs of eigenvectors of
$X$, two-level spectra). With Haar-random starts only, a first run for $C=32$ sectors of size 2 fell short of the true
envelopes by up to $35\%$. With structured starts the searches of the companion paper~\cite{nn_trace} reproduce the
known envelopes of the classical inequalities to $10^{-7}$, and every closed form proved there to $3\times10^{-4}$ or better. (b) Where the
objective has a kink at the maximiser (a trace norm at a degenerate point), the remaining gap decreases with the step
size. Applied to the three functionals, the optimiser located their envelopes for sectors of every size; the closed
forms it suggested are proved in the companion paper~\cite{nn_trace}.

\paragraph{Channels, generators and measurements as parameters.} The same machinery applies to maps. In
\texttt{optimize.StinespringParam} every input sector $c$ carries an isometry $V_c=Z_c(Z_c^*Z_c)^{-1/2}$ into an
environment of dimension $r$ times the output sectors, and the Kraus operators are its blocks; every value of the free
matrices $Z_c$ is a trace-preserving channel, within one algebra or between two algebras with a prescribed set of
allowed sector transitions, and Gaussian $Z_c$ give Haar-random isometries. \texttt{LindbladParam} does the same for
GKSL generators (a self-adjoint Hamiltonian and free jump operators inside and between sectors), and \texttt{fit}
minimises a loss over such parameters and leaves them at the fitted point. With the module \texttt{measurement}
(POVMs, sampling of counts, likelihoods, instruments, Fisher information) this turns learning a channel or a
generator from data into a call of the optimiser. In the test suite a channel of $M_1\oplus M_3\oplus M_2$ of Kraus
rank two is recovered from twelve input--output pairs to $10^{-5}$, and the generator of a damped and dephased qubit
from two time points of six trajectories, predicting an unseen trajectory to $10^{-4}$. These are small checks, not
studies; tomography in symmetry algebras is the subject of ongoing work.

\subsection{Rediscovering the data-processing thresholds of the Rényi divergences}\label{sec:renyi}

A further test of the method is to ask it where the range of validity of a known inequality ends, without telling it.
The module \texttt{divergences} evaluates the Petz divergences
$D_\alpha(\rho\|\sigma)=\frac{1}{\alpha-1}\log\Tr(\rho^\alpha\sigma^{1-\alpha})$~\cite{petz1986} and the sandwiched
divergences $\widetilde D_\alpha(\rho\|\sigma)=\frac{1}{\alpha-1}\log\Tr(\sigma^{\frac{1-\alpha}{2\alpha}}\rho\,
\sigma^{\frac{1-\alpha}{2\alpha}})^{\alpha}$~\cite{mullerlennert2013,wilde2014}, batched and differentiable, with exact
support conventions and as logarithms of sums over sectors; it also provides the Petz recovery
map~\cite{petz1986suff,petz1988chan} and the universal recovery map of~\cite{junge2018}. The data-processing
inequality $D(\Phi(\rho)\|\Phi(\sigma))\le D(\rho\|\sigma)$ holds for the Petz divergences for
$\alpha\in(0,1)\cup(1,2]$~\cite{petz1986} and for the sandwiched ones for
$\alpha\ge\frac12$~\cite{mullerlennert2013,wilde2014,beigi2013,frank2013}, and fails in general outside these
ranges~\cite{mullerlennert2013,berta2017,zhang2020}. \texttt{scripts/renyi\_threshold.py} maximises the violation
$v(\alpha)=D(\Phi(\rho)\|\Phi(\sigma))-D(\rho\|\sigma)$ on a grid of $\alpha\in[0.2,3]$, over states on $M_d$ and over
the complete dephasing and all channels with a $d$-dimensional environment, with all values of $\alpha$ and all starts
in one batch. Every reported violation is attained at an explicit point and recomputed in mpmath with
\G{renyi_threshold}{\RenyiDigits}{50} significant digits, so a certified violation proves that the inequality fails at
that $\alpha$; the absence of a violation is evidence only.
\ifdefined\RenyiSandLow The sandwiched inequality is violated (certified) at
$\alpha=\RenyiSandLow$ with $v=\RenyiSandLowV$, and no violation above $\RenyiTol$ was found at
$\alpha=\RenyiSandHigh$; the Petz inequality is violated at $\alpha=\RenyiPetzHigh$ with $v=\RenyiPetzHighV$, and no
violation was found at $\alpha=\RenyiPetzLow$. The thresholds are thus bracketed by $[\RenyiSandLow,\RenyiSandHigh]$ and
$[\RenyiPetzLow,\RenyiPetzHigh]$, in agreement with $\frac12$ and $2$ ($d\in\{\RenyiDims\}$, \RenyiStarts\ starts and
\RenyiSteps\ steps per $\alpha$; \RenyiMode\ run on \RenyiDevice). Table~\ref{tab:renyi} and Figure~\ref{fig:renyi}
show $v(\alpha)$ on the grid.

\begin{table}[t]
\centering\footnotesize
\begin{tabular}{rcccc}
\toprule
 & \multicolumn{2}{c}{sandwiched $\widetilde D_\alpha$} & \multicolumn{2}{c}{Petz $D_\alpha$}\\
$\alpha$ & dephasing & channels & dephasing & channels\\
\midrule
\RenyiRows
\bottomrule
\end{tabular}
\caption[]{Largest certified violation $v(\alpha)$ of the data-processing inequality on $M_2$ (``--'': no violation
above the tolerance $\RenyiTol$ found; \RenyiMode\ run on \RenyiDevice). The theorems guarantee $v\le0$ for
$\alpha\ge\frac12$ (sandwiched) and $\alpha\le2$ (Petz).}
\label{tab:renyi}
\end{table}

\begin{figure}[t]
\centering\fig{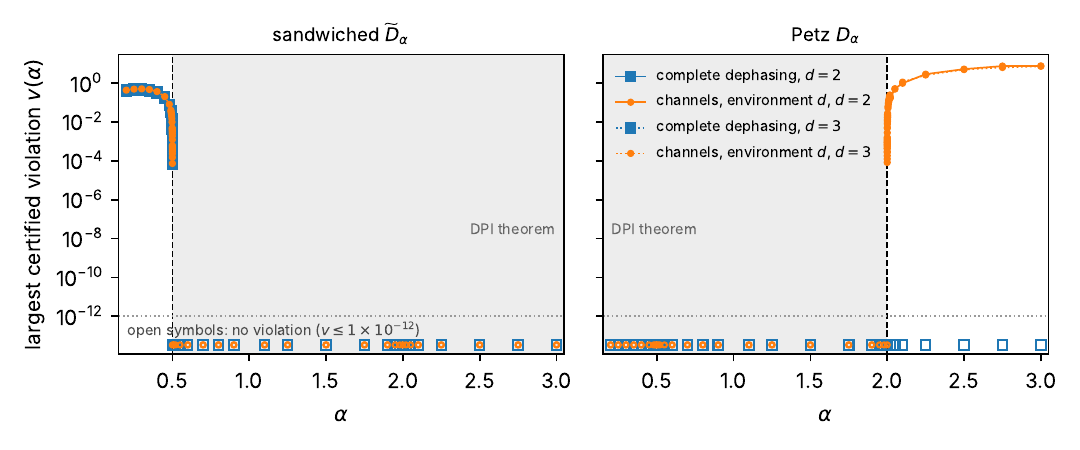}{\textwidth}
\caption{Largest certified violation $v(\alpha)$ for the sandwiched (left) and the Petz (right) divergences; dashed:
the known thresholds; shaded: the range in which the data-processing inequality is a theorem.}
\label{fig:renyi}
\end{figure}
\else The results of this search are not part of the present build (run \texttt{scripts/renyi\_threshold.py}).\fi

\section{Symmetric tensor products and spin chains}\label{sec:chains}

\begin{figure}[t]
\centering\fig{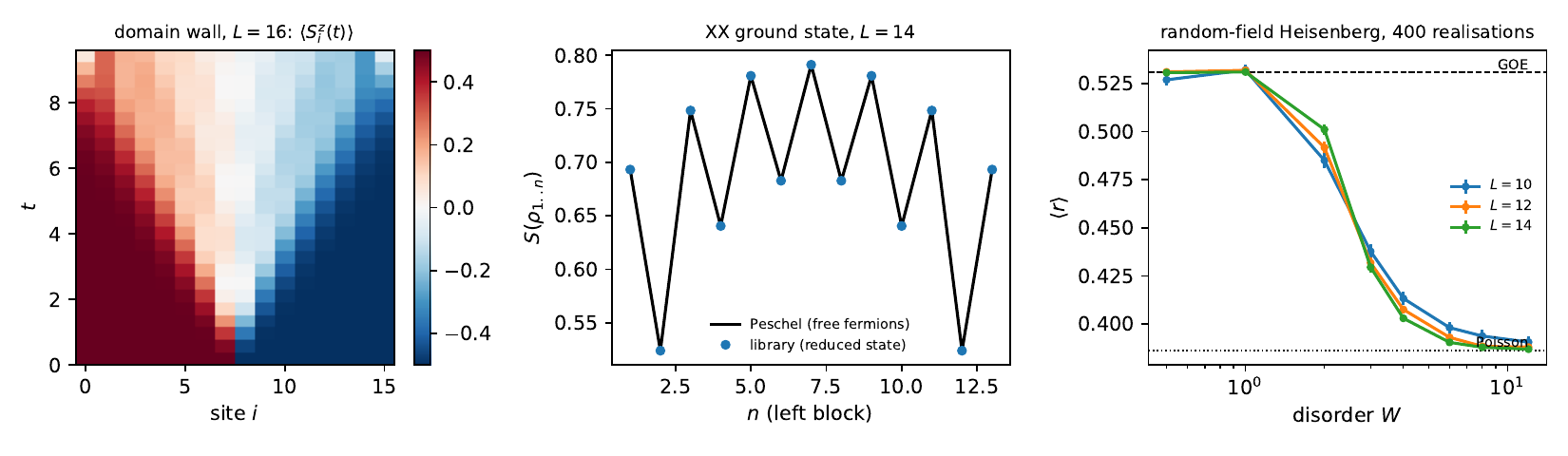}{0.88\textwidth}
\caption{Left: $\langle S^z_i(t)\rangle$ after a domain-wall quench of the XX chain (light cone).
Middle: entanglement entropy of the XX ground state, library vs.\ Peschel's formula. Right: mean
level-spacing ratio of the random-field Heisenberg chain, with the GOE and Poisson values.}
\label{fig:chains}
\end{figure}

The basis of \texttt{SpinChain} is, by construction, that of the iterated fused product of single sites. Integrable
chains provide exact results for every part of the module, and Table~\ref{tab:chainchecks} lists those reproduced by
\texttt{scripts/chains.py}; Table~\ref{tab:ring} and Figure~\ref{fig:chains} show some of them. Three entries need explanation. Peschel's formula gives the entanglement entropy of a block of the XX ground state from the
eigenvalues $\nu$ of the restricted correlation matrix, $S=-\sum_\nu[\nu\ln\nu+(1-\nu)\ln(1-\nu)]$~\cite{peschel2003},
including the parity oscillations of open chains. Momentum sectors of periodic chains use the basis
$|a,k\rangle=N_a^{-1/2}\sum_{j=0}^{L-1}e^{-ikj}T^j|a\rangle$ over orbit representatives $a$ of the translation $T$
(orbit length $R_a$, $N_a=L^2/R_a$; $a$ contributes to $k$ iff $kR_a\in2\pi\mathbb Z$). The ground state of the
antiferromagnetic Heisenberg ring has $k=0$ for even and $k=\pi$ for odd $L/2$ (Marshall's sign rule). Blocks with
$k\ne0,\pi$ are complex in this basis even for real Hamiltonians, one of the cases where the field is fixed by the
problem. With particle loss $S^-_i$ on every site, any $N$-conserving Hamiltonian leaves the particle number a pure
death process, so the sector distribution must be binomial.

For the Heisenberg chain with random fields $h_i\in[-W,W]$, the mean ratio of consecutive level spacings in the middle
of the spectrum crosses over from the GOE value $0.5307$ to the Poisson value $2\ln2-1\approx0.3863$ as $W$
grows~\cite{oganesyan2007,atas2013} (Figure~\ref{fig:chains}, right). At these sizes this is the finite-size crossover
used as a spectral diagnostic in studies of many-body localisation, and the batch dimension holds the
\G{chains}{\ChainReal}{many} disorder realisations per point.

\begin{table}[t]
\centering\footnotesize
\setlength{\tabcolsep}{4pt}
\begin{tabular}{>{\raggedright\arraybackslash}p{0.28\textwidth}>{\raggedright\arraybackslash}p{0.34\textwidth}>{\raggedright\arraybackslash}p{0.17\textwidth}c}
\toprule
computation & exact reference & size & max.\ deviation\\
\midrule
\ifdefined\ChainFFErr
open XX chain, all $2^L$ levels & free fermions, $\varepsilon_k=J\cos\frac{\pi k}{L+1}$ & $L=\ChainFFL$ & $\ChainFFErr$\\
Heisenberg ring, levels of sector $N<L/2$ & reappear in sector $N+1$ (SU(2) multiplets) & & $\ChainSUErr$\\
domain-wall melting, XX chain (dense) & single-particle $\langle n_i(t)\rangle=[e^{iht}C(0)e^{-iht}]_{ii}$ &
$L=\ChainDWL$, dim $\ChainDWDim$ & $\ChainDWErr$\\
domain-wall melting (Krylov) & same & $L=\ChainKDWL$, dim $\ChainKDWDim$ & $\ChainKDWErr$\\
block entropy of the XX ground state & Peschel's formula & $L=\ChainPesL$ & $\ChainPesErr$\\
momentum sectors, $N=L/2$ & union of spectra = spectrum of the sector & $L=\ChainMomL$ & $\ChainMomErr$\\
particle loss $S^-_i$, any $N$-conserving $H$ & binomial $p_n(t)=\binom{N_0}{n}e^{-n\gamma t}(1-e^{-\gamma t})^{N_0-n}$ & & $\ChainLossErr$\\
\else\multicolumn{4}{c}{\emph{run scripts/chains.py}}\\
\fi
Heisenberg ring ground states & Bethe ansatz ($L=4$), exact diagonalisation & Table~\ref{tab:ring} & nine digits\\
\bottomrule
\end{tabular}
\caption[]{Exact results reproduced by the chain module (\texttt{scripts/chains.py}\G{chains}{; \ChainMode\ run on
\ChainDevice}{}).}
\label{tab:chainchecks}
\end{table}

\paragraph{Beyond dense matrices.} Lanczos with full re-orthogonalisation and restarts, batched over Hamiltonians,
reproduces free-fermion and Heisenberg ground-state energies in sectors up to the sizes of Table~\ref{tab:krylov}. For
a single Hamiltonian on a CPU, ARPACK~\cite{lehoucq1998} (\texttt{scipy.sparse.linalg.eigsh}) is the reference. Our
solver re-orthogonalises fully for robustness and is batched over Hamiltonians, which is what lets disorder averages and
GPUs pay off (Table~\ref{tab:arpack}).

\begin{table}[t]
\centering\small
\begin{tabular}{lrrrrr}
\toprule
model & sector dim & $E_0$ (Lanczos) & $E_0$ (exact) & difference & time [s]\\
\midrule
\ChainKrylovRows
\bottomrule
\end{tabular}
\caption[]{Batched Lanczos on sparse sector Hamiltonians (time includes building the sparse Hamiltonian):
ground-state energies against free fermions and the tabulated Heisenberg value\G{chains}{; \ChainMode\ run on
\ChainDevice}{}.}
\label{tab:krylov}
\end{table}

\begin{table}[t]
\centering\small
\ifnum\BaseNew=1
\begin{tabular}{rrrrrrr}
\toprule
$L$ & sector dim & Hamiltonians & \texttt{eigsh} [s] & library, CPU [s] & library, GPU [s] & $\max|\Delta E_0|$\\
\else
\begin{tabular}{rrrrrr}
\toprule
$L$ & sector dim & Hamiltonians & \texttt{eigsh} [s] & library, GPU [s] & $\max|\Delta E_0|$\\
\fi
\midrule
\BaseGSRows
\bottomrule
\end{tabular}
\caption[]{Ground states of Heisenberg rings (one Hamiltonian) and random-field chains ($W=3$, several Hamiltonians) in
the sector $N=L/2$: ARPACK, one call per Hamiltonian, against the batched Lanczos solver; total time for all
Hamiltonians (\texttt{scripts/baselines.py}\G{baselines}{; \BaseMode\ run on \BaseDevice}{}).}
\label{tab:arpack}
\end{table}

\begin{table}[t]
\centering\small
\begin{tabular}{rrrrrr}
\toprule
$L$ & $\dim$ ($N=L/2$) & $E_0$ (library) & $E_0$ (exact) & difference & $E_0/L$\\
\midrule
\ChainRingRows
\bottomrule
\end{tabular}
\caption[]{Ground-state energy of the Heisenberg ring $H=\sum_i\mathbf S_i\cdot\mathbf S_{i+1}$; exact values to
nine digits; $E_0/L$ approaches the Bethe-ansatz value $\tfrac14-\ln2\approx-0.4431$\G{chains}{ (\ChainMode\ run on
\ChainDevice)}{}.}
\label{tab:ring}
\end{table}

\section{Subalgebras, inclusions and symmetries}\label{sec:symmetries}

Three modules relate algebras to each other and to the symmetries that produce them: \texttt{inclusions} treats a
subalgebra $\mathcal N\subset\M$ with its conditional expectations and indices, \texttt{decompose} computes the algebra
generated by given matrices or by a finite group, and \texttt{su2} uses the full SU(2) symmetry of spin-$\frac12$
chains. Table~\ref{tab:tests} lists the identities against which the test suite checks them.

\paragraph{Inclusions and conditional expectations.} Let $\mathcal N=\bigoplus_{c=1}^CM_{k_c}$ and
$\M=\bigoplus_{d=1}^DM_{m_d}$. Every unital inclusion $\mathcal N\to\M$ is, up to an inner automorphism $\operatorname{Ad}u$
of $\M$, of the form $\iota(x)_d=\bigoplus_cx_c\otimes1_{\lambda_{dc}}$. It is given by the inclusion matrix
$\Lambda=(\lambda_{dc})\in\mathbb Z_{\ge0}^{D\times C}$, $m_d=\sum_c\lambda_{dc}k_c$, whose entry $\lambda_{dc}$ is
the multiplicity of the summand $c$ of $\mathcal N$ in the summand $d$ of $\M$~\cite{ghj1989}. \texttt{Inclusion}
stores $\Lambda$ and an optional batched unitary $u$; with the isometries $V_{dc,i}$ onto the copies of $\C^{k_c}$,
$\iota$ is an \texttt{InterSectorChannel}. A faithful trace $\tau=\sum_dw_d\Tr_d$ on $\M$ restricts to $\mathcal N$
with weights $\nu=\Lambda^Tw$, and
\[
  E(y)_c=\nu_c^{-1}\sum_{d,i}w_d\,V_{dc,i}^*u_d^*\,y_d\,u_dV_{dc,i}
\]
is the unique $\tau$-preserving conditional expectation onto $\mathcal N$, the adjoint of $\iota$ with respect to
$\tau$. For a connected inclusion the index is $[\M:\mathcal N]=\|\Lambda\|^2$~\cite{jones1983,ghj1989}, e.g.\
$(m/k)^2$ for $M_k\subset M_m$ and $n$ for the diagonal subalgebra $\C^n\subset M_n$. The \emph{Markov trace} is the
trace whose weight vector is the Perron--Frobenius vector of $\Lambda\Lambda^T$; it is the trace for which the
conditional expectation behaves as in a subfactor. The Watatani index is $\operatorname{Ind}E=\sum_ju_ju_j^*$ over a
\emph{quasi-basis}, a finite family $\{u_j\}\subset\M$ with $\sum_ju_jE(u_j^*y)=y$ for all
$y$~\cite{watatani1990}. It is central, $(\operatorname{Ind}E)_d=(\Lambda\Lambda^Tw)_d/w_d$, and equals $[\M:\mathcal
N]\,1$ exactly for the Markov trace. The Pimsner--Popa constant $\lambda(E)=\max\{\lambda:E(y)\ge\lambda y\ \text{for}\
y\ge0\}$ equals $\bigl(\max_d\sum_c(\nu_c/w_d)\min(k_c,\lambda_{dc})\bigr)^{-1}$~\cite[Theorem~6.1]{pimsner1986}. For
subfactors of II$_1$ factors it is $[\M:\mathcal N]^{-1}$, but for Type~I inclusions it differs; the library also
computes the complete version $\max\{\lambda:E-\lambda\,\mathrm{id}\ \text{completely positive}\}=1/\|\operatorname{Ind}E\|$~\cite{bdh1988,frank1998}.
For $\C\subset M_n$ the two are $1/n$ and $1/n^2$. Basic construction, Jones projection, Jones tower and relative
commutant are also available, all batched and differentiable (\repofile{inclusions.py}).

\paragraph{The spin-chain tower.} The algebras $A_L=\bigoplus_{N=0}^LM_{\binom LN}$ of Section~\ref{sec:chains} form the
tower $A_L\subset A_{L+1}$, $x\mapsto x\otimes1_2$, whose Bratteli diagram is Pascal's triangle. Its inclusion matrix is
the incidence matrix of the path with $2L+3$ vertices, so
\[
  [A_{L+1}:A_L]=\|\Lambda\|^2=4\cos^2\frac{\pi}{2L+4}\;\nearrow\;4 ,
\]
the values $4\cos^2(\pi/n)$, $n=4,6,8,\dots$, of Jones' discrete series below 4~\cite{jones1983}.

\paragraph{From a symmetry to the algebra: Wedderburn decomposition.} Given matrices $S_1,\dots,S_J\in M_n(\C)$, or a
unitary representation $\pi$ of a finite group, \texttt{decompose.wedderburn} returns a unitary $V$ with
\[
  V^*\mathcal AV=\bigoplus_cM_{k_c}(\C)\otimes1_{m_c},\qquad V^*\mathcal A'V=\bigoplus_c1_{k_c}\otimes M_{m_c}(\C)
\]
for the von Neumann algebra $\mathcal A$ generated by the matrices, together with maps into the algebras of the
library. The algorithm combines two randomised methods of Murota and coauthors~\cite{murota2010,maehara2011}. The
eigenspaces of a random Hermitian element of the commutant $\mathcal A'$ are, generically, irreducible
$\mathcal A$-submodules. For a second random $Y\in\mathcal A'$ and orthonormal bases $Q_i,Q_j$ of two submodules,
$Q_i^*YQ_j$ intertwines them; by Schur's lemma it vanishes for inequivalent submodules and is a multiple of a unitary
otherwise, which gives the equivalence classes and aligns their bases. Random elements of $\mathcal A'$ come from the
group average $|G|^{-1}\sum_g\pi(g)X\pi(g)^*$ or from the null space of $X\mapsto([H_j,X])_j$. Every result is verified
before it is returned (generators in the decomposed algebra, fresh commutant elements in its commutant, and
$\sum_cm_c^2=\dim\mathcal A'$, which together prove equality by the double commutant theorem); a failed check triggers a
new draw. For real generators, \texttt{real\_structure} classifies every summand as real, complex or quaternionic from
the sign of $J_c\overline{J_c}$, which has a gap of 2, and \texttt{real\_basis} returns an adapted real
basis~\cite{maehara2010}. \repofile{symmetry.py} recovers Schur--Weyl duality~\cite{fulton1991} for four spins.

\paragraph{SU(2) symmetry of spin chains.} The Heisenberg Hamiltonian $H=\sum J_{ij}\,\mathbf S_i\cdot\mathbf S_j$
commutes with all of SU(2). On $(\C^2)^{\otimes L}=\bigoplus_S\C^{m_S}\otimes V_S$ the commutant is
$\bigoplus_SM_{m_S}(\C)$, $m_S=\binom{L}{L/2-S}-\binom{L}{L/2-S-1}$, and an invariant Hamiltonian is
$\bigoplus_SH_S\otimes1_{2S+1}$: every multiplet is computed once, and the blocks shrink from $\binom L{L/2+S}$ to
$m_S$. \texttt{SU2Chain} realises $\C^{m_S}$ as the highest-weight vectors in the $U(1)$ sector $N=L/2+S$, in the
\emph{Yamanouchi basis}: built by adding sites one at a time with exact Clebsch--Gordan coefficients (Racah's
formula~\cite{racah1942}) and labelled by the sequence of intermediate total spins. It is orthonormal by construction and
expressed in the $U(1)$ basis, where the block of an invariant operator is $A_S=B_S^*A_NB_S$; spin correlations follow
from reduced matrix elements (Wigner--Eckart theorem~\cite{edmonds1957}). For $L=14$ the largest block shrinks from
$3432$ to $1001$ and the cubic cost of diagonalising all blocks by a factor of about 48; the bases themselves
($1.0\cdot10^8$ numbers for $L=16$) limit dense SU(2) computations to about $L=16$.

\begin{table}[t]
\centering\footnotesize
\begin{tabular}{>{\raggedright\arraybackslash}p{0.16\textwidth}>{\raggedright\arraybackslash}p{0.78\textwidth}}
\toprule
module & identities checked by the test suite\\
\midrule
\texttt{chains} & $S^z_i$ of \texttt{SpinChain(L)} equals $S^z_i\otimes1$ formed with \texttt{kron} in the fused product of
\texttt{SpinChain(L-1)} and a site\\
\texttt{inclusions} & $E$ unital, completely positive, $E\circ\iota=\mathrm{id}$, bimodule property, adjoint of $\iota$ (three
traces, Markov trace, random weights); Pimsner--Popa constants against brute force (bisection over rank-one $y$, Choi
matrices); $e_{\mathcal N}ye_{\mathcal N}=E(y)e_{\mathcal N}$, $e_{\mathcal N}\in\mathcal N'$, Markov property of the
basic construction; tower: path-graph spectrum and index for $L\le8$, embedding $x\otimes1_2$, conditional
expectation = normalised partial trace\\
\texttt{decompose} & hidden block structures under random unitaries; Fourier basis of the regular representation of
$\mathbb Z_n$; Schur--Weyl duality on $(\C^2)^{\otimes L}$, $L\le6$ (same central projections, $k_c$ and $m_c$
exchanged); Kramers degeneracy in quaternionic sectors\\
\texttt{su2} & full spectra of open and periodic chains with field and next-nearest-neighbour couplings against dense
diagonalisation ($L\le12$, $10^{-10}$); ring energies of Table~\ref{tab:ring} ($L\le14$); Lieb--Mattis
theorem~\cite{lieb1962}; the two degenerate singlets at the Majumdar--Ghosh point~\cite{majumdar1969,majumdar1969b};
Clebsch--Gordan coefficients against SymPy\\
\texttt{divergences} & dense references; independence of the trace convention; data processing inside the proven
ranges; Petz's equality condition; the recovery bound of~\cite[Theorem~2.1]{junge2018}\\
\bottomrule
\end{tabular}
\caption{Identities checked by the test suite for the modules of Sections~\ref{sec:renyi}--\ref{sec:symmetries}.}
\label{tab:tests}
\end{table}

\section{Use case: noisy circuits with a conserved charge}\label{sec:noisy}

\begin{figure}[t]
\centering\fig{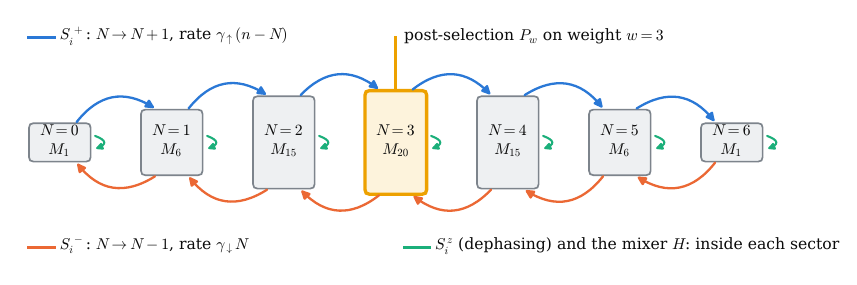}{0.85\textwidth}
\caption{The sector ladder of a noisy register of $n=6$ qubits. Sector $N$ (Hamming weight) is the block $M_{\binom
nN}$. Excitation $S^+_i$ moves up, amplitude damping $S^-_i$ down, with total rates $\gamma_\uparrow(n-N)$ and
$\gamma_\downarrow N$ that depend only on $N$. Dephasing and the mixer act inside the sectors, and post-selection
projects onto the initial weight $w$.}
\label{fig:ladder}
\end{figure}

\paragraph{Context.} Quantum algorithms that preserve a charge offer a cheap test for errors. If the ideal evolution
keeps the charge, a measurement of the charge at the end detects every error that changed it, and discarding those
runs improves the accepted results. This \emph{symmetry verification}~\cite{bonetmonroig2018,mcardle2019} is one of
the standard error-mitigation techniques reviewed in~\cite{cai2023}. Post-selection on the particle number was used,
for instance, in Google's Hartree--Fock experiment~\cite{arute2020}. Botelho et al.~\cite{botelho2022} proposed
mid-circuit post-selection on Hamming-weight subspaces for QAOA. The XY mixers of the quantum alternating operator
ansatz keep the Hamming weight~\cite{hadfield2019,wang2020}; Streif et al.~\cite{streif2021} derived the probability
that such circuits stay in the feasible subspace under local depolarising noise and proposed codes that correct
symmetry-breaking errors. In the language of open systems, the noisy register has a weak $U(1)$ symmetry in the sense of
Buča and Prosen~\cite{buca2012} and Albert and Jiang~\cite{albert2014}. The Lindbladian commutes with conjugation by
$e^{i\phi N}$, which says exactly that block-diagonal states stay block diagonal, i.e.\ that the dynamics never leaves
$\M=\bigoplus_{N=0}^nM_{\binom nN}$. Without jumps that change $N$ the symmetry is strong and every sector is conserved
separately. In this language the measurement of the weight is a central projection $P_w$ and post-selection is the
Lüders conditioning on it, so the whole question is a computation in the algebra (Figure~\ref{fig:ladder}).

\paragraph{Model.} A ring of $n$ qubits evolves under the XY mixer
$H=\tfrac14\sum_{\langle ij\rangle}(X_iX_j+Y_iY_j)=\tfrac12\sum_{\langle ij\rangle}(S^+_iS^-_j+S^-_iS^+_j)$
for a time $T=\NXYT$ from $|1^w0^{n-w}\rangle$, $w=n/2$. Three kinds of local noise act on every site: amplitude damping $S^-_i$
(rate $\gamma_\downarrow$), thermal excitation $S^+_i$ ($\gamma_\uparrow$) and dephasing $S^z_i$ ($\gamma_\varphi$).
With $\psi$ the noiseless state at time $T$ we compute the acceptance probability $p=\Tr(P_w\rho(T))$, the fidelity
$F=\langle\psi|\rho(T)|\psi\rangle$ and the fidelity after post-selection $F_{\mathrm{post}}=F/p$. Let
$\Gamma_w=\gamma_\downarrow w+\gamma_\uparrow(n-w)$ be the total rate of leaving sector $w$ and
$I=\int_0^T\sum_i\bigl(\frac14-\langle S^z_i\rangle_{\psi(t)}^2\bigr)\,dt$ the dephasing integral along the noiseless
trajectory.

\begin{proposition}\label{prop:noisy}
Let $H$ (possibly time dependent) commute with $N$, and let the register start in a pure state of weight $w$.
\begin{enumerate}
\item[(a)] The weight distribution $p_N(t)=\Tr(P_N\rho(t))$ is that of the classical birth--death process with death rate
$\gamma_\downarrow N$ and birth rate $\gamma_\uparrow(n-N)$; it does not depend on $H$ or $\gamma_\varphi$.
\item[(b)] If $\gamma_\uparrow=0$: $P_w\rho(T)P_w=e^{-\gamma_\downarrow wT}\rho_\varphi(T)$, with $\rho_\varphi$ the
solution for $\gamma_\downarrow=0$. In particular $p=e^{-\gamma_\downarrow wT}$, and $F_{\mathrm{post}}=1$ if also
$\gamma_\varphi=0$.
\item[(c)] If $\gamma_\varphi=0$: $0\le1-F_{\mathrm{post}}\le1-e^{-\Gamma_wT}/p
=\gamma_\downarrow\gamma_\uparrow T^2\bigl[w(n-w)+\tfrac n2\bigr]+O\bigl((n\gamma T)^3\bigr)$,
$\gamma=\max(\gamma_\downarrow,\gamma_\uparrow)$.
\item[(d)] If $\gamma_\downarrow=\gamma_\uparrow=0$: $p=1$ and $1-F_{\mathrm{post}}=1-F=\gamma_\varphi
I+O(\gamma_\varphi^2)$.
\item[(e)] To first order in all rates, $1-F=\Gamma_wT+\gamma_\varphi I$ and $1-F_{\mathrm{post}}=\gamma_\varphi I$; for
$\gamma_\varphi>0$ the gain $(1-F)/(1-F_{\mathrm{post}})$ tends to $1+\Gamma_wT/(\gamma_\varphi I)$ as the rates go to
zero at fixed ratios.
\end{enumerate}
\end{proposition}
\begin{proof}
Since $S^\pm_iP_N=P_{N\pm1}S^\pm_i$ and $\sum_iS^+_iS^-_i=N$, $\sum_iS^-_iS^+_i=n-N$ are central, $\dot p_N=
\gamma_\downarrow[(N+1)p_{N+1}-Np_N]+\gamma_\uparrow[(n-N+1)p_{N-1}-(n-N)p_N]$; $H$ and the dephasing jumps commute
with $P_N$ and do not contribute, which is (a). Expand $\rho(T)$ in the jumps $S^\pm_i$ (Dyson series). The no-jump
generator differs from the Lindbladian with $\gamma_\downarrow=\gamma_\uparrow=0$ by
$-\frac12\{\gamma_\downarrow N+\gamma_\uparrow(n-N),\cdot\}$, which on sector $w$ is the scalar $-\Gamma_w$. Hence the
part of $\rho(T)$ in sector $w$ is $e^{-\Gamma_wT}\rho_\varphi(T)$ plus terms with $m\ge1$ up and $m$ down jumps. (b)
Without up jumps only the first term remains. (c) For $\gamma_\varphi=0$, $\rho_\varphi(T)=|\psi\rangle\langle\psi|$,
and the remaining terms form $J\ge0$ with $\Tr J=p-e^{-\Gamma_wT}$. Then $F_{\mathrm{post}}=(e^{-\Gamma_wT}+\langle\psi|
J|\psi\rangle)/p\ge e^{-\Gamma_wT}/p$. The leading terms of $\Tr J$ are one down and one up jump in either order, with
probabilities $\gamma_\downarrow w\,\gamma_\uparrow(n-w+1)T^2/2$ and $\gamma_\uparrow(n-w)\,\gamma_\downarrow(w+1)T^2/2$.
(d) Dephasing keeps the sector, and the first-order term gives $1-F=\gamma_\varphi\int_0^T\sum_i(\langle(S^z_i)^2\rangle
-\langle S^z_i\rangle^2)\,dt$ with $(S^z_i)^2=\frac14$. (e) Every first-order jump $S^\pm_i$ leaves sector $w$ and
contributes $0$ to $F$, so $F=e^{-\Gamma_wT}(1-\gamma_\varphi I)+O(\gamma^2)$; with post-selection only the dephasing
term remains at first order, by (c) and (d).
\end{proof}

\noindent None of these statements is new in substance. Statement (b) is the standard reason why particle-number
post-selection removes pure loss: $N$ is central, so damping commutes with the post-selection. Statements (c)--(e)
make explicit what remains. Thermal noise is suppressed from first to second order with a coefficient
$w(n-w)+\frac n2=n(n+2)/4$ at half filling, and dephasing is not detected at all. We state the proposition because it
fixes what an exact computation must reproduce and isolates the two numbers it cannot predict: the integral $I$, and the
fraction of the bound in (c) that is actually lost.

\begin{figure}[t]
\centering\fig{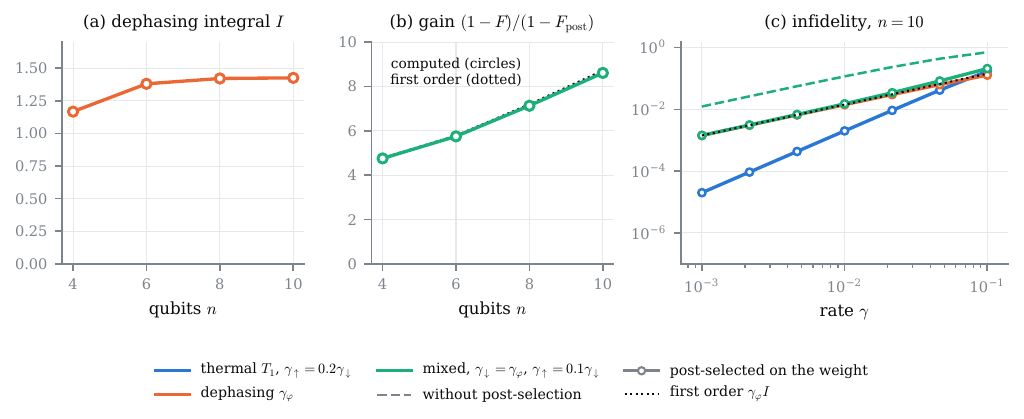}{\textwidth}
\caption{Noisy XY mixer, $T=2$. (a) The dephasing integral $I$ saturates with $n$. (b) The gain of post-selection for
the mixed model at $\gamma=10^{-3}$ grows with $n$ and follows the first-order value $1+\Gamma_wT/(\gamma_\varphi I)$
of Proposition~\ref{prop:noisy}(e). (c) Infidelities for $n=10$: post-selection suppresses thermal noise to second order
and leaves dephasing, which follows $\gamma_\varphi I$; for zero-temperature damping (not shown) the post-selected
state is exact.}
\label{fig:noisy}
\end{figure}

\paragraph{Computation.} Listing~\ref{lst:noisy} shows the core of the computation for zero temperature. All noise
models and rates form one batch: $\NXYBatch$ settings per $n$ (four noise models, seven rates between $10^{-3}$ and
$10^{-1}$), with the rates folded into the coefficients of the jumps. One call of \texttt{lindblad\_evolve} with the
completely positive splitting (\texttt{method='kraus'}, $\NXYSubsteps$ steps) evolves the batch. The jumps are
\texttt{SectorJump}s (Section~\ref{sec:speed}); as dense maps between sectors they would need $3n$ tensors of
$(n+1)^2\binom n{n/2}^2$ entries, $2.3$\,GB each for $n=12$. Table~\ref{tab:noisy} and Figure~\ref{fig:noisy} show the
results for $n=4,\dots,10$ (script \texttt{scripts/noisy\_xy\_mixer.py}).

\begin{lstlisting}[float=t, caption={Amplitude damping of a six-qubit XY mixer for seven rates in one batch, and
post-selection on the Hamming weight (Proposition~\ref{prop:noisy}(b)): both printed deviations are at the level of the
time-step error and of round-off.}, label=lst:noisy]
import torch
from torch_vn_algebra import SpinChain, dynamics, states

n, w, T = 6, 3, 2.0
q = SpinChain(n, 'periodic', precision='double')   # sectors: Hamming weight N = 0..n
H = q.xxz(J=1.0, Delta=0.0)                         # XY mixer
rho0 = q.basis_state('111000')
psi = dynamics.von_neumann(rho0, H, [T])[0]         # noiseless reference state
g = torch.logspace(-3, -1, 7, dtype=torch.float64)  # seven damping rates: one batch
jumps = [q.sector_jump(i, 'lower').scaled(g.sqrt()) for i in range(n)]   # S^-_i
rho = dynamics.lindblad_evolve(rho0, H, jumps, [0.0, T], substeps=40, method='kraus')[-1]
P = q.algebra.central([float(N == w) for N in range(n + 1)])  # weight-w projection
post, p = rho.condition_on(P)                       # post-selection (Lueders rule)
F_post = states.trace_of_product(post, psi).real
print(p / torch.exp(-w * g * T) - 1, 1 - F_post)   # both ~ 0
\end{lstlisting}

\begin{table}[t]
\centering\scriptsize
\setlength{\tabcolsep}{2.4pt}
\begin{tabular}{rrrrrrrrrrrr}
\toprule
 & & & & & \multicolumn{2}{c}{(b) $T_1$, zero temp.} & (c) thermal & \multicolumn{2}{c}{(d) dephasing} &
 \multicolumn{2}{c}{mixed, $\gamma\approx0.022$}\\
$n$ & $\binom n{w}$ & $\sum_N\binom nN^2$ & $4^n$ & time [s] & err.\ $p$ & $|1-F_{\mathrm{post}}|$ & slope & $I$ &
$\frac{1-F_{\mathrm{post}}}{\gamma_\varphi I}$ & $1-F$ & $1-F_{\mathrm{post}}$\\
\midrule
\NXYRows
\bottomrule
\end{tabular}
\caption[]{Noisy XY mixer on a ring of $n$ qubits, $T=\NXYT$ (CPU, double precision). Dimensions: the weight sector, the
algebra and a dense density matrix; time for all $\NXYBatch$ settings. (b) Relative deviation of $p$ from $e^{-w\gamma
T}$ (the error of the second-order splitting) and $|1-F_{\mathrm{post}}|$, maxima over the rates. (c) Slope of
$\log(1-F_{\mathrm{post}})$ against $\log\gamma$ for $\gamma_\uparrow=0.2\gamma_\downarrow=0.2\gamma$, at small
$\gamma$. (d) The integral $I$ and the ratio of the computed infidelity to its first-order value at
$\gamma_\varphi=10^{-3}$. Mixed: $\gamma_\downarrow=\gamma_\varphi=\gamma$, $\gamma_\uparrow=0.1\gamma$ at
$\gamma=10^{-5/3}\approx0.022$, without and with post-selection. Full sizes of the study, CPU (4 threads).}
\label{tab:noisy}
\end{table}

\paragraph{Results.} The computation confirms the proposition in every case. At zero temperature the post-selected
state agrees with the noiseless one to $\NXYTOnePost$, and the acceptance agrees with $e^{-w\gamma T}$ to
$\NXYAccErr$, the error of the time steps. For all noise models the acceptance agrees with the birth--death process of
(a) to $3\times10^{-4}$. The thermal infidelity after post-selection has slope $2.00$. For $n=4,6,8,10$ it is $0.73$,
$0.76$, $0.81$ and $0.84$ times the bound of (c), independently of the rate. The complementary fraction is the overlap
with $\psi$ that a state retains after one up and one down jump. The dephasing infidelity agrees with $\gamma_\varphi I$
to $0.1\%$. The integral $I$ saturates with $n$ at fixed $T$ ($1.17$ for $n=4$, $1.43$ for $n=10$): by time $T$ only
the qubits near the two domain walls of $|1^w0^{n-w}\rangle$ have moved, and dephasing hurts only where the state is not
a product of $S^z$ eigenstates. The detected error grows with the weight $w$ and the undetected one does not, so
post-selection gains more for larger registers. At $\gamma=10^{-3}$ the gain for the mixed model is $4.75$ for $n=4$ and
$8.61$ for $n=10$, within $1.3\,\%$ of the first-order value of (e). At $\gamma\approx0.022$ it is $4.4$ and $6.8$
(Table~\ref{tab:noisy}). For algorithm design this quantifies the value of the symmetry: it protects against energy
relaxation, not against dephasing, and the remaining error is predicted to first order by one integral over the
noiseless trajectory.

\paragraph{What the computation adds, and what it does not cover.} The computation gives exact values, up to the
time-step error, for all noise models and rates in one batched run per register. It also gives the integral $I$ and its
saturation in $n$, the fraction of the thermal bound that is realised, and a gain that grows with $n$; none of these
follows from the proposition alone. It does not model a QAOA circuit as practised. There is no phase separator, no
layering, a single evolution time and a product initial state rather than a Dicke state. Phase separators are diagonal
and commute with $N$, so the proposition still holds for layered circuits, but $I$ and the gain depend on the circuit
and would have to be recomputed. Two-qubit gate noise, readout errors and coherent errors are not modelled.
Single-qubit depolarising noise does stay in the algebra: its Pauli-$X$ and $Y$ parts combine into $S^\pm$ jumps of equal
rate, $\mathcal D[X]+\mathcal D[Y]=2(\mathcal D[S^+]+\mathcal D[S^-])$, so it is the thermal model at
$\gamma_\uparrow=\gamma_\downarrow$ plus dephasing. Noise that is not $U(1)$-covariant leaves the algebra: bit flips
$X_i$ alone, unequal $X$ and $Y$ rates, or coherent $X$ rotations. For instance $X\rho X$ contains $S^+\rho S^+$, which
maps the block $(w,w)$ to the off-diagonal block $(w+1,w-1)$; such noise cannot be represented in~\eqref{eq:M} and
needs the full $M_{2^n}$, which is a limitation of the approach. Finally, we make no claim of speed for this example.
The run for $n=10$ took 930\,s on a CPU for 28 settings, dominated by dense products of the padded Hamiltonian
(Section~\ref{sec:limitations}). A single density matrix of dimension $2^{10}$ is within easy reach of dedicated
solvers~\cite{lambert2026,mercurio2025,dynamiqs}; comparisons with them are left for future work.

%
\ifdefined\NCRows\else\relax\fi{}
\providecommand{\NCSegments}{4}\providecommand{\NCSettings}{12}\providecommand{\NCSteps}{--}\providecommand{\NCStepsZero}{--}\providecommand{\NCSubsteps}{2}
\providecommand{\NCStarts}{--}\providecommand{\NCWarm}{--}\providecommand{\NCWarmLarge}{--}\providecommand{\NCLambda}{0.05}\providecommand{\NCEvalSubsteps}{20}\providecommand{\NCnMax}{6}
\providecommand{\NCRows}{\multicolumn{9}{c}{\emph{run scripts/noisy\_xy\_control.py}}\\}
\subsection{Noise-aware state preparation with post-selection}\label{sec:control}

Section~\ref{sec:noisy} evaluated a fixed evolution. In practice the evolution is designed, and the design can take
into account the noise and the fact that the Hamming weight will be measured. XY-QAOA needs an initial state in the
feasible subspace, and the Dicke state $|D^n_w\rangle$, the uniform superposition of all strings of weight $w$, is the
standard choice~\cite{hadfield2019,wang2020}. Streif et al.~\cite{streif2021} computed how likely such circuits are
to stay in the feasible subspace under local noise. Botelho et al.~\cite{botelho2022} proposed post-selection on
Hamming-weight subspaces for QAOA, an instance of symmetry verification~\cite{cai2023}. This raises a design question. Does a preparation that will be
post-selected need different controls from one that will not, and how much does designing for the noise gain over
a noiseless design? The question is an optimisation over the controls of a Lindblad evolution in
$\M=\bigoplus_NM_{\binom nN}$, and its objective uses the centre of $\M$. It needs gradients through open-system
dynamics in the sectors, the central projection $P_w$, and many noise settings at once. These are the three things the
library provides.

\paragraph{Parametrisation.} A ring of $n$ qubits starts in $|1^w0^{n-w}\rangle$, $w=n/2$, and is driven during a time
$T$ by $S=\NCSegments$ piecewise-constant segments of length $T/S$ with Hamiltonians
\begin{equation}\label{eq:control}
H_s=\sum_{\langle ij\rangle}J_{ij,s}\,\tfrac12\bigl(S^+_iS^-_j+S^-_iS^+_j\bigr)+\sum_ih_{i,s}S^z_i
+V_s\sum_{\langle ij\rangle}S^z_iS^z_j,\qquad |J_{ij,s}|,|h_{i,s}|,|V_s|\le1,\quad 0<T\le2n.
\end{equation}
These are bond-resolved XY couplings, local fields and a uniform phase-separator term. Every term commutes with $N$,
so the controls stay in $\M$. The bond terms are \texttt{SpinChain.bond(i, j, Jxy, Jz)}, and $H_s$ is a linear
combination of them with trainable coefficients. Without the fields the target cannot be reached. The global spin
flip $\prod_iX_i$ commutes with all bond terms and maps sector $n/2$ to itself, the Dicke state is even under it, and
$|1^w0^{n-w}\rangle$ has equal weight on both parities, so $F\le\frac12$. With $V_s=0$ as well, the evolution is
quadratic in Jordan--Wigner fermions and cannot leave the Slater determinants. The noise consists of the four models of
Section~\ref{sec:noisy} ($T_1$ at zero temperature, thermal, dephasing and mixed), each at the rates $\gamma=0.003$,
$0.01$ and $0.03$, so the batch has $K=\NCSettings$ settings. Each segment is integrated over unit time with $H_s$
replaced by $(T/S)H_s$ and jump coefficients $\sqrt{\gamma T/S}$ (\texttt{SectorJump.scaled}), so $\rho(T)$ is
differentiable in all controls and in $T$.

\paragraph{Objectives.} With $F_k=\langle D^n_w|\rho_k(T)|D^n_w\rangle$ for noise setting $k$,
$p_k=\Tr(P_w\rho_k(T))$ the acceptance and $F_{\mathrm{post},k}=F_k/p_k$ the fidelity of the Lüders-conditioned state
(\texttt{condition\_on}), we compare three designs:
(i) the noiseless design, which maximises $F_0-\lambda T/T_{\max}$ without noise, $\lambda=\NCLambda$. The penalty
pushes $T$ close to the shortest time at which the controls reach $F_0\approx1$, which is the most favourable choice
for a design that ignores the noise;
(ii) the noise-aware design without post-selection, which maximises $\bar F=\frac1K\sum_kF_k$; and
(iii) the noise-aware design with post-selection, which maximises $\bar F_{\mathrm{post}}=\frac1K\sum_kF_{\mathrm{post},k}$.
By Proposition~\ref{prop:noisy}(a), $p_k$ depends on $T$ and the rates but not on the controls. At fixed $T$,
therefore, $F_k=p_k(T)F_{\mathrm{post},k}$, and (ii) and (iii) differ only in the weights $p_k$ of the settings and in
the choice of $T$: design (ii) pays for the detected loss $\Gamma_wT$, design (iii) does not. Each objective is
maximised by \texttt{optimize.extremize} (Adam, \NCSteps\ steps; \NCStepsZero\ for (i)). Every evaluation is a
single batched \texttt{lindblad\_evolve} per segment for all starts and all $K$ settings (\texttt{method='kraus'},
\NCSubsteps\ steps per segment). The noiseless design is the best of \NCStarts\ random starts. Designs (ii) and (iii)
start from the \NCWarm\ ($n=4$) or \NCWarmLarge\ ($n=6$) best noiseless solutions, and the starts of (iii) also include
the solution of (ii).

\begin{table}[t]
\centering\footnotesize
\setlength{\tabcolsep}{3.2pt}
\begin{tabular}{rlrrrrrrr}
\toprule
$n$ & design & $T$ & $F_0$ & $\bar p$ & $\bar F$ & $\bar F_{\mathrm{post}}$ & $\min_kF_k$ &
$\min_kF_{\mathrm{post},k}$\\
\midrule
\NCRows
\bottomrule
\end{tabular}
\caption[]{Dicke-state preparation on a ring of $n$ qubits by the three designs, evaluated over the $\NCSettings$ noise
settings (four models, $\gamma\in\{0.003,0.01,0.03\}$). $F_0$: fidelity without noise. Means and minima over the
settings; every value is from an independent evaluation with $\NCEvalSubsteps$ steps per segment. Full run, CPU
(4 threads).}
\label{tab:control}
\end{table}

\begin{figure}[t]
\centering\fig{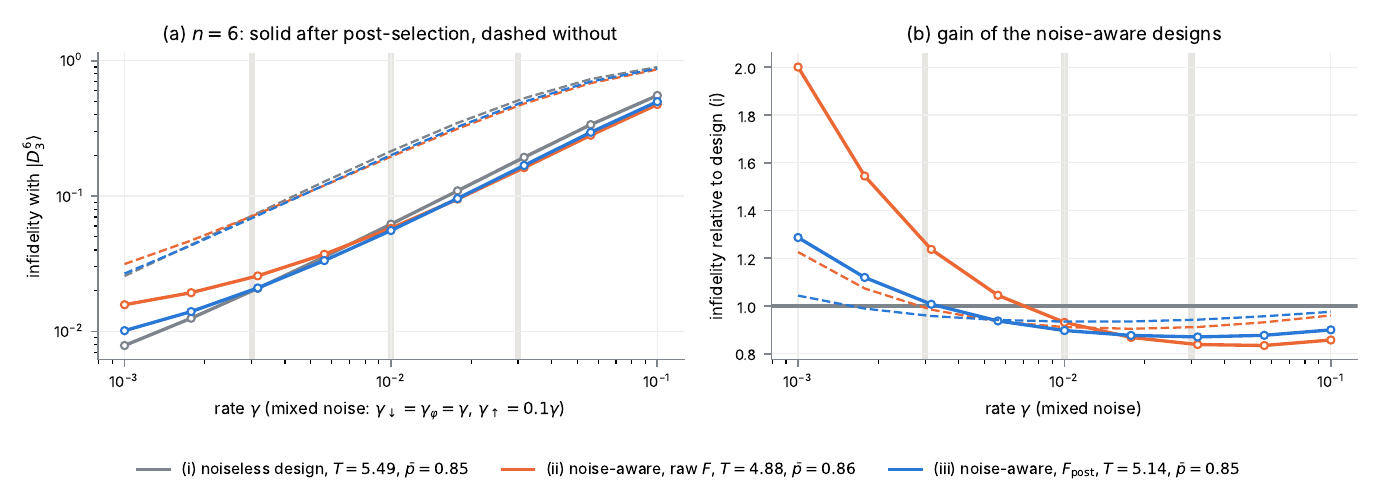}{\textwidth}
\caption{Designs (i)--(iii) for $n=\NCnMax$ under mixed noise
($\gamma_\downarrow=\gamma_\varphi=\gamma$, $\gamma_\uparrow=0.1\gamma$) at rates inside and outside the training batch
(grey bars). (a) Infidelity with $|D^n_w\rangle$ with (solid) and without (dashed) post-selection. (b) The same,
relative to the noiseless design (i).}
\label{fig:control}
\end{figure}

\paragraph{Results.} Table~\ref{tab:control} and Figure~\ref{fig:control} show the outcome. For $n=6$ the three
designs use the times $T=\G{noisy_control}{\NCSixTOne}{--}$, $\G{noisy_control}{\NCSixTTwo}{--}$ and
$\G{noisy_control}{\NCSixTThree}{--}$. Design (ii) is the fastest and gives up the most noiseless fidelity
($F_0=\G{noisy_control}{\NCSixFZeroTwo}{--}$ against $\G{noisy_control}{\NCSixFZeroOne}{--}$ for (i)), because every
unit of time costs it the detected loss as well. Design (iii) ignores that loss, so it spends more time than (ii) but
still less than (i), because the undetected part of the noise (dephasing, and pairs of an up and a down jump) also
grows with $T$. The order of the times is
the same for $n=4$. Each noise-aware design wins on its own objective. Design (ii) has the largest mean raw fidelity,
$\bar F=\G{noisy_control}{\NCSixFTwo}{--}$ against $\G{noisy_control}{\NCSixFOne}{--}$ for (i), and design (iii) the
largest mean post-selected fidelity, $\bar F_{\mathrm{post}}=\G{noisy_control}{\NCSixFPostThree}{--}$ against
$\G{noisy_control}{\NCSixFPostOne}{--}$. As factors on the infidelity, the noise-aware gains are modest:
\G{noisy_control}{\NCSixGainRaw}{--} for (ii) and \G{noisy_control}{\NCSixGainPost}{--} for (iii) at $n=6$, and
\G{noisy_control}{\NCFourGainRaw}{--} and \G{noisy_control}{\NCFourGainPost}{--} at $n=4$. In the worst setting of
the batch the post-selected fidelity rises from \G{noisy_control}{\NCSixMinFPostOne}{--} for (i) to
\G{noisy_control}{\NCSixMinFPostThree}{--} for (iii). Figure~\ref{fig:control}(b) shows the price. The noise-aware
designs are worse than (i) at rates below the training batch (factor \G{noisy_control}{\NCSixRatioLow}{--} at
$\gamma=10^{-3}$ for (iii)) and better above it (\G{noisy_control}{\NCSixRatioHigh}{--} at $\gamma=0.1$). Post-selection
itself is far stronger than any choice of design. For the noiseless design it lowers the mean infidelity by a factor
\G{noisy_control}{\NCSixPostFactor}{--} ($n=6$) and \G{noisy_control}{\NCFourPostFactor}{--} ($n=4$), and from the best
design without post-selection, (ii), to design (iii) with it the factor is \G{noisy_control}{\NCSixGainTotal}{--} at
$n=6$. The price is the acceptance, $\bar p=\G{noisy_control}{\NCSixPThree}{--}$ for (iii) against
$\G{noisy_control}{\NCSixPTwo}{--}$ for (ii); it differs between the designs only through $T$. In short, once the
noiseless design is as fast as the controls allow, little is left for noise-aware shaping. The centre does most of the
work: the measurement of the weight removes the detected errors, and the optimiser then trades the remaining,
undetected ones against time.

\paragraph{Checks.} Every number in the table comes from a separate forward evaluation, not from the optimiser. It
integrates in physical time, passes the rates as \texttt{rates} rather than as scaled coefficients, and uses
\G{noisy_control}{\NCEvalSubsteps}{20} steps per segment. The optimiser's own values differ from it by at most
\G{noisy_control}{$\NCRecheck$}{--}, the error of the coarse time steps. The acceptances agree with the birth--death
process of Proposition~\ref{prop:noisy}(a) to \G{noisy_control}{$\NCBirthDeath$}{--}. At one point (design (iii),
mixed noise, $\gamma=0.03$, $n=\G{noisy_control}{\NCnMax}{6}$), \texttt{method='exact'} reproduces $F$,
$F_{\mathrm{post}}$ and $p$ to \G{noisy_control}{$\NCExactDiff$}{--}. A directional derivative of the objective of
(iii) agrees with central differences to a relative \G{noisy_control}{$\NCGradErr$}{--}. The whole study ($n=4$ and
$6$) took \G{noisy_control}{\NCSeconds\,s}{--} on a CPU with \G{noisy_control}{\NCThreads}{several} threads
(script \texttt{scripts/noisy\_xy\_control.py}).

\paragraph{What this shows, and what it does not.} The study shows that the library handles the whole loop of
noise-aware design: a Lindblad evolution in symmetry sectors with sparse jumps, differentiated with respect to
Hamiltonian controls and the duration; an objective defined through a central projection; and a batch of noise models,
rates and starts in a single evolution. It also gives a quantitative answer for this model. With a time-optimal
noiseless design as the reference, the noise-aware designs gain only a few per cent of the infidelity, while
post-selection gains a factor of several. Proposition~\ref{prop:noisy}(a) explains why the two noise-aware objectives
give such similar results: post-selection changes the objective only through $T$ and the weights $p_k$. The comparison
depends on the reference. With a weaker time penalty $\lambda$, the noiseless design stays well above the shortest
time, and the same optimisation then reports larger noise-aware gains, which are really gains in speed. The search is
local: there are few starts (\G{noisy_control}{\NCWarmLarge}{four} for the noise-aware designs at $n=6$), $S$ is
small, the controls are bounded and piecewise constant, and the reported designs are good solutions, not proven optima.
The sizes are small ($n=4,6$). For $n=8$ the padded $70\times70$ blocks of all nine sectors make every gradient step
much more expensive, and we did not attempt that size on a CPU. The noise models are the idealised local ones of
Section~\ref{sec:noisy}, with rates measured in units of the coupling bound. Gate-level circuits, readout errors and
non-covariant noise are not modelled, and nothing here is a claim about particular hardware.

\section{Quantum channels as trainable networks}\label{sec:learn}

The previous sections optimise a handful of parameters. The same differentiable objects can be stacked into networks
whose layers are channels, and trained on data. This section describes the module \texttt{learn} that does this, the
dictionary behind it, and two experiments that show what the algebra contributes: tomography in the algebra as
training (Section~\ref{sec:learn-tomo}), and residual networks of channels as Markovian evolutions
(Section~\ref{sec:learn-markov}).

\paragraph{The dictionary.} A channel is a linear map of density matrices, so a network of channels is a deep
\emph{linear} network with a constraint on every layer, and the analogy with neural networks is precise only where
the constraint is respected. By Stinespring's theorem~\cite{stinespring1955} a channel is an isometry into a larger
space followed by a partial trace: a linear layer followed by a contraction, whose width is the Kraus rank. Composition
of channels is depth; a Lindblad evolution $e^{t\mathcal L}$ is a block of continuous depth, the open-system form of a
neural ODE~\cite{chen2018,e2017}; a conditional expectation onto a subalgebra is a pooling layer. The sector structure
of~\eqref{eq:M} is a structured sparsity of the weights: a channel that respects the sectors has Kraus operators with
blocks only on the allowed transitions, and correspondingly fewer parameters. Table~\ref{tab:learn} lists the layers.

\begin{table}[t]
\centering\footnotesize
\begin{tabular}{p{0.2\textwidth}p{0.34\textwidth}p{0.36\textwidth}}
\toprule
layer & map & network analogue\\
\midrule
\texttt{ChannelLayer} & Stinespring isometry per input sector, within one algebra or between algebras with allowed
sector transitions & linear layer plus contraction; width = Kraus rank; sparsity = sectors\\
\texttt{UnitaryLayer} & $\rho\mapsto U\rho U^*$, $U=e^{-iH}$, $H=H^*\in\M$ & unitary layer~\cite{arjovsky2016}\\
\texttt{LindbladLayer} & $\rho\mapsto e^{t\mathcal L}\rho$, $\mathcal L$ a free GKSL generator & continuous depth~\cite{chen2018}\\
\texttt{Residual} & $\rho\mapsto(1-p)\rho+p\,\Phi(\rho)$ & skip connection~\cite{he2016}\\
\texttt{PostSelect} & $\rho\mapsto P\rho P/\Tr(P\rho P)$, $P\in\M$ a projection (central: a set of sectors) &
nonlinear activation; cost $\Tr(P\rho P)$\\
\texttt{FeedForward} & instrument, outcome in the centre, branch per outcome & fan-out skip, classical only\\
\bottomrule
\end{tabular}
\caption{Layers of \texttt{torch\_vn\_algebra.learn}. Every layer is a \texttt{torch.nn.Module} that maps a batch of
states to a batch of states; every value of its parameters is a physical map.}
\label{tab:learn}
\end{table}

\paragraph{Every parameter value is a channel.} A \texttt{ChannelLayer} holds free matrices $Z_c$ and uses the
isometries $V_c=Z_c(Z_c^*Z_c)^{-1/2}$ (from \texttt{StinespringParam} of the module \texttt{optimize}); a
\texttt{LindbladLayer} holds a free Hamiltonian and free jump operators, i.e.\ a Cholesky-type factor of the
Kossakowski matrix (\texttt{LindbladParam}). No constraint is imposed and no projection is needed, so any
optimiser of \texttt{torch.optim} trains the network without leaving the set of channels, and gradients go through $(Z_c^*Z_c)^{-1/2}$ and
through the dynamics by the Daleckii--Krein backward pass of Section~\ref{sec:calculus}. Listing~\ref{lst:learn} shows
a small network trained on input--output pairs; the losses provided are the infidelity, the trace distance, the
relative entropy and the negative log-likelihood of measurement counts (from \texttt{POVM} of the module
\texttt{measurement}).

\begin{lstlisting}[float=t, caption={A network of channels of $\M=M_1\oplus M_3\oplus M_2$, trained end to end with
\texttt{torch.optim}. Every iterate is a channel.}, label=lst:learn]
import torch
from torch import nn
from torch_vn_algebra import TypeIAlgebra, channels, learn, states

alg = TypeIAlgebra([1, 3, 2], [1, 3, 2], precision='double')
X = states.random_density_matrix(alg, batch_size=16)           # inputs
Y = channels.random_channel(alg, kraus_rank=3)(X)                # outputs of an unknown channel
net = nn.Sequential(learn.UnitaryLayer(alg),
                    learn.Residual(learn.ChannelLayer(alg, kraus_rank=2), p=0.5, learn_p=True),
                    learn.LindbladLayer(alg, n_jumps=1))
opt = torch.optim.Adam(net.parameters(), lr=0.05)
for step in range(200):
    loss = learn.infidelity(net(X), Y).mean()
    opt.zero_grad(); loss.backward(); opt.step()
\end{lstlisting}

\paragraph{Where the nonlinearity comes from.} A composition of channels is again a channel, so depth adds no new
maps: one Stinespring layer of sufficient width is universal, and a deep network of channels is a reparametrisation of
that set, as a deep linear network is of the linear maps. As a function of the parameters the output is nonlinear, as a
function of the input state it is linear; on encoded classical data this makes such a model a kernel
method~\cite{schuld2021}. A
genuinely nonlinear layer needs one of three physical resources: post-selection, whose renormalisation by the
acceptance $\Tr(P\rho P)$ is the activation (\texttt{PostSelect} records the acceptance, whose logarithm is the natural
penalty, since acceptances multiply along the network); several copies of the state; or a mean-field limit of many
symmetric systems, whose finite-size dynamics the module \texttt{permutational} computes exactly in its total-spin
sectors.

\paragraph{Skip connections.} The residual block of a ResNet~\cite{he2016}, $x\mapsto x+f(x)$, becomes
$\mathrm{id}+\Phi$ for a channel $\Phi$, which is completely positive but doubles the trace. Its normalised counterpart \texttt{Residual} is the convex combination $(1-p)\,\mathrm{id}+p\,\Phi$: one
step of a jump process that applies $\Phi$ at rate $p$, so that $n$ blocks with $p=t/n$ tend to
$e^{t(\Phi-\mathrm{id})}$, the semigroup of the GKSL generator whose jump operators are the Kraus operators of $\Phi$
(verified in the test suite). ResNets become Lindblad evolutions, with the consequence studied in
Section~\ref{sec:learn-markov}. Skips that copy activations forward (DenseNet, U-Net) have no counterpart for quantum
states, which can neither be cloned nor broadcast~\cite{barnum1996}; only the commutative part can be copied. In the
algebra this is the centre: an instrument writes its outcome into the centre of $\bigoplus_j\M$
(\texttt{measurement.Instrument.as\_channel}), and \texttt{FeedForward} applies a branch per outcome, the quantum form of
a fan-out connection. A block-diagonal channel $\bigoplus_k\Phi_k$ is likewise a mixture of experts whose router, the
measurement of the sector, does not disturb the state.

\paragraph{Relation to other work.} Feed-forward networks whose layers are completely positive maps were trained on
quantum data by Beer et al.~\cite{beer2020}; process tomography with learned models is done with tensor
networks~\cite{torlai2023}, and state tomography with neural-network states~\cite{torlai2018}; networks with
parameters in a $C^*$-algebra were proposed in~\cite{hashimoto2022}, and equivariant linear layers are elements of a
commutant~\cite{cohen2016,finzi2021}. What the library adds is the algebra: layers between algebras with different
sector structures, the centre as the carrier of classical information, losses defined through it, and the batching and
gradients of the previous sections. The plan for this direction, with further experiments, is kept in the repository
(\texttt{ROADMAP\_learning.md}).

%
\ifdefined\CLRows\else\relax\fi{}
\providecommand{\CLRows}{\multicolumn{10}{c}{\emph{run scripts/channel\_learning.py}}\\}
\providecommand{\CLT}{1}\providecommand{\CLGdown}{0.1}\providecommand{\CLGup}{0.02}\providecommand{\CLGphi}{0.05}
\providecommand{\CLnMin}{3}\providecommand{\CLnMax}{4}\providecommand{\CLProbes}{--}\providecommand{\CLProbesMin}{--}
\providecommand{\CLTest}{--}\providecommand{\CLSettingsAlg}{--}\providecommand{\CLSettingsDense}{--}
\providecommand{\CLDimAlg}{--}\providecommand{\CLDimDense}{--}\providecommand{\CLSetAlg}{--}
\providecommand{\CLSetDense}{--}\providecommand{\CLIdentDense}{--}\providecommand{\CLDenseCheck}{--}
\providecommand{\CLFastPath}{--}\providecommand{\CLAdam}{--}\providecommand{\CLLbfgs}{--}
\providecommand{\CLStartsRates}{--}\providecommand{\CLStartsLindblad}{--}\providecommand{\CLStartsAlgebra}{--}
\providecommand{\CLStartsDense}{--}\providecommand{\CLStartSpread}{--}\providecommand{\CLRepeats}{--}
\providecommand{\CLRepeatShots}{--}\providecommand{\CLRuntime}{--}\providecommand{\CLWorkers}{4}
\providecommand{\CLHScale}{--}\providecommand{\CLInitConverged}{--}\providecommand{\CLInitStarts}{--}
\providecommand{\CLInitLoss}{--}\providecommand{\CLParLindblad}{--}\providecommand{\CLParAlgebra}{--}
\providecommand{\CLParDense}{--}\providecommand{\CLTDRatesTwo}{--}\providecommand{\CLTDRatesFour}{--}
\providecommand{\CLTDLindbladFour}{--}\providecommand{\CLTDAlgebraFour}{--}\providecommand{\CLTDDenseFour}{--}
\providecommand{\CLTDLindbladExact}{--}\providecommand{\CLTDAlgebraExact}{--}\providecommand{\CLTDDenseExact}{--}
\providecommand{\CLTDAlgebraFive}{--}\providecommand{\CLOffDense}{--}\providecommand{\CLTarget}{--}
\providecommand{\CLNeedLindblad}{--}\providecommand{\CLNeedAlgebra}{--}\providecommand{\CLNeedDense}{--}
\providecommand{\CLNeedRatio}{--}\providecommand{\CLResidual}{--}\providecommand{\CLParRatio}{--}
\providecommand{\CLSlopeRates}{--}\providecommand{\CLSlopeLindblad}{--}\providecommand{\CLSlopeAlgebra}{--}\providecommand{\CLSlopeDense}{--}
\providecommand{\CLRatesJ}{--}\providecommand{\CLRatesJSd}{--}\providecommand{\CLCRJ}{--}
\providecommand{\CLRatesDown}{--}\providecommand{\CLRatesDownSd}{--}\providecommand{\CLCRDown}{--}
\providecommand{\CLRatesUp}{--}\providecommand{\CLRatesUpSd}{--}\providecommand{\CLCRUp}{--}
\providecommand{\CLRatesPhi}{--}\providecommand{\CLRatesPhiSd}{--}\providecommand{\CLCRPhi}{--}
\providecommand{\CLLindbladDown}{--}\providecommand{\CLLindbladDownSd}{--}
\providecommand{\CLLindbladUp}{--}\providecommand{\CLLindbladUpSd}{--}
\providecommand{\CLLindbladPhi}{--}\providecommand{\CLLindbladPhiSd}{--}
\subsection{Tomography in the algebra as training}\label{sec:learn-tomo}

Process tomography by maximum likelihood is training in the sense of this section: the model is a parametrised
channel, the data are measurement counts on the outputs of known inputs, and the loss is the negative
log-likelihood. Learned models of this kind are used for process tomography with tensor
networks~\cite{torlai2023} and for state tomography with neural-network states~\cite{torlai2018}. The algebra enters
twice. First, it fixes what has to be measured. If $\M\subset B(\mathcal H)$ is a unital $*$-subalgebra, then
$\Tr(\rho A)=\Tr(E_\M(\rho)A)$ for $A\in\M$, where $E_\M$ is the trace-preserving conditional expectation, and the
pairing $(\sigma,A)\mapsto\Tr(\sigma A)$ is nondegenerate on $\M$ (take $A=\sigma^*$). Measurements of observables of
$\M$ therefore determine exactly $E_\M(\rho)$, and a process that maps $\M$ into $\M$, such as the noisy register of
Section~\ref{sec:noisy}, is determined on $\M$ by inputs in $\M$ and measurements in $\M$. Permutationally invariant
tomography~\cite{toth2010} is the case of the commutant of the permutations. Second, the model can respect the
sectors. A channel of $\M=\bigoplus_NM_{\binom nN}$ ranges over a set of dimension $(\dim\M)^2-\dim\M$, which is
$\CLSetAlg$ for $n=\CLnMax$, against $4^{2n}-4^n=\CLSetDense$ for a channel of $M_{2^n}$. A GKSL generator whose jumps
change $N$ by at most one has fewer parameters still. We measure what each level of structure buys, with the same
simulated experiment for all models.

\paragraph{Setup.} The register is the ring of Section~\ref{sec:noisy} with $n=\CLnMin$ and $n=\CLnMax$ qubits. The
XY mixer acts for a time $T=\CLT$, together with amplitude damping, thermal excitation and dephasing at the rates
$\gamma_\downarrow=\CLGdown$, $\gamma_\uparrow=\CLGup$ and $\gamma_\varphi=\CLGphi$. The true process
$\Phi=e^{T\mathcal L}$ is computed by \texttt{lindblad\_evolve} with the exact exponential. A dense Lindblad evolution in
$M_{2^n}$, with $H$ and the jumps converted by \texttt{interop}, agrees with it to $\CLDenseCheck$. The training
inputs are the $2^n$ basis states and $\binom nN^2$ Haar-random pure states in every sector with $\binom nN>1$. They
span $\M$, and
there are $\CLProbes$ of them for $n=\CLnMax$. The held-out inputs are $\CLTest$ further states: three pure states in each
of these sectors and eight random mixed states with weight on all sectors. Each output is measured in $\CLSettingsAlg$
Haar-random bases of $\M$ (\texttt{POVM.from\_unitary} of block-diagonal unitaries, combined with equal weights by
\texttt{POVM.combine}). Every outcome of such a basis also reveals $N$, so the charge measurement
\texttt{POVM.sectors} is a coarse-graining of each setting. The counts are sampled with $10^2$ to $10^5$ shots per
input in total, and the exact probabilities serve as the limit of infinitely many shots. Four models are fitted:
\begin{itemize}
\item[(i)] \emph{rates}: the noise model with known jump operators and four unknowns $J,\gamma_\downarrow,
\gamma_\uparrow,\gamma_\varphi$ (the coupling $J$ multiplies the mixer);
\item[(ii)] \emph{GKSL layer}: a \texttt{LindbladLayer} with a free self-adjoint $H\in\M$, $n-1$ free jumps inside the
sectors and $n$ free jumps between neighbouring sectors $N\to N\pm1$; it contains the true generator;
\item[(iii)] \emph{channel of $\M$}: a \texttt{ChannelLayer} from $\M$ to $\M$ with all sector transitions and Kraus
rank $k_{\max}^2$, enough for every channel of $\M$;
\item[(iv)] \emph{channel of $M_{2^n}$}: a \texttt{ChannelLayer} on $M_{2^n}$ with Kraus rank $4^n$, enough for every
channel. This is the experimenter who ignores the symmetry: the same inputs, embedded by \texttt{interop.to\_dense},
are measured in $\CLSettingsDense$ Haar-random bases of $M_{2^n}$, with the same number of shots per input. The model
has to learn that the process creates no coherences between sectors.
\end{itemize}
The raw parameter counts (\texttt{learn.count\_parameters}, Table~\ref{tab:learn-tomo}) include the Kraus gauge; the
sets of channels that models (iii) and (iv) can reach have dimensions $\CLSetAlg$ and $\CLSetDense$, and the inputs
in $\M$ identify $\CLIdentDense$ of the latter. Each model is fitted by $\CLAdam$ Adam steps followed by at most
$\CLLbfgs$ iterations of L-BFGS (both from \texttt{torch.optim}), with $\CLStartsRates$, $\CLStartsLindblad$,
$\CLStartsAlgebra$ and $\CLStartsDense$ random starts for models (i)--(iv), and the start with the smallest loss is
kept. The losses of the starts agree to a relative $\CLStartSpread$ in every run with finite shots. During training the
outputs are computed by equivalent fast paths: the exponential of the generator as a matrix on $\M$, of size
$\dim\M=\CLDimAlg$, and Kraus operators applied to the pure input vectors. These agree with the forward passes of the
layers to $\CLFastPath$. Evaluation uses the layers themselves. The whole study takes $\CLRuntime$ minutes on
$\CLWorkers$ CPU cores.

\begin{table}[t]
\centering\scriptsize
\setlength{\tabcolsep}{3pt}
\begin{tabular}{rlrrrrrrrr}
\toprule
 & & & \multicolumn{5}{c}{held-out trace distance, shots per input} & Choi error & time\\
$n$ & model & parameters & $10^2$ & $10^3$ & $10^4$ & $10^5$ & exact & $10^4$ shots & [s]\\
\midrule
\CLRows
\bottomrule
\end{tabular}
\caption[]{Learning the noisy XY register from simulated counts. Parameters: real parameters of the model
(\texttt{learn.count\_parameters}, Kraus gauge included). Trace distance: mean over the held-out inputs of
$\frac12\|\hat\Phi(\rho)-\Phi(\rho)\|_1$, one data set per entry; ``exact'': fitted to the exact probabilities, which
shows the optimisation error. Choi error: trace distance of the normalised Choi matrices of the restrictions to the
input sectors, averaged over the sectors. Time: wall-clock seconds per random start, averaged over the shot budgets
(one CPU core per fit).}
\label{tab:learn-tomo}
\end{table}

\begin{figure}[t]
\centering\fig{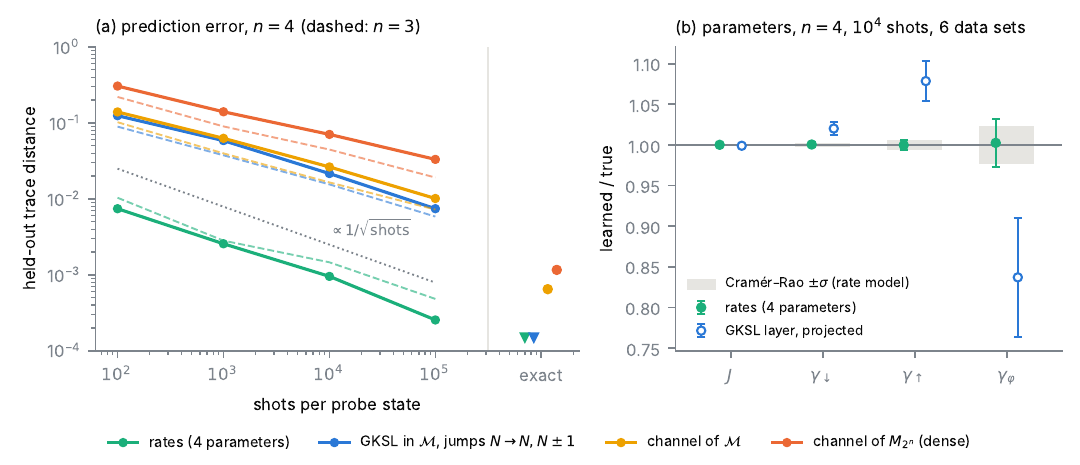}{\textwidth}
\caption{(a) Held-out trace distance against the number of shots per input for the four models, $n=\CLnMax$ (solid)
and $n=\CLnMin$ (dashed); at the right, the fit to the exact probabilities (triangles: below $10^{-4}$). (b) Learned
coupling and rates for $n=\CLnMax$ at $\CLRepeatShots$ shots per input: mean and standard deviation over
$\CLRepeats$ independent data sets, for the rate model and for the least-squares projection of the learned generator
of the GKSL layer onto $-i[H,\cdot\,]$ and the three dissipators. Grey: the Cram\'er--Rao standard deviation of the
rate model, from the Fisher information of the counts.}
\label{fig:learn-tomo}
\end{figure}

\paragraph{Results.} The models differ by large factors only where the structure is strong
(Figure~\ref{fig:learn-tomo}a). At $10^4$ shots per input and $n=\CLnMax$, the held-out trace distance is
$\CLTDRatesFour$ for the rate model, $\CLTDLindbladFour$ for the GKSL layer, $\CLTDAlgebraFour$ for the channel of $\M$
and $\CLTDDenseFour$ for the dense channel. The rate model follows the law $\text{shots}^{-1/2}$ (fitted slope
$\CLSlopeRates$ over the four budgets). The larger models fall more slowly in this range, with slopes $\CLSlopeLindblad$,
$\CLSlopeAlgebra$ and $\CLSlopeDense$, so they are not yet in their asymptotic regime and the gaps between them grow
with the number of shots. The structure of the algebra is worth more than an order of magnitude in shots: the dense
model needs $\CLNeedDense$ shots per input to reach the error $\CLTarget$ that the channel of $\M$ reaches with
$\CLNeedAlgebra$, a factor $\CLNeedRatio$. A large part of its error has a simple origin. Its predictions contain
coherences between sectors of trace norm $\CLOffDense$ (at $10^4$ shots), which the true process never creates and
which the channel of $\M$ excludes by construction. The GKSL layer gains much less over the channel of $\M$ than its
parameter count suggests: it needs $\CLNeedLindblad$ shots for the same error although it has $\CLParRatio$ times
fewer raw parameters. For so small a register, Markovianity and nearest-neighbour jumps constrain little. Knowing the
jump operators is what pays: the rate model is below the target error already at $10^2$ shots ($\CLTDRatesTwo$). The
fits to the exact probabilities reach $\CLTDLindbladExact$, $\CLTDAlgebraExact$ and $\CLTDDenseExact$ for models
(ii)--(iv), far below the statistical errors, so the comparison at finite shots is not limited by the optimiser. The
same ordering holds for the Choi error and for $n=\CLnMin$, where the GKSL layer and the channel of $\M$ tie at
$10^4$ shots.

\paragraph{Learned rates.} For the rate model the learned values at $\CLRepeatShots$ shots are $J=\CLRatesJ\pm
\CLRatesJSd$, $\gamma_\downarrow=\CLRatesDown\pm\CLRatesDownSd$, $\gamma_\uparrow=\CLRatesUp\pm\CLRatesUpSd$ and
$\gamma_\varphi=\CLRatesPhi\pm\CLRatesPhiSd$ (mean and standard deviation over $\CLRepeats$ data sets). The
Cram\'er--Rao standard deviations, from the Fisher information of the counts at the true parameters
(\texttt{measurement.fisher\_information}), are $\CLCRJ$, $\CLCRDown$, $\CLCRUp$
and $\CLCRPhi$, so the maximum-likelihood estimator is unbiased and close to efficient here (with six data sets the
empirical spread is itself uncertain by about a third). The GKSL layer is not a rate estimator. Its generator,
projected onto the physical terms, gives $\gamma_\downarrow=\CLLindbladDown\pm\CLLindbladDownSd$,
$\gamma_\uparrow=\CLLindbladUp\pm\CLLindbladUpSd$ and $\gamma_\varphi=\CLLindbladPhi\pm\CLLindbladPhiSd$. These are
biased by more than their spread, and a fraction $\CLResidual$ of the learned generator lies outside the physical span:
the free jumps absorb statistical noise as extra dissipation, which the projection attributes in part to the rates.
This fraction falls with the number of shots (from about $0.19$ at $10^2$ to $0.01$ at $10^5$ shots per input, one data set
each), and from the exact probabilities the GKSL layer recovers the rates to a relative $10^{-4}$ or better.

\paragraph{Scope.} The data are simulated without state-preparation and measurement errors, the inputs and
measurement bases are known exactly, and the true process lies inside every model class: it is Markovian and
respects the sectors exactly. The experiment therefore measures statistical efficiency, not robustness to a wrong
model. With noise that breaks the $U(1)$ symmetry, models (i)--(iii) would be biased and the dense model would not,
which is where the dense model would win; we did not simulate this case. The dense comparison changes the measurement
together with the model, as an experimenter without the symmetry would; we did not separate the two effects. The
registers are small ($n\le\CLnMax$, one evolution time), and every fit is a local optimisation. For the GKSL layer the
start matters: with the library's default initial Hamiltonian of unit scale, only $\CLInitConverged$ of
$\CLInitStarts$ starts reached the optimum for the exact data at $n=\CLnMax$, and the others stopped at losses up to
$\CLInitLoss$. We therefore start it near the identity channel, with the Hamiltonian scaled by $\CLHScale$, and with
this choice all starts agree. The full-rank models cost one parameter per entry of a Stinespring isometry. At
$n=5$ the dense model would have $4^{10}\approx10^6$ complex Kraus entries, which is why the study stops at
$n=\CLnMax$; the GKSL layer and the channel of $\M$ would still be practical there.

%
\ifdefined\RMSteps\else\relax\fi{}
\providecommand{\RMSteps}{--}\providecommand{\RMRestarts}{--}\providecommand{\RMp}{0.2}\providecommand{\RMDepthMax}{32}
\providecommand{\RMDepths}{1, 2, 4, 8, 16, 32}\providecommand{\RMSeconds}{--}\providecommand{\RMWorkers}{4}
\providecommand{\RMMarkovA}{--}\providecommand{\RMMarkovHS}{--}\providecommand{\RMMarkovTD}{--}
\providecommand{\RMGkslChannel}{--}\providecommand{\RMGkslLindblad}{--}\providecommand{\RMGkslResOne}{--}
\providecommand{\RMGkslResDeep}{--}\providecommand{\RMGkslDet}{--}
\providecommand{\RMDampedChannel}{--}\providecommand{\RMDampedLindblad}{--}\providecommand{\RMDampedResOne}{--}
\providecommand{\RMDampedResDeep}{--}\providecommand{\RMDampedDet}{--}
\providecommand{\RMPauliChannel}{--}\providecommand{\RMPauliLindblad}{--}\providecommand{\RMPauliResOne}{--}
\providecommand{\RMPauliResDeep}{--}\providecommand{\RMPauliDet}{--}
\providecommand{\RMPauliResMin}{--}\providecommand{\RMPauliResMinTD}{--}\providecommand{\RMPauliLindbladTD}{--}
\providecommand{\RMPauliResDetMin}{--}\providecommand{\RMPauliResDetMax}{--}\providecommand{\RMPauliLindbladDet}{--}
\providecommand{\RMPauliChannelDet}{--}\providecommand{\RMDephasedHS}{--}\providecommand{\RMPauliResTDAll}{--}
\providecommand{\RMPauliResTDAllDepth}{--}\providecommand{\RMSemigroupKmax}{--}\providecommand{\RMDampedSemigroupViolation}{--}
\providecommand{\RMRestartSpread}{--}\providecommand{\RMSemigroupBound}{--}
\subsection{Residual channel networks and Markovianity}\label{sec:learn-markov}

A stack of residual blocks $(1-p_l)\,\mathrm{id}+p_l\Phi_l$ with small $p_l$ is a product of channels close to the
identity. Wolf and Cirac~\cite{wolfcirac2008} call a channel \emph{infinitesimal divisible} if it is a limit of
channels that are, for every $\varepsilon>0$, products of channels $\varepsilon$-close to the identity; these
include all products of Markovian channels $e^{\mathcal L}$, and their superoperators have nonnegative determinant.
For residual blocks the determinant condition has a two-line proof, and it gives a quantitative obstruction.

\paragraph{The determinant obstruction.} Write the superoperator $T$ of a channel of $M_d$ in a basis of
self-adjoint matrices: a real $d^2\times d^2$ matrix with spectrum in the closed unit disc. If $p<\tfrac12$, the
eigenvalues $1-p+p\lambda$ of $(1-p)\,\mathrm{id}+p\,T$ have modulus at least $1-2p$; the real ones are positive and
the others come in conjugate pairs, so its determinant is at least $(1-2p)^{d^2}>0$. Determinants multiply and
$\det e^{t\mathcal L}=e^{t\operatorname{tr}\mathcal L}>0$, so every residual network with $p<\tfrac12$ and every
Lindblad layer has positive determinant, at any depth. The Pauli channel
$\Phi_P(\rho)=\tfrac13(X\rho X+Y\rho Y+Z\rho Z)$ has Pauli transfer matrix ($R_{ab}=\tfrac12\Tr(\sigma_a\Phi(\sigma_b))$,
$\sigma=1,X,Y,Z$) equal to $\operatorname{diag}(1,-\tfrac13,-\tfrac13,-\tfrac13)$, with determinant $-\tfrac1{27}$.
A trace-preserving map has $R=\bigl(\begin{smallmatrix}1&0\\t&M\end{smallmatrix}\bigr)$, so $\det R=\det M$. If
$\det M\ge0$, the segment from $-\tfrac13\mathbf 1$ to $M$ contains a singular matrix, which by Eckart--Young is at
distance at least $\tfrac13$ from $-\tfrac13\mathbf 1$ in the Frobenius and in the operator norm. Since the output
Bloch vectors differ by $t+(M+\tfrac13\mathbf 1)n$, every such map $T$ satisfies
\begin{equation}\label{eq:det-floor}
\|R_T-R_{\Phi_P}\|_F\ge\tfrac13,\qquad\sup_\psi\tfrac12\bigl\|T(\psi)-\Phi_P(\psi)\bigr\|_1\ge\tfrac16 .
\end{equation}
The second bound is sharp for infinitesimal divisible channels: a rotation by $\pi$ about $x$ after the Pauli semigroup
with eigenvalues $(\varepsilon,\varepsilon,\tfrac13)$ is a product of two Markovian channels, and as
$\varepsilon\to0$ it tends to $\operatorname{diag}(1,0,0,-\tfrac13)$, at worst-case trace distance $\tfrac16$ (and
Frobenius distance $\RMDephasedHS$). The best single semigroup we know, a rotation by $\pi$ about $z$ composed with
a Pauli semigroup, $R=\operatorname{diag}(1,-a,-a,a^2)$, $a=\RMMarkovA$, is at Frobenius distance $\RMMarkovHS$.

\paragraph{Experiment.} In $M_2$ we fit three targets: (i) $e^{\mathcal L}$ for a random GKSL generator (a random
Hamiltonian and two random jump operators); (ii) amplitude damping ($\gamma=0.5$) after a rotation by $2\pi/3$, a
product of two Markovian channels that is not itself a semigroup. Its spectrum is non-degenerate, so by the test of
Wolf, Eisert, Cubitt and Cirac~\cite{wolf2008} it is $e^{\mathcal L}$ only if a branch of its logarithm is
conditionally completely positive. The principal branch misses by $\RMDampedSemigroupViolation$, and on the branch $k$
the compressed Choi matrix is $J_0+kG$, so a vector with $\langle v,Gv\rangle<0$ (for $k>0$) or $>0$ (for $k<0$)
excludes all branches with $|k|$ above an explicit bound; here the bound is $\RMSemigroupBound$, so no branch passes;
and (iii) $\Phi_P$. The models are (a) one \texttt{ChannelLayer} of Kraus rank four, which contains every qubit
channel; (b) $L$ blocks \texttt{Residual(ChannelLayer(alg, 4), p)} with $p=\RMp$, $L\in\{\RMDepths\}$ (a learned
weight $p'\le p$ would add nothing, since the inner channel may itself be $(1-s)\,\mathrm{id}+s\Psi$); and (c) one
\texttt{LindbladLayer} with a Hamiltonian and three jump operators, which contains every generator of $M_2$. The data
are the outputs on $|0\rangle,|1\rangle,|{+}\rangle,|{+i}\rangle$, the loss is $\|R_{\rm net}-R\|_F^2$, and every
model is trained by Adam for \RMSteps\ steps from \RMRestarts\ random starts. The starts run in parallel as the
sectors of $\bigoplus_cM_2$: every layer is sector preserving with independent blocks, so each sector is an independent
network. The study takes \RMSeconds\,s on \RMWorkers\ CPU cores (\texttt{scripts/residual\_markov.py}).

\begin{figure}[t]
\centering
\fig{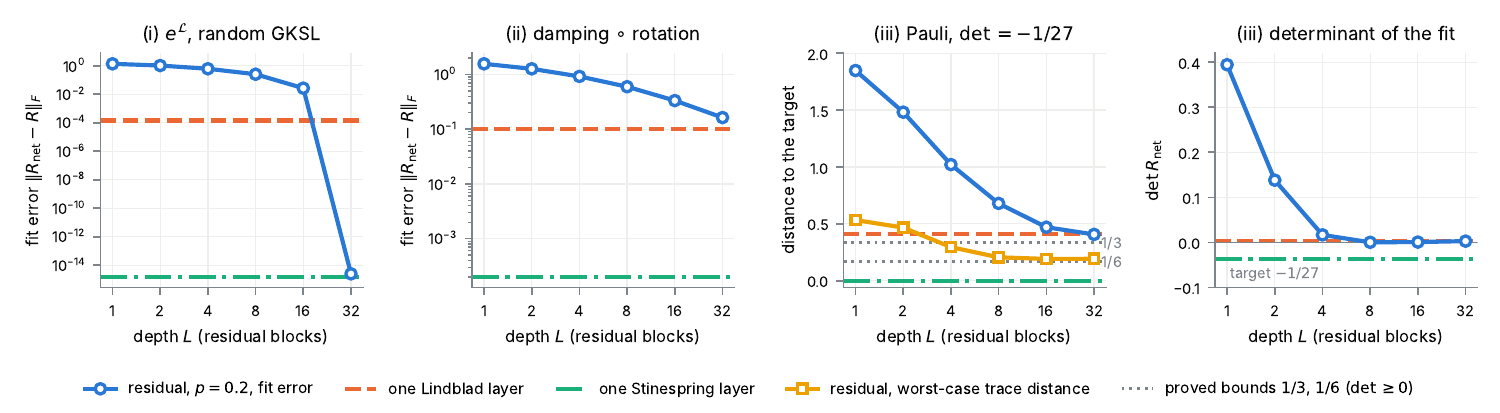}{\textwidth}
\caption{Fit error $\|R_{\rm net}-R\|_F$ (best of \RMRestarts\ starts) against the depth $L$ of a residual network
with blocks $(1-p)\,\mathrm{id}+p\,\Phi_l$, $p=\RMp$ (circles), one Lindblad layer (dashed) and one Stinespring layer
(dash-dotted), for (i) $e^{\mathcal L}$ with a random GKSL generator, (ii) amplitude damping after a rotation and
(iii) the Pauli channel $\Phi_P$, $\det=-\tfrac1{27}$. In (iii), squares: smallest worst-case trace distance among
the fitted networks; dotted: the bounds $\tfrac13$ and $\tfrac16$ of~\eqref{eq:det-floor}. Right: determinant of the
fitted maps for (iii).}
\label{fig:residual-markov}
\end{figure}

\paragraph{Results.} The Stinespring layer fits all three targets (errors $\RMGkslChannel$, $\RMDampedChannel$,
$\RMPauliChannel$), Figure~\ref{fig:residual-markov}. On the Markovian targets the residual error decreases with
depth, from $\RMGkslResOne$ at $L=1$ to $\RMGkslResDeep$ at $L=\RMDepthMax$ for (i), and from $\RMDampedResOne$ to
$\RMDampedResDeep$ for (ii). The Lindblad layer fits (i) to $\RMGkslLindblad$ but stops at $\RMDampedLindblad$ on
(ii), which is not a semigroup. On the Pauli channel every fitted residual network and the Lindblad layer have
positive determinant, the residual error never falls below $\RMPauliResMin$, and the Lindblad layer stops at
$\RMPauliLindblad$, the semigroup above. The smallest worst-case trace distance among the residual fits is
$\RMPauliResTDAll$ (at $L=\RMPauliResTDAllDepth$): training minimises the Frobenius error, not this distance, whose
infimum over infinitesimal divisible channels is $\tfrac16$. At every depth the \RMRestarts\ starts of the residual
network end within $\RMRestartSpread$ of each other. Proved are~\eqref{eq:det-floor} and the sign of the determinant; the
floor of the Frobenius error, between $\tfrac13$ and $\RMPauliResMin$, is observed, not proved.

\paragraph{Reading.} The three models are nested: semigroups, limits of residual networks with small blocks, and all
channels. On the same data the two gaps between their errors measure two kinds of non-Markovianity: dependence on
time, as for (ii), which a time-dependent master equation can still produce; and no Markovian master equation at all,
not even a time-dependent one, of which $\det R<0$ is an exact certificate, as for (iii). A residual fit error that stays away from zero as the depth grows is
therefore a learned witness of non-Markovianity, comparable with the test of~\cite{wolf2008}. The witness is
one-sided, since a large error can also be a failure of the optimiser; restarts and the Stinespring fit, which
reaches the data, control this.


\section{Applications in the companion papers}\label{sec:usecases}

The library was built for, and used in, four further studies, which are reported with proofs, validation and figures
in companion papers; here we only name the parts of the library they use.

%
\paragraph{How strongly trace characterisations fail}\label{sec:tc} (with S.~A.~Abed~\cite{nn_trace}). For
\TCItems{} inequalities that characterise the trace, the violation
by $\varphi_W=\Tr(W\,\cdot\,)$ is maximised over all operators at every contrast $\Delta(W)$, in one batched optimisation per
inequality (\texttt{optimize.SpectrumParam}, \texttt{PositiveParam}, \texttt{UnitaryParam}; lattice operations
$P\wedge Q$, $P\vee Q$ by functional calculus with Daleckii--Krein gradients), and the best points are recomputed in mpmath
with \TCDigits{} digits. The computed envelopes led to closed forms,
proved in~\cite{nn_trace}, that vanish at the trace to the orders $\delta$, $\delta^2$, $\delta^4$ and $\delta^8$.

%
\paragraph{The Connes--St{\o}rmer entropy of two MASAs}\label{sec:cs} \cite{nn_cs}. The relative entropy
$H(\Delta_n|U\Delta_nU^*)$ of the diagonal algebra and a rotated copy is a supremum over partitions of unity; with the
conditional expectations of \texttt{Inclusion} and partitions parametrised by isometries, many unitaries and starts form
one batch. For $n=2$ the computation suggested a closed form, proved in~\cite{nn_cs}; for $n=3,4,5$ it finds partitions
that beat every orthonormal basis found.

%
\paragraph{A boundary time crystal under local dissipation}\label{sec:btc} \cite{nn_btc}. Local emission and dephasing of
$N$ two-level systems commute with the permutations of the sites, so the state stays in the Schur--Weyl algebra
$\bigoplus_jM_{2j+1}\otimes1_{m_j}$, and the local dissipators are maps between its sectors
(\texttt{PermutationalLindbladian}). With all sectors kept, the lifetimes of the oscillations follow for
$N\le\PMNmax{}$ without truncation, ten rates in one batch (a check-mode
run on one CPU thread; the full mode, $N\le200$, is designed for a GPU and was not run).

%
\paragraph{Symmetry-resolved entanglement in the XXZ chain}\label{sec:sre} \cite{nn_btc}. The reduced state of a
charge-conserving state is an operator of $\bigoplus_qM_{d_q}$, so resolved entropies and charge statistics are read off
its blocks. For $L=\SRL{}$ batched Lanczos runs, four anisotropies each, give
the ground states for eight anisotropies, and the charge statistics give the Luttinger parameter to about one per cent.

\section{Performance}\label{sec:performance}

Table~\ref{tab:benchmark} reports the wall-clock time per sample of library operations
(not of isolated kernels), in single precision, with batch $B=\min\bigl(4096,\,\G{benchmark}{2^{\BenchTargetLog}}{2^{22}}/(Ck^2)\bigr)$\G{benchmark}{ (\BenchMode\ run on \BenchDevice;
CPU with \BenchCpuThreads\ threads; PyTorch \BenchTorch, CUDA \BenchCuda, cuSOLVER \BenchCusolver)}{}. The operations
are: generating random operators with prescribed spectra, $f(A)$ by eigendecomposition, $|A|$, the Michelson contrast,
applying a channel, the Lindblad channel $e^{t\mathcal L}$, and one gradient step of the optimisation.

Library versions matter: batched Hermitian eigensolvers on GPUs cover all block sizes only in recent cuSOLVER and
PyTorch releases, and batched SVD on GPUs is limited to blocks of size at most 32, which is why the library uses the
Hermitian eigensolver wherever it can. Batched QR on GPUs factorises small matrices one at a time, so Haar sampling on
GPUs uses a batched Gram--Schmidt orthonormalisation with reorthogonalisation (the same matrices as QR with the phase
correction). Table~\ref{tab:benchmark} was measured before this change; its rows ``generate'' on the GPU therefore show
the earlier per-sector QR and understate the GPU. On the CPU the library splits batched decompositions into one chunk per thread, so all cores are used.

GPUs need enough independent work per kernel launch: many samples, many sectors or large blocks, and matrix products
rather than decompositions of small matrices. The design follows from this. Everything is batched, so the batch
dimension supplies the parallel work, and the cost module warns before configurations that would be slow. The structure
of the computation is also checked directly (\texttt{scripts/gpu\_check.py} on a GPU, \texttt{tests/test\_sync\_free.py}
on a CPU): the operations read no values back to the host, and their number of kernel launches does not grow with $C$
at fixed total size. For workloads that are small or dominated by a single eigenproblem, the CPU, or SciPy, is the
right tool, and the library runs there unchanged.

\begin{table}[tbp]
\centering\scriptsize\renewcommand{\arraystretch}{0.85}
\ifnum\BenchHasGPU=1
\begin{tabular}{lrrrrrr}
\toprule
operation & $k$ & $C$ & batch & GPU [$\mu$s] & CPU [$\mu$s] & speedup\\
\else
\begin{tabular}{lrrrr}
\toprule
operation & $k$ & $C$ & batch & time per sample [$\mu$s]\\
\fi
\midrule
\BenchmarkRows
\bottomrule
\end{tabular}
\caption[]{Wall-clock time per sample of library operations for $C$ sectors of size $k$ and the batch size
shown\G{benchmark}{ (\BenchMode\ run on \BenchDevice, CPU with \BenchCpuThreads\ threads)}{} (single precision;
``exp($tL$)'' uses a smaller batch because its superoperators have $k^4$ entries). With a GPU the table also gives the
CPU time and the speedup CPU/GPU. The run predates the batched Gram--Schmidt generator of Haar unitaries on GPUs.}
\label{tab:benchmark}
\end{table}

\section{Limitations and roadmap}\label{sec:limitations}

\begin{itemize}\itemsep1pt
\item \emph{Layout.} Blocks are padded to $k_{\max}$. For very different sector sizes, such as the binomial sectors of
spin chains, one works in single sectors, and \texttt{SU2Chain} builds the block of one total spin without padding; a
bucketed layout for operators on all sectors is not implemented yet. Matrix products of padded operators also multiply
the padding: for the ten-qubit register of Section~\ref{sec:noisy} the dense Hamiltonian part of a time step costs
about four times $\sum_Nk_N^3$, and it, not the jumps, dominates the run time.
\item \emph{Symmetries.} Abelian charges with an arbitrary fusion rule, lattice translations and, for spin-$\frac12$
chains, SU(2) multiplets are supported; SU(2) with momentum, larger spins, reflections, spin flip and sparse solvers in
SU(2) blocks are not. The Wedderburn decomposition uses dense linear algebra. Dense methods reach about 16--18 spins in
the $U(1)$ sectors, the sparse solvers the half-filled sector of 24 spins (dimension $2.7\cdot10^6$).
\item \emph{Noise.} Only noise that commutes with the symmetry, in the weak sense of Section~\ref{sec:noisy}, can be
represented; bit flips and other symmetry-breaking errors leave the algebra.
\item \emph{Optimisation.} The search is local, so envelopes are lower bounds for suprema. Gradients of the functional
calculus are finite at degenerate spectra, but gradients through the eigenvectors returned by \texttt{eigh} are not,
and second derivatives of $f(A)$ are correct at non-degenerate spectra only.
\item \emph{Divergences and scale.} For small $\alpha$ the sandwiched divergence of ill-conditioned states is not
accurate (hence $\alpha\ge0.2$ in Section~\ref{sec:renyi}). Choi matrices, superoperators and $e^{t\mathcal L}$ work
with $k_c^2\times k_c^2$ matrices and are limited to blocks of a few tens; RK4 and the splitting scale further.
\item \emph{Learning.} The layers of Section~\ref{sec:learn} are small: dense Kraus blocks and dense generators, trained
on simulated data for registers of a few qubits. Encoders into smaller algebras, tomography of larger registers with
sparse measurements, training viewed as a dynamics of the channel and certified upper bounds for the learned maps are
not implemented; the plan is kept in the repository (\texttt{ROADMAP\_learning.md}).
\item \emph{Scope.} Finite dimensions and Type~I only; no tensor-network methods. Sparse methods exist only for sector
Hamiltonians and for jumps that map basis vectors to basis vectors; other jumps between sectors are stored densely.
Results are cross-checked against QuTiP and QuSpin in the test suite, and the permutational Lindblad solver is timed
against QuTiP's PIQS in~\cite{nn_btc}; systematic benchmarks against QuTiP, \texttt{dynamiqs}, QuantumToolbox.jl, QuSpin
and XDiag are future work.
\end{itemize}

\section{Availability and reproducibility}\label{sec:availability}

The code is available under the MIT licence at \url{https://github.com/andre-a-no/von_neumann_type_i}; this paper
describes version~1.2.0 (tag \texttt{v1.2.0}), and its scripts need the repository version of the library
(\texttt{pip install -e ".[scripts]"}) rather than an older release from PyPI. The command
\texttt{bash scripts/run\_all.sh full cuda} regenerates every table and figure on a GPU; without the mode argument it
runs a reduced version on the default device, which takes well under an hour on a four-core CPU. The script
\texttt{scripts/\allowbreak gpu\_check.py} checks a GPU, and \texttt{scripts/\allowbreak library\_paper\_schematics.py}
draws the schematic figures and the summary figure of the noisy register. The tables state the device and, except the
use-case tables of Sections~\ref{sec:noisy} and~\ref{sec:control}, which are single full-size runs, the mode they were
produced with; the JSON files written by the current versions of the scripts also record the commit of the code.

\appendix
{\sloppy
}


\small
\begin{thebibliography}{99}\setlength{\itemsep}{1pt}
\bibitem{nn_trace} S.~A.~Abed, I.~A.~Nikolaeva, A.~A.~Novikov, ``How characterisations of the trace fail: sharp
  violation bounds in terms of Michelson contrasts'', preprint (2026); source in \texttt{man/trace\_envelopes} of the
  repository.
\bibitem{nn_cs} I.~Nikolaeva, A.~Novikov, ``The Connes--St{\o}rmer entropy of two maximal abelian subalgebras of $M_2$:
  an explicit variational formula, and partitions that beat bases in $M_n$'', in preparation; source in
  \texttt{man/connes\_stormer} of the repository.
\bibitem{nn_btc} I.~Nikolaeva, A.~Novikov, ``Exact symmetry sectors at work: finite-$N$ lifetimes of a boundary time
  crystal under local dissipation and symmetry-resolved entanglement in the XXZ chain'', in preparation; source in
  \texttt{man/btc\_sre} of the repository.
\bibitem{v1} I.~Nikolaeva, A.~Novikov, ``Finite-dimensional Type I von Neumann algebras in PyTorch: a
  GPU-accelerated framework for random block-diagonal operators'', arXiv:2606.15882v1 (2026).
\bibitem{gardner1979} L.~T.~Gardner, ``An inequality characterizes the trace'', Canad.\ J.\ Math.\ 31
  (1979), no.~6, 1322--1328.
\bibitem{petz1988} D.~Petz, J.~Zem\'anek, ``Characterizations of the trace'', Linear Algebra Appl.\ 111
  (1988), 43--52.
\bibitem{novikov2015} A.~A.~Novikov, O.~E.~Tikhonov, ``Characterization of central elements of operator
  algebras by inequalities'', Lobachevskii J.\ Math.\ 36 (2015), no.~2, 208--210; arXiv:1502.01267.
\bibitem{virosztek2017} D.~Virosztek, ``Connections between centrality and local monotonicity of certain
  functions on C*-algebras'', J.\ Math.\ Anal.\ Appl.\ 453 (2017), 221--226.
\bibitem{abed2024} S.~A.~Abed, I.~A.~Nikolaeva, A.~A.~Novikov, ``Generalisation of Michelson contrast for
  operators and its properties'', Lobachevskii J.\ Math.\ 45 (2024), no.~8, 3835--3848; arXiv:2010.10130.
\bibitem{carlen2010} E.~Carlen, ``Trace inequalities and quantum entropy: an introductory course'', in
  \emph{Entropy and the Quantum}, Contemp.\ Math.\ 529, Amer.\ Math.\ Soc., 2010, 73--140.
\bibitem{connes1975} A.~Connes, E.~St{\o}rmer, ``Entropy for automorphisms of II$_1$ von Neumann algebras'', Acta
  Math.\ 134 (1975), 289--306.
\bibitem{watrous2009} J.~Watrous, ``Semidefinite programs for completely bounded norms'', Theory Comput.\ 5 (2009),
  217--238.
\bibitem{fletcher2007} A.~S.~Fletcher, P.~W.~Shor, M.~Z.~Win, ``Optimum quantum error recovery using semidefinite
  programming'', Phys.\ Rev.\ A 75 (2007), 012338.
\bibitem{hastings2009} M.~B.~Hastings, ``Superadditivity of communication capacity using entangled inputs'', Nature
  Phys.\ 5 (2009), 255--257.
\bibitem{khaneja2005} N.~Khaneja, T.~Reiss, C.~Kehlet, T.~Schulte-Herbr\"uggen, S.~J.~Glaser, ``Optimal control of
  coupled spin dynamics: design of NMR pulse sequences by gradient ascent algorithms'', J.\ Magn.\ Reson.\ 172 (2005),
  296--305.
\bibitem{koch2016} C.~P.~Koch, ``Controlling open quantum systems: tools, achievements, and limitations'', J.\ Phys.:
  Condens.\ Matter 28 (2016), 213001.
\bibitem{cerezo2021} M.~Cerezo, A.~Arrasmith, R.~Babbush, S.~C.~Benjamin, S.~Endo, K.~Fujii, J.~R.~McClean, K.~Mitarai,
  X.~Yuan, L.~Cincio, P.~J.~Coles, ``Variational quantum algorithms'', Nat.\ Rev.\ Phys.\ 3 (2021), 625--644.
\bibitem{johansson2013} J.~R.~Johansson, P.~D.~Nation, F.~Nori, ``QuTiP 2: A Python framework for the
  dynamics of open quantum systems'', Comput.\ Phys.\ Commun.\ 184 (2013), 1234--1240.
\bibitem{lambert2026} N.~Lambert et al., ``QuTiP~5: The Quantum Toolbox in Python'', Phys.\ Rep.\ 1153 (2026),
  1--62; arXiv:2412.04705.
\bibitem{qiskit2019} A.~Javadi-Abhari et al., ``Quantum computing with Qiskit'', arXiv:2405.08810 (2024).
\bibitem{dynamiqs} P.~Guilmin, A.~Bocquet, \'E.~Genois, D.~Weiss, R.~Gautier, ``Dynamiqs: an open-source Python
  library for GPU-accelerated and differentiable simulation of quantum systems'', in preparation (2025);
  software version 0.3.7, \url{https://github.com/dynamiqs/dynamiqs} (accessed October 2026).
\bibitem{mezzadri2007} F.~Mezzadri, ``How to generate random matrices from the classical compact
  groups'', Notices Amer.\ Math.\ Soc.\ 54 (2007), 592--604.
\bibitem{mehta2004} M.~L.~Mehta, \emph{Random Matrices}, 3rd ed., Elsevier, 2004.
\bibitem{page1993} D.~N.~Page, ``Average entropy of a subsystem'', Phys.\ Rev.\ Lett.\ 71 (1993),
  1291--1294.
\bibitem{foong1994} S.~K.~Foong, S.~Kanno, ``Proof of Page's conjecture on the average entropy of a subsystem'',
  Phys.\ Rev.\ Lett.\ 72 (1994), 1148--1151.
\bibitem{sanchezruiz1995} J.~S\'anchez-Ruiz, ``Simple proof of Page's conjecture on the average entropy of a
  subsystem'', Phys.\ Rev.\ E 52 (1995), 5653--5655.
\bibitem{lubkin1978} E.~Lubkin, ``Entropy of an $n$-system from its correlation with a $k$-reservoir'', J.\ Math.\ Phys.\ 19 (1978), 1028--1031.
\bibitem{collins2006} B.~Collins, P.~\'Sniady, ``Integration with respect to the Haar measure on unitary,
  orthogonal and symplectic group'', Commun.\ Math.\ Phys.\ 264 (2006), 773--795.
\bibitem{zyczkowski2001} K.~\.Zyczkowski, H.-J.~Sommers, ``Induced measures in the space of mixed quantum
  states'', J.\ Phys.\ A 34 (2001), 7111--7125.
\bibitem{peschel2003} I.~Peschel, ``Calculation of reduced density matrices from correlation functions'',
  J.\ Phys.\ A 36 (2003), L205--L208.
\bibitem{oganesyan2007} V.~Oganesyan, D.~A.~Huse, ``Localization of interacting fermions at high temperature'',
  Phys.\ Rev.\ B 75 (2007), 155111.
\bibitem{atas2013} Y.~Y.~Atas, E.~Bogomolny, O.~Giraud, G.~Roux, ``Distribution of the ratio of consecutive
  level spacings in random matrix ensembles'', Phys.\ Rev.\ Lett.\ 110 (2013), 084101.
\bibitem{takesaki2002} M.~Takesaki, \emph{Theory of Operator Algebras I}, Springer, 2002.
\bibitem{wick1952} G.~C.~Wick, A.~S.~Wightman, E.~P.~Wigner, ``The intrinsic parity of elementary particles'',
  Phys.\ Rev.\ 88 (1952), 101--105.
\bibitem{bartlett2007} S.~D.~Bartlett, T.~Rudolph, R.~W.~Spekkens, ``Reference frames, superselection rules, and
  quantum information'', Rev.\ Mod.\ Phys.\ 79 (2007), 555--609.
\bibitem{zurek2003} W.~H.~Zurek, ``Decoherence, einselection, and the quantum origins of the classical'',
  Rev.\ Mod.\ Phys.\ 75 (2003), 715--775.
\bibitem{sandvik2010} A.~W.~Sandvik, ``Computational studies of quantum spin systems'', AIP Conf.\ Proc.\ 1297
  (2010), 135--338.
\bibitem{weinberg2017} P.~Weinberg, M.~Bukov, ``QuSpin: a Python package for dynamics and exact diagonalisation of
  quantum many body systems part I: spin chains'', SciPost Phys.\ 2 (2017), 003.
\bibitem{goldstein2018} M.~Goldstein, E.~Sela, ``Symmetry-resolved entanglement in many-body systems'',
  Phys.\ Rev.\ Lett.\ 120 (2018), 200602.
\bibitem{xavier2018} J.~C.~Xavier, F.~C.~Alcaraz, G.~Sierra, ``Equipartition of the entanglement entropy'',
  Phys.\ Rev.\ B 98 (2018), 041106.
\bibitem{zanardi1997} P.~Zanardi, M.~Rasetti, ``Noiseless quantum codes'', Phys.\ Rev.\ Lett.\ 79 (1997), 3306--3309.
\bibitem{knill2000} E.~Knill, R.~Laflamme, L.~Viola, ``Theory of quantum error correction for general noise'',
  Phys.\ Rev.\ Lett.\ 84 (2000), 2525--2528.
\bibitem{beny2007} C.~B\'eny, A.~Kempf, D.~W.~Kribs, ``Generalization of quantum error correction via the Heisenberg
  picture'', Phys.\ Rev.\ Lett.\ 98 (2007), 100502.
\bibitem{bonetmonroig2018} X.~Bonet-Monroig, R.~Sagastizabal, M.~Singh, T.~E.~O'Brien, ``Low-cost error mitigation
by symmetry verification'', Phys.\ Rev.\ A \textbf{98} (2018), 062339.
\bibitem{mcardle2019} S.~McArdle, X.~Yuan, S.~Benjamin, ``Error-mitigated digital quantum simulation'', Phys.\ Rev.\
Lett.\ \textbf{122} (2019), 180501.
\bibitem{hadfield2019} S.~Hadfield, Z.~Wang, B.~O'Gorman, E.~G.~Rieffel, D.~Venturelli, R.~Biswas, ``From the quantum
  approximate optimization algorithm to a quantum alternating operator ansatz'', Algorithms 12 (2019), 34.
\bibitem{dyson1962} F.~J.~Dyson, ``The threefold way. Algebraic structure of symmetry groups and ensembles in quantum
  mechanics'', J.\ Math.\ Phys.\ 3 (1962), 1199--1215.
\bibitem{cohen2016} T.~Cohen, M.~Welling, ``Group equivariant convolutional networks'', Proc.\ ICML 2016, PMLR 48,
  2990--2999.
\bibitem{finzi2021} M.~Finzi, M.~Welling, A.~G.~Wilson, ``A practical method for constructing equivariant multilayer
  perceptrons for arbitrary matrix groups'', Proc.\ ICML 2021, PMLR 139, 3318--3328.
\bibitem{harris2020} C.~R.~Harris et al., ``Array programming with NumPy'', Nature 585 (2020), 357--362.
\bibitem{virtanen2020} P.~Virtanen et al., ``SciPy 1.0: fundamental algorithms for scientific computing in Python'',
  Nat.\ Methods 17 (2020), 261--272.
\bibitem{lehoucq1998} R.~B.~Lehoucq, D.~C.~Sorensen, C.~Yang, \emph{ARPACK Users' Guide}, SIAM, 1998.
\bibitem{fishman2022} M.~Fishman, S.~R.~White, E.~M.~Stoudenmire, ``The ITensor software library for tensor network
  calculations'', SciPost Phys.\ Codebases 4 (2022).
\bibitem{rams2025} M.~M.~Rams, G.~W\'ojtowicz, A.~Sinha, J.~Hasik, ``YASTN: Yet another symmetric tensor networks; A
  Python library for Abelian symmetric tensor network calculations'', SciPost Phys.\ Codebases 52 (2025);
  arXiv:2405.12196.
\bibitem{hauschild2018} J.~Hauschild, F.~Pollmann, ``Efficient numerical simulations with tensor networks: Tensor
  Network Python (TeNPy)'', SciPost Phys.\ Lect.\ Notes 5 (2018).
\bibitem{higham2008} N.~J.~Higham, \emph{Functions of Matrices: Theory and Computation}, SIAM, 2008.
\bibitem{trefethen2014} L.~N.~Trefethen, J.~A.~C.~Weideman, ``The exponentially convergent trapezoidal rule'',
SIAM Rev. 56 (2014) 385--458.
\bibitem{daleckii1965} Ju.~L.~Daleckii, S.~G.~Krein, ``Integration and differentiation of functions of Hermitian
operators and applications to the theory of perturbations'', Amer. Math. Soc. Transl. (2) 47 (1965) 1--30.
\bibitem{bhatia1997} R.~Bhatia, \emph{Matrix Analysis}, Graduate Texts in Mathematics 169, Springer, 1997.
\bibitem{kasim2020} M.~F.~Kasim, ``Derivatives of partial eigendecomposition of a real symmetric matrix for
degenerate cases'', arXiv:2011.04366 (2020).
\bibitem{petz1986} D.~Petz, ``Quasi-entropies for finite quantum systems'', Rep.\ Math.\ Phys.\ 23 (1986), 57--65.
\bibitem{petz1986suff} D.~Petz, ``Sufficient subalgebras and the relative entropy of states of a von Neumann
  algebra'', Commun.\ Math.\ Phys.\ 105 (1986), 123--131.
\bibitem{petz1988chan} D.~Petz, ``Sufficiency of channels over von Neumann algebras'', Quart.\ J.\ Math.\ Oxford 39
  (1988), no.~1, 97--108.
\bibitem{mullerlennert2013} M.~M\"uller-Lennert, F.~Dupuis, O.~Szehr, S.~Fehr, M.~Tomamichel, ``On quantum R\'enyi
  entropies: a new generalization and some properties'', J.\ Math.\ Phys.\ 54 (2013), 122203.
\bibitem{wilde2014} M.~M.~Wilde, A.~Winter, D.~Yang, ``Strong converse for the classical capacity of
  entanglement-breaking and Hadamard channels via a sandwiched R\'enyi relative entropy'', Commun.\ Math.\ Phys.\ 331
  (2014), 593--622.
\bibitem{frank2013} R.~L.~Frank, E.~H.~Lieb, ``Monotonicity of a relative R\'enyi entropy'', J.\ Math.\ Phys.\ 54
  (2013), 122201.
\bibitem{beigi2013} S.~Beigi, ``Sandwiched R\'enyi divergence satisfies data processing inequality'', J.\ Math.\
  Phys.\ 54 (2013), 122202.
\bibitem{berta2017} M.~Berta, O.~Fawzi, M.~Tomamichel, ``On variational expressions for quantum relative
  entropies'', Lett.\ Math.\ Phys.\ 107 (2017), 2239--2265.
\bibitem{zhang2020} H.~Zhang, ``From Wigner--Yanase--Dyson conjecture to Carlen--Frank--Lieb conjecture'', Adv.\
  Math.\ 365 (2020), 107053.
\bibitem{junge2018} M.~Junge, R.~Renner, D.~Sutter, M.~M.~Wilde, A.~Winter, ``Universal recovery maps and approximate
  sufficiency of quantum relative entropy'', Ann.\ Henri Poincar\'e 19 (2018), 2955--2978.
\bibitem{jones1983} V.~F.~R.~Jones, ``Index for subfactors'', Invent.\ Math.\ 72 (1983), 1--25.
\bibitem{ghj1989} F.~M.~Goodman, P.~de~la~Harpe, V.~F.~R.~Jones, \emph{Coxeter Graphs and Towers of Algebras}, MSRI
  Publications 14, Springer, 1989.
\bibitem{pimsner1986} M.~Pimsner, S.~Popa, ``Entropy and index for subfactors'', Ann.\ Sci.\ \'Ecole Norm.\ Sup.\ (4)
  19 (1986), 57--106.
\bibitem{watatani1990} Y.~Watatani, ``Index for $C^*$-subalgebras'', Mem.\ Amer.\ Math.\ Soc.\ 83 (1990), no.~424.
\bibitem{frank1998} M.~Frank, E.~Kirchberg, ``On conditional expectations of finite index'', J.\ Operator Theory 40
  (1998), 87--111.
\bibitem{bdh1988} M.~Baillet, Y.~Denizeau, J.-F.~Havet, ``Indice d'une esp\'erance conditionnelle'', Compositio
  Math.\ 66 (1988), 199--236.
\bibitem{murota2010} K.~Murota, Y.~Kanno, M.~Kojima, S.~Kojima, ``A numerical algorithm for block-diagonal
  decomposition of matrix $*$-algebras with application to semidefinite programming'', Japan J.\ Indust.\ Appl.\
  Math.\ 27 (2010), 125--160.
\bibitem{maehara2010} T.~Maehara, K.~Murota, ``A numerical algorithm for block-diagonal decomposition of matrix
  $*$-algebras with general irreducible components'', Japan J.\ Indust.\ Appl.\ Math.\ 27 (2010), 263--293.
\bibitem{maehara2011} T.~Maehara, K.~Murota, ``Algorithm for error-controlled simultaneous block-diagonalization of
  matrices'', SIAM J.\ Matrix Anal.\ Appl.\ 32 (2011), 605--620.
\bibitem{fulton1991} W.~Fulton, J.~Harris, \emph{Representation Theory: A First Course}, Graduate Texts in
  Mathematics 129, Springer, 1991.
\bibitem{racah1942} G.~Racah, ``Theory of complex spectra.~II'', Phys.\ Rev.\ 62 (1942), 438--462.
\bibitem{edmonds1957} A.~R.~Edmonds, \emph{Angular Momentum in Quantum Mechanics}, Princeton University Press, 1957.
\bibitem{lieb1962} E.~Lieb, D.~Mattis, ``Ordering energy levels of interacting spin systems'', J.\ Math.\ Phys.\ 3
  (1962), 749--751.
\bibitem{majumdar1969} C.~K.~Majumdar, D.~K.~Ghosh, ``On next-nearest-neighbor interaction in linear chain.~I'',
  J.\ Math.\ Phys.\ 10 (1969), 1388--1398.
\bibitem{majumdar1969b} C.~K.~Majumdar, D.~K.~Ghosh, ``On next-nearest-neighbor interaction in linear chain.~II'',
  J.\ Math.\ Phys.\ 10 (1969), 1399--1402.
\bibitem{paszke2019} A.~Paszke et al., ``PyTorch: An imperative style, high-performance deep learning
  library'', NeurIPS 2019.
\bibitem{shammah2018} N.~Shammah, S.~Ahmed, N.~Lambert, S.~De~Liberato, F.~Nori, ``Open quantum systems with local
  and collective incoherent processes: Efficient numerical simulations using permutational invariance'',
  Phys.\ Rev.\ A 98 (2018), 063815.
\bibitem{mercurio2025} A.~Mercurio, Y.-T.~Huang, L.-X.~Cai, Y.-N.~Chen, V.~Savona, F.~Nori, ``QuantumToolbox.jl: An
  efficient Julia framework for simulating open quantum systems'', Quantum 9 (2025), 1866;
  doi:10.22331/q-2025-09-29-1866.
\bibitem{qiskitaer} Qiskit Aer, high-performance simulators for Qiskit, documentation at
  \url{https://qiskit.github.io/qiskit-aer/} (accessed October 2026).
\bibitem{quest2019} T.~Jones, A.~Brown, I.~Bush, S.~C.~Benjamin, ``QuEST and high performance simulation of quantum
  computers'', Sci.\ Rep.\ 9 (2019), 10736; doi:10.1038/s41598-019-47174-9.
\bibitem{bayraktar2023} H.~Bayraktar et al., ``cuQuantum SDK: A high-performance library for accelerating quantum
  science'', in 2023 IEEE International Conference on Quantum Computing and Engineering (QCE), pp.~1050--1061;
  doi:10.1109/QCE57702.2023.00119; arXiv:2308.01999.
\bibitem{cudensitymat} NVIDIA, ``cuDensityMat: a high-performance library for analog quantum dynamics computations'',
  cuQuantum SDK documentation, \url{https://docs.nvidia.com/cuda/cuquantum/latest/cudensitymat/index.html} (accessed
  October 2026).
\bibitem{bergholm2018} V.~Bergholm et al., ``PennyLane: Automatic differentiation of hybrid quantum-classical
  computations'', arXiv:1811.04968 (2018).
\bibitem{puzzuoli2023} D.~Puzzuoli, C.~J.~Wood, D.~J.~Egger, B.~Rosand, K.~Ueda, ``Qiskit Dynamics: A Python package
  for simulating the time dynamics of quantum systems'', J.\ Open Source Softw.\ 8 (2023), no.~90, 5853.
\bibitem{diamond2016} S.~Diamond, S.~Boyd, ``CVXPY: A Python-embedded modeling language for convex optimization'',
  J.\ Mach.\ Learn.\ Res.\ 17 (2016), no.~83, 1--5.
\bibitem{sagnol2022} G.~Sagnol, M.~Stahlberg, ``PICOS: A Python interface to conic optimization solvers'', J.\ Open
  Source Softw.\ 7 (2022), no.~70, 3915.
\bibitem{qetlab} N.~Johnston, QETLAB: A MATLAB toolbox for quantum entanglement, version 1.0 (2025),
  \url{https://qetlab.com} (accessed October 2026).
\bibitem{townsend2016} J.~Townsend, N.~Koep, S.~Weichwald, ``Pymanopt: A Python toolbox for optimization on manifolds
  using automatic differentiation'', J.\ Mach.\ Learn.\ Res.\ 17 (2016), no.~137, 1--5.
\bibitem{kochurov2020} M.~Kochurov, R.~Karimov, S.~Kozlukov, ``Geoopt: Riemannian optimization in PyTorch'',
  arXiv:2005.02819 (2020).
\bibitem{meghwanshi2018} M.~Meghwanshi, P.~Jawanpuria, A.~Kunchukuttan, H.~Kasai, B.~Mishra, ``McTorch, a manifold
  optimization library for deep learning'', arXiv:1810.01811 (2018).
\bibitem{bergmann2022} R.~Bergmann, ``Manopt.jl: Optimization on manifolds in Julia'', J.\ Open Source Softw.\ 7
  (2022), no.~70, 3866.
\bibitem{axen2023} S.~D.~Axen, M.~Baran, R.~Bergmann, K.~Rzecki, ``Manifolds.jl: An extensible Julia framework for
  data analysis on manifolds'', ACM Trans.\ Math.\ Softw.\ 49 (2023), no.~4, 1--23.
\bibitem{weinberg2019} P.~Weinberg, M.~Bukov, ``QuSpin: a Python package for dynamics and exact diagonalisation of
  quantum many body systems. Part II: bosons, fermions and higher spins'', SciPost Phys.\ 7 (2019), 020;
  doi:10.21468/SciPostPhys.7.2.020.
\bibitem{wietek2026} A.~Wietek, L.~Staszewski, M.~Ulaga, P.~L.~Ebert, H.~Karlsson, S.~Sarkar, L.~Shackleton,
  A.~Sinha, R.~D.~Soares, ``XDiag: Exact diagonalization for quantum many-body systems'', SciPost Phys.\ Codebases 70
  (2026); doi:10.21468/SciPostPhysCodeb.70; arXiv:2505.02901.
\bibitem{devos2025} L.~Devos, J.~Haegeman, ``TensorKit.jl: A Julia package for large-scale tensor computations, with a
  hint of category theory'', arXiv:2508.10076 (2025).
\bibitem{gardner2018} J.~R.~Gardner, G.~Pleiss, D.~Bindel, K.~Q.~Weinberger, A.~G.~Wilson, ``GPyTorch: Blackbox
  matrix-matrix Gaussian process inference with GPU acceleration'', in Advances in Neural Information Processing
  Systems 31 (NeurIPS 2018), pp.~7576--7586; arXiv:1809.11165.
\bibitem{linearoperator} \texttt{linear\_operator}: a PyTorch package for structured matrices, from the GPyTorch
  project, \url{https://github.com/cornellius-gp/linear_operator} (accessed October 2026).
\bibitem{potapczynski2023} A.~Potapczynski, M.~Finzi, G.~Pleiss, A.~G.~Wilson, ``CoLA: Exploiting compositional
  structure for automatic and efficient numerical linear algebra'', in Advances in Neural Information Processing
  Systems 36 (NeurIPS 2023); arXiv:2309.03060.
\bibitem{orlitzky2022} M.~Orlitzky, ``Rank computation in Euclidean Jordan algebras'', J.\ Symbolic Comput.\ 113
  (2022), 181--192; doi:10.1016/j.jsc.2022.04.003; software at \url{https://codeberg.org/mjo/sage.d}.
\bibitem{alco} B.~Nasmith, ALCO: tools for algebraic combinatorics, GAP package, version 1.1.2 (2025),
  \url{https://github.com/BNasmith/alco}.
\bibitem{hankin2023} R.~K.~S.~Hankin, ``Jordan algebra in R'', arXiv:2303.06062 (2023).
\bibitem{brooksbank2012} P.~A.~Brooksbank, J.~B.~Wilson, ``Computing isometry groups of Hermitian maps'', Trans.\ Amer.\
  Math.\ Soc.\ 364 (2012), 1975--1996; doi:10.1090/S0002-9947-2011-05388-2.
\bibitem{deklerk2011} E.~de~Klerk, C.~Dobre, D.~V.~Pasechnik, ``Numerical block diagonalization of matrix $*$-algebras
  with application to semidefinite programming'', Math.\ Program.\ 129 (2011), 91--111; doi:10.1007/s10107-011-0461-3.
\bibitem{permenter2020} F.~Permenter, P.~A.~Parrilo, ``Dimension reduction for semidefinite programs via Jordan
  algebras'', Math.\ Program.\ 181 (2020), 51--84; doi:10.1007/s10107-019-01372-5.
\bibitem{rosset2021} D.~Rosset, F.~Montealegre-Mora, J.-D.~Bancal, ``RepLAB: a computational/numerical approach to
  representation theory'', in \emph{Quantum Theory and Symmetries}, CRM Series in Mathematical Physics, Springer, 2021,
  pp.~643--653; doi:10.1007/978-3-030-55777-5\_60; arXiv:1911.09154.
\bibitem{symbolicwedderburn} M.~Kaluba, T.~Weisser, SymbolicWedderburn.jl, Julia package,
  \url{https://github.com/kalmarek/SymbolicWedderburn.jl} (accessed October 2026).
\bibitem{goulart2026} P.~J.~Goulart, Y.~Chen, ``Clarabel: An interior-point solver for conic programs with quadratic
  objectives'', Math.\ Program.\ Comput.\ (2026); doi:10.1007/s12532-026-00320-7.
\bibitem{coey2023} C.~Coey, L.~Kapelevich, J.~P.~Vielma, ``Conic optimization with spectral functions on Euclidean
  Jordan algebras'', Math.\ Oper.\ Res.\ 48 (2023), no.~4, 1906--1933; doi:10.1287/moor.2022.1324.
\bibitem{morrison2012} S.~Morrison, N.~Snyder, ``Subfactors of index less than 5, Part 1: The principal graph
  odometer'', Commun.\ Math.\ Phys.\ 312 (2012), 1--35; doi:10.1007/s00220-012-1426-y.
\bibitem{jones2014} V.~F.~R.~Jones, S.~Morrison, N.~Snyder, ``The classification of subfactors of index at most 5'',
  Bull.\ Amer.\ Math.\ Soc.\ 51 (2014), 277--327; doi:10.1090/S0273-0979-2013-01442-3.
\bibitem{dedecker2025} A.~Dedecker, J.~Loreaux, ``The continuous functional calculus in Lean'', arXiv:2501.15639
  (2025).
\bibitem{wang2020} Z.~Wang, N.~C.~Rubin, J.~M.~Dominy, E.~G.~Rieffel, ``$XY$ mixers: Analytical and numerical results
  for the quantum alternating operator ansatz'', Phys.\ Rev.\ A 101 (2020), 012320; doi:10.1103/PhysRevA.101.012320.
\bibitem{streif2021} M.~Streif, M.~Leib, F.~Wudarski, E.~Rieffel, Z.~Wang, ``Quantum algorithms with local
  particle-number conservation: Noise effects and error correction'', Phys.\ Rev.\ A 103 (2021), 042412;
  doi:10.1103/PhysRevA.103.042412.
\bibitem{botelho2022} L.~Botelho, A.~Glos, A.~Kundu, J.~A.~Miszczak, \"O.~Salehi, Z.~Zimbor\'as, ``Error mitigation for
  variational quantum algorithms through mid-circuit measurements'', Phys.\ Rev.\ A 105 (2022), 022441;
  doi:10.1103/PhysRevA.105.022441.
\bibitem{cai2023} Z.~Cai, R.~Babbush, S.~C.~Benjamin, S.~Endo, W.~J.~Huggins, Y.~Li, J.~R.~McClean, T.~E.~O'Brien,
  ``Quantum error mitigation'', Rev.\ Mod.\ Phys.\ 95 (2023), 045005; doi:10.1103/RevModPhys.95.045005.
\bibitem{arute2020} Google AI Quantum and Collaborators (F.~Arute et al.), ``Hartree-Fock on a superconducting qubit
  quantum computer'', Science 369 (2020), 1084--1089; doi:10.1126/science.abb9811.
\bibitem{buca2012} B.~Bu\v{c}a, T.~Prosen, ``A note on symmetry reductions of the Lindblad equation: transport in
  constrained open spin chains'', New J.\ Phys.\ 14 (2012), 073007; doi:10.1088/1367-2630/14/7/073007.
\bibitem{albert2014} V.~V.~Albert, L.~Jiang, ``Symmetries and conserved quantities in Lindblad master equations'',
  Phys.\ Rev.\ A 89 (2014), 022118; doi:10.1103/PhysRevA.89.022118.
\bibitem{chen2018} R.~T.~Q.~Chen, Y.~Rubanova, J.~Bettencourt, D.~Duvenaud, ``Neural ordinary differential
  equations'', Advances in Neural Information Processing Systems 31 (2018), 6571--6583.
\bibitem{arjovsky2016} M.~Arjovsky, A.~Shah, Y.~Bengio, ``Unitary evolution recurrent neural networks'', Proc.\ 33rd
  Int.\ Conf.\ Machine Learning, PMLR 48 (2016), 1120--1128.
\bibitem{beer2020} K.~Beer, D.~Bondarenko, T.~Farrelly, T.~J.~Osborne, R.~Salzmann, D.~Scheiermann, R.~Wolf,
  ``Training deep quantum neural networks'', Nat.\ Commun.\ 11 (2020), 808.
\bibitem{hashimoto2022} Y.~Hashimoto, Z.~Wang, T.~Matsui, ``$C^*$-algebra net: a new approach generalizing neural
  network parameters to $C^*$-algebra'', Proc.\ 39th Int.\ Conf.\ Machine Learning, PMLR 162 (2022), 8523--8534.
\bibitem{stinespring1955} W.~F.~Stinespring, ``Positive functions on $C^*$-algebras'', Proc.\ Amer.\ Math.\ Soc.\ 6
  (1955), 211--216.
\bibitem{e2017} W.~E, ``A proposal on machine learning via dynamical systems'', Commun.\ Math.\ Stat.\ 5 (2017), 1--11.
\bibitem{he2016} K.~He, X.~Zhang, S.~Ren, J.~Sun, ``Deep residual learning for image recognition'', Proc.\ IEEE CVPR
  2016, 770--778.
\bibitem{barnum1996} H.~Barnum, C.~M.~Caves, C.~A.~Fuchs, R.~Jozsa, B.~Schumacher, ``Noncommuting mixed states cannot
  be broadcast'', Phys.\ Rev.\ Lett.\ 76 (1996), 2818--2821.
\bibitem{torlai2018} G.~Torlai, G.~Mazzola, J.~Carrasquilla, M.~Troyer, R.~Melko, G.~Carleo, ``Neural-network quantum
  state tomography'', Nat.\ Phys.\ 14 (2018), 447--450.
\bibitem{torlai2023} G.~Torlai, C.~J.~Wood, A.~Acharya, G.~Carleo, J.~Carrasquilla, L.~Aolita, ``Quantum process
  tomography with unsupervised learning and tensor networks'', Nat.\ Commun.\ 14 (2023), 2858.
\bibitem{toth2010} G.~T\'oth, W.~Wieczorek, D.~Gross, R.~Krischek, C.~Schwemmer, H.~Weinfurter, ``Permutationally
  invariant quantum tomography'', Phys.\ Rev.\ Lett.\ 105 (2010), 250403.
\bibitem{wolfcirac2008} M.~M.~Wolf, J.~I.~Cirac, ``Dividing quantum channels'', Commun.\ Math.\ Phys.\ 279 (2008),
  147--168.
\bibitem{schuld2021} M.~Schuld, ``Supervised quantum machine learning models are kernel methods'',
  arXiv:2101.11020 (2021).
\bibitem{wolf2008} M.~M.~Wolf, J.~Eisert, T.~S.~Cubitt, J.~I.~Cirac, ``Assessing non-Markovian quantum dynamics'',
  Phys.\ Rev.\ Lett.\ 101 (2008), 150402.
\end{thebibliography}
\end{document}